\documentclass[review,3p]{elsarticle}

\usepackage{amsfonts,amsmath,amssymb}
\usepackage{color}
\usepackage{graphicx}
\usepackage{dcolumn}
\usepackage{bm}
\usepackage{multirow}
\usepackage{soul}
\usepackage{comment}
\usepackage{lineno}
\usepackage{float}
\usepackage[colorlinks=true,citecolor=blue,linkcolor=blue,urlcolor=blue]{hyperref}
\usepackage{url,placeins}

\usepackage{longtable}
\usepackage{array}
\usepackage{booktabs}
\usepackage{tikz}

\usepackage{wrapfig}

\newcolumntype{M}[1]{>{\centering\arraybackslash}m{#1}}

\usepackage{pgfplots} 
\usetikzlibrary{arrows.meta,arrows,shapes,backgrounds,fit,decorations.pathreplacing,chains,snakes,positioning,angles,quotes,decorations.pathmorphing,decorations.markings}  
\usetikzlibrary{mindmap} 
\usepackage{subfig}
\usepackage{xcolor}
\usepackage{contour}
\usetikzlibrary{fadings}

\usepackage{color}                    			     
\definecolor{RoyalAzure}{rgb}{0.0, 0.22, 0.66}	
\usepackage{braket,graphicx}
\pgfplotsset{compat=newest}

\def\exd{\mathrm{d}}

\graphicspath{{imma/}}

\newcommand{\rC}{r_\text{\tiny C}}
\newcommand{\x}{{\bf x}}
\newcommand{\y}{{\bf y}}
\newcommand{\q}{{\bf q}}
\newcommand{\be}{\begin{equation}}
\newcommand{\ee}{\end{equation}}

\biboptions{sort&compress}
\begin{document}

\begin{frontmatter}

\title{Quantum Physics in Space}

\author[TUB,QUB]{Alessio Belenchia\corref{myequalauthor}}
\author[QUB,Trieste1,Trieste2]{Matteo Carlesso\corref{myequalauthor}}
\cortext[mycorrespondingauthor]{Corresponding author, abassi@units.it }
\cortext[myequalauthor]{These authors contributed equally to this work}
\author[Erl1,Erl2]{\"{O}mer Bayraktar}
\author[ASI]{Daniele Dequal}
\author[Pal]{Ivan Derkach}
\author[UAB,SOH]{Giulio Gasbarri}
\author[DLRSI,Han]{Waldemar Herr}
\author[UCL]{Ying Lia Li }
\author[UCL]{Markus Rademacher}
\author[SUPA]{Jasminder Sidhu}
\author[SUPA]{Daniel KL Oi}
\author[ADS]{Stephan T. Seidel}
\author[Liu,IQOQI]{Rainer Kaltenbaek}
\author[Erl1,Erl2]{Christoph Marquardt}
\author[SOH]{Hendrik Ulbricht}
\author[Pal]{Vladyslav C. Usenko}
\author[ZARM,DLR]{Lisa W\"{o}rner}
\author[Malta]{Andr\'{e} Xuereb}
\author[QUB]{Mauro Paternostro}
\author[Trieste1,Trieste2]{Angelo Bassi\corref{mycorrespondingauthor}}

\address[TUB]{Institut f\"{u}r Theoretische Physik, Eberhard-Karls-Universit\"{a}t T\"{u}bingen, 72076 T\"{u}bingen, Germany}
\address[QUB]{Centre  for  Theoretical  Atomic,Molecular,  and  Optical  Physics, School  of  Mathematics  and  Physics,  Queen’s  University, Belfast  BT7  1NN, United Kingdom}

\address[Trieste1]{Department  of  Physics,University  of  Trieste, Strada  Costiera  11,  34151  Trieste,Italy}
\address[Trieste2]{Istituto  Nazionale  di  Fisica  Nucleare, Trieste  Section,  Via  Valerio  2,  34127  Trieste, Italy} 

\address[Erl1]{Max Planck  Institute  for  the  Science  of  Light, Staudtstra\ss{}e  2,  91058  Erlangen, Germany}
\address[Erl2]{Institute  of  Optics, Information  and  Photonics, Friedrich-Alexander  University  Erlangen-N\"{u}rnberg, Staudtstra\ss{}e  7  B2,  91058  Erlangen, Germany}
\address[ASI]{Scientific Research Unit, Agenzia  Spaziale  Italiana,  Matera, Italy}
\address[Pal]{Department  of  Optics, Palacky  University, 17.    listopadu  50,772  07  Olomouc,Czech Republic}

\address[UAB]{F\'isica Te\`orica: Informaci\'o i Fen\`omens Qu\`antics, Department de F\'isica, Universitat Aut\`onoma de Barcelona, 08193 Bellaterra (Barcelona), Spain}
\address[SOH]{Department  of  Physics  and  Astronomy, University  of  Southampton, Highfield  Campus,  SO17  1BJ, United Kingdom}
\address[DLRSI]{Deutsches Zentrum für Luft- und Raumfahrt e. V. (DLR), Institut für Satellitengeodäsie und Inertialsensorik (SI), Vorläufige Anschrift: DLR-SI, c/o Leibniz Universität Hannover, Callinstraße 36, 30167 Hannover}
\address[Han]{Institut für Quantenoptik, Leibniz Universität Hannover, Am Welfengarten 1, 30167 Hannover, Germany}
\address[UCL]{Department  of  Physics  \&  Astronomy,University  College  London,  WC1E  6BT, United Kingdom}
\address[SUPA]{SUPA  Department  of  Physics, University  of  Strathclyde,  Glasgow,
United Kingdom}
\address[ADS]{Airbus Defence and Space GmbH, Robert-Koch-Straße 1, 82024 Taufkirchen}
\address[Liu]{Faculty  of  Mathematics  and  Physics, University  of  Ljubljana, Jadranska  ulica  19,  1000  Ljubljana,Slovenia}
\address[IQOQI]{Institute  for  Quantum  Optics  and  Quantum  Information,  Vienna, Austria}
\address[ZARM]{ZARM,  University  of  Bremen, Am  Fallturm  2,  28359  Bremen, Germany}
\address[DLR]{Deutsches Zentrum für Luft- und Raumfahrt e. V. (DLR), Institut für Quantentechnologie (QT), S\"{o}flinger Strasse 100, 89077 Ulm, Germany}
\address[Malta]{Department  of  Physics, University  of  Malta, Msida MSD~2080, Malta}

\begin{abstract}
Advances in quantum technologies are giving rise to a revolution in the way fundamental physics questions are explored at the empirical level. At the same time, they are the seeds for future disruptive technological applications of quantum physics. Remarkably, a space-based environment may open many new avenues for exploring and employing quantum physics and technologies. Recently, space missions employing quantum technologies for fundamental or applied studies have been proposed and implemented  with stunning results. The combination of quantum physics and its space application is the focus of this review: we cover both the fundamental scientific questions that can be tackled with quantum technologies in space and the possible implementation of these technologies for a variety of academic and commercial purposes. 
\end{abstract}

\begin{keyword}
Quantum Technology, Space, Quantum Physics
\end{keyword}

\end{frontmatter}

\tableofcontents
    \makeatletter
    \let\toc@pre\relax
    \let\toc@post\relax
    \makeatother

\section{Introduction}\label{introduction}

Space provides the means to enhance the potential of quantum technological platforms and to challenge some of the most fundamental open problems in modern physics.
Driven by remarkable recent technological progress,
 many proposals for space-based applications of quantum technologies have been put forward and implemented for academic and commercial purposes. 
The hybridisation between space science and quantum physics is  attracting a growing attention for the many possibilities it has to offer, from quantum-enhanced satellite-based communication and Earth observation, to the leverage of quantum sensing and interferometry, to the exploration of elusive physical phenomena, including dark matter and quantum gravity. This review addressed precisely such exciting opportunities for scientific and technological development, focusing on the three main physical platforms -- cold atoms, photonics and optomechanical setups -- that have been considered for space applications.

Cold atoms interferometry and atomic clocks are among the most sensitive quantum sensing devices for both applied and fundamental physics~\cite{RevModPhys.90.025008,bongs2019taking}. Experiments based on such platforms rely on the coherent property of quantum matter-waves and have been the subject of fast-paced progress in the last three decades, leading to results that have demonstrated unambiguous space readiness. Among them, it is worth mentioning the interference experiments on-board of sounding rockets~\cite{becker2018space} and the observation of Bose-Einstein condensation on the International Space Station~\cite{aveline2020observation}.

Photonic technology 
is central to tests of the foundations of quantum mechanics and the development of architectures for (secure) long-haul quantum communication. 
The proved potential of photonic systems has been recently used to demonstrate the viability of 
satellite-based quantum key and entanglement distribution~\cite{yin2017satellite,liao2017satellite}, which are key stepping stones towards the construction of a quantum internet. 

Quantum optomechanics 
holds the promise to be the enabling technology for the generation of quantum superposition states of macroscopic objects~\cite{delic2020cooling}. 
Several proposals have advanced the idea of performing quantum optomechanical experiments in space~\cite{kaltenbaek2012macroscopic} 
 in light of their sensitivity, and they are emerging as key components of future quantum sensing technologies.

Despite the developments of space-oriented applications of these platforms is only relatively recent and the achievement of full space-translation readiness requires further developments, the demonstration of their viability for space-based scientific and technological endeavours is emerging strongly and unquestionably. By providing a technical, science-oriented overview of the most significant achievements gathered through the use of these physical platforms, the future prospects for their implementations in space,
and their potential for testing fundamental physical questions beyond the capabilities of ground-based experiments, this review aims to embody a useful {\it compendium} for both the non-initiated and the expert reader. 

The paper is organised as follows. In Sec.~\ref{Fundamental} we consider the application of quantum technologies in space to fundamental physics studies: from the interface between gravity and quantum mechanics to quantum foundations, from the detection of gravitational waves to searches for dark matter and dark energy. We focus on the advantages provided to such quests by a space-based environment 
and describe the technological platforms that have been proposed to perform them so far. In Sec.~\ref{Applications} we discuss new  applications of current quantum technologies for the assessment of  fundamental-physics tests. We address the long sight-lines and reduced losses (compared with fibre-based arrangements) of free-space optical transmission lines, which are the driving forces for satellite-based quantum-communication applications. We  discuss how the vantage point of low Earth orbit is valuable for remote sensing and remote observation and then tackle space-based quantum clocks for distributed  time-reference frames and synchronisation towards enhanced global systems for navigation. 
In Sec.~\ref{ProofOfPrincipleImlementation} we  delve into state-of-the-art proof-of-principle quantum experiments and implementations of quantum technologies in space that have been performed to date.
Section~\ref{WishList} highlights the technical hurdles that should be faced to mature technology for space applications. Due to a wealth of prototyping activities -- led by the cold-atom and photonics community -- many of such challenges are no longer insurmountable. We describe how to build upon already interconnected scientific and industrial communities and which steps are needed to enable further cross-disciplinary networks and collaborations. Expanding the diversity of quantum experiments being performed in the promising conditions of reduced environmental effects and micro-gravity has clear benefits for the advances of fundamental understanding and technological developments. 
We provide a technical wish-list detailing the  steps forward that should be made to enable space-based experiments with quantum systems.
Section~\ref{Conclusions} puts forward a realistic outlook for the problems tackled in this review.   

\section{Fundamental Tests with Quantum Technologies in Space}\label{Fundamental}
\begin{figure*}[b!]
\centering
\includegraphics[width=0.8\linewidth]{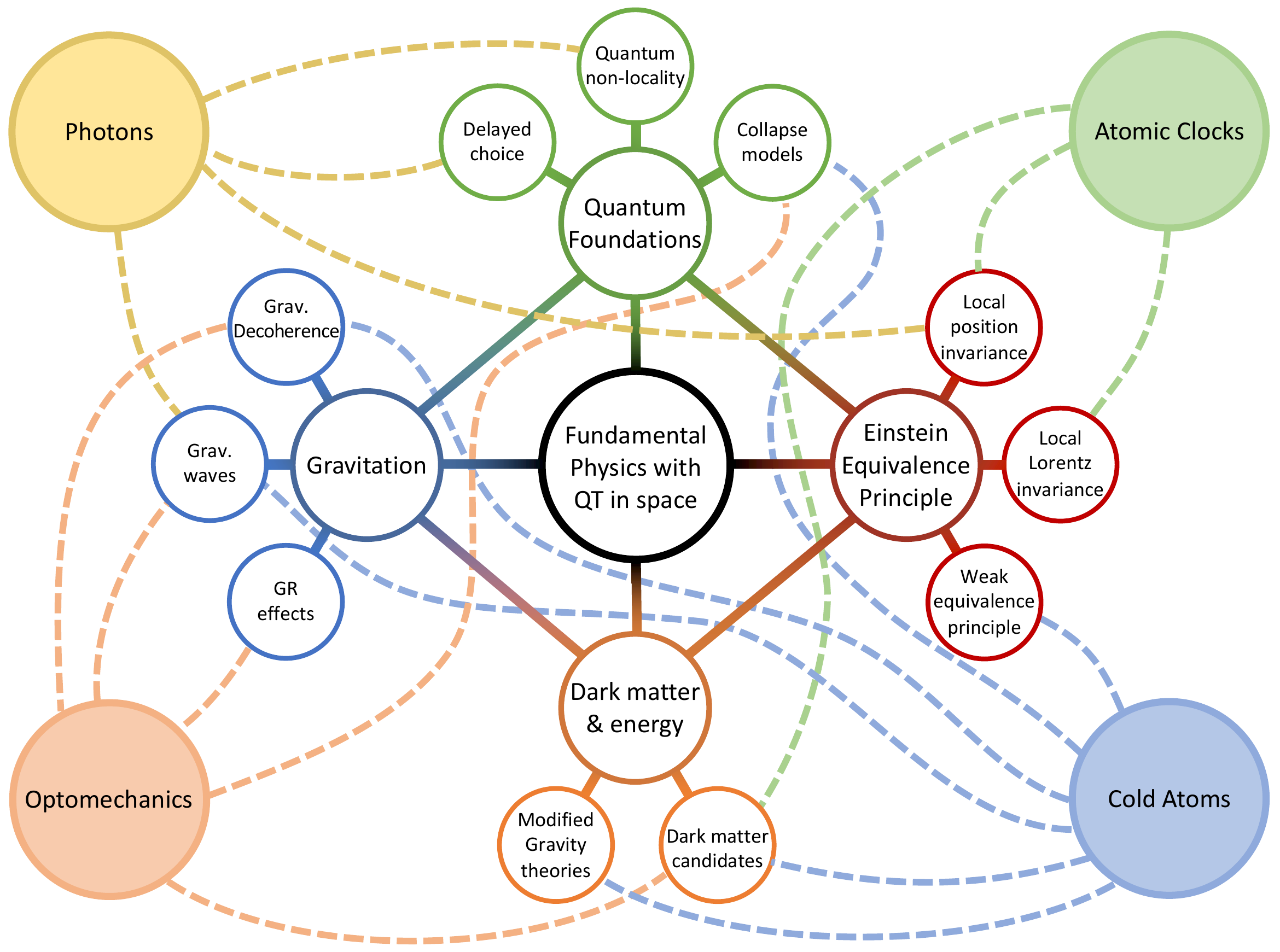}
\caption{Mindmap for Fundamental Physics tests with QT in space. Four main subjects are covered in this section: the Einstein Equivalence Principle, Dark Matter and Dark Energy, Gravitation, and Quantum Foundations. Each of these subjects contains different topics sprouting from them. Full circles represent the different quantum technological platforms that we consider in this review (cf. Sec.~\ref{Applications} and \ref{ProofOfPrincipleImlementation}). Their relation to the fundamental physics subjects described in this section is highlighted by the dashed lines.
}
\label{Sec2:FigTopics}
\end{figure*}

Space offers the ideal environment to fully employ the potential of quantum platforms and challenge some of the most fundamental open problems in modern physics (see Fig.~\ref{Sec2:FigTopics}). Recently, many proposals for space-based fundamental physics tests resorting to the possibilities offered by quantum mechanics have appeared driven by incredible technological advancements in different areas of physics. In this section, we review the plethora of different fundamental questions which can be addressed by combining space science and quantum technologies. We will focus on the advantages in performing fundamental tests in space, with respect to analogous ground-based ones, and on which platforms have been proposed to perform them. 

\subsection{Einstein Equivalence Principle}
The equivalence principle, whose history dates back to Galileo and Newton, is at the basis of the modern geometrical description of spacetime epitomized by Einstein's General Relativity (GR)~\cite{will2018theory}. In modern terms, the Einstein Equivalence Principle (EEP) is a combination of three different ingredients (see also~\cite{will2006confrontation,will2018theory})
\begin{itemize}
\item\textbf{Weak Equivalence principle (WEP):} The equivalence between the gravitational and inertial mass of a system, independently of its composition, has far-reaching physical consequences and constitutes the weak equivalence principle. An alternative statement of the WEP is that the trajectory of a freely falling test particle\footnote{The definition of test body is in general not a trivial matter~\cite{giulini2012equivalence}. In the following, if not specified otherwise, we consider only objects with negligible self-gravitational effects.} is independent of the composition of the particle itself. In the paradigmatic example of weights dropped from a tower, the WEP implies that their acceleration is the same; This is the so called universality of free fall. 
\item\textbf{Local Position Invariance (LPI):} The results of any non-gravitational experiment are independent on where and when such experiments are performed in the Universe.  
\item\textbf{Local Lorentz Invariance (LLI):} The results of any non-gravitational experiment are independent of the velocity of the local freely-falling frame in which they are performed.
\end{itemize}
The EEP has the crucial implication that, locally, the gravitational field is indistinguishable from a uniform acceleration. Far from being only a curiosity, it suggests that we should interpret gravity not as a force 
as an effect of the curvature of spacetime~\cite{PhysRevD.7.3563,will2018theory}. Indeed, the EEP is satisfied not only by GR but also by a wide range of gravitational theories, namely the {metric theories of gravity}~\cite{belenchia2016higher}. In these theories, as for GR, gravity has a geometrical interpretation as the effect of the curvature of the spacetime manifold. More in general, these theories are based on the assumptions of 
having a symmetric metric $g_{\mu\nu}$ endowing spacetime and 
that test bodies follow the geodesics of this metric. Moreover, in these theories the laws of physics reduce to those of special relativity in local freely-falling reference frames. 

\paragraph{Strong Equivalence Principle} 
The EEP is formulated assuming 
non-self-gravitating test bodies and non-gravitational experiments. The extension to self-gravitating test-bodies and non-gravitational experiments of the EEP is referred to as the strong equivalence principle. Whereas, the extension to self-gravitating test-bodies of the sole WEP is sometimes referred to as the gravitational WEP. While the EEP is respected by all metric theories of gravity, the strong equivalence principle is more restrictive. Thus, its tests can provide additional information on the underlying gravitational theory~\cite{sotiriou2008theory,di2015nonequivalence}. 

\paragraph{Quantum Equivalence principle}

The EEP also assumes 
classical test-bodies. The discussion about the validity and formulation of the EEP 
in the case of quantum systems is the  
subject of current theoretical and experimental investigations~\cite{greenberger1968role,davies1982quantum,lammerzahl1996equivalence,viola1997testing,alvarez1997testing,adunas2001probing,davies2004quantum,kajari2010inertial,okon2011does,schlippert2014quantum,rosi2017quantum,giacomini2020einstein,albers2020quantum}. 
This is particularly relevant for 
tests of the EEP employing quantum systems \cite{altschul2015quantum}, as well as for regimes where one expects quantum effects to become relevant. 
A quantum formulation of the EEP was introduced and analyzed in~\cite{zych2018quantum}, furnishing a phenomenological 
framework for tests of the quantum EEP. It is crucial to notice that 
experiments with quantum objects are a necessary but not sufficient condition for testing the quantum version of the EEP. Nevertheless, proposals in this direction are already present~\cite{orlando2016test,geiger2018proposal}, and a first experiment~\cite{rosi2017quantum} employing atom-interferometry has been performed bounding violations of the quantum EEP to the order of one part in $10^{8}$. 

\subsubsection{Experimental tests and proposals for space}
Given the deep physical implications of the EEP for our understanding of the structure of spacetime and of gravity, it is not surprising that considerable efforts were made to experimentally verify this principle. Experimental tests in favour of the validity of the EEP at the classical level are quite a few~\cite{will2006confrontation}. However, the ongoing efforts to cast more stringent constraints and to search for possible violations are still numerous. In particular, the search for violations of the EEP is mainly driven by their implications for new physics beyond GR and the Standard Model of particle physics (SM). Indeed, several models of quantum gravity predict violations of the EEP in one or more of its components~\cite{amelino2013quantum}. Models of dark matter and dark energy contain violations of the EEP due to the non-universal coupling of SM's fields with those originating dark matter and dark energy~\cite{wolf2019exploring}. Violations of the EEP are also present, by construction, in all proposals with varying fundamental constants motivated by high-energy physics' fine-tuning problems~\cite{uzan2011varying}. 
It is clear that tests of the EEP and of its violations have far-reaching ramifications in different fields of modern physics and offer the possibility to unveil new physics beyond GR and the SM. As we will see, technological advancements and  quantum technologies in space play a prominent role in this endeavour.  

Tests of the EEP are usually designed to probe one or more components of the principle. Thus, in the following, we report the current constraints on EEP violations following this same logic. We then delve into the panorama of proposals for quantum experiments in space.

\paragraph{Tests of WEP}
The weak equivalence principle can be directly tested by comparing the accelerations of two test bodies with different composition in an external gravitational field. This is the equivalent of Galileo's alleged idea of throwing boulders from the Leaning Tower of Pisa. Possible violations of the WEP are then parametrised by the E\"otv\"os ratio $\eta=2|a_1-a_2|/|a_1+a_2|$, which is the fractional difference between the accelerations $a_i$ ($i=1,2$) of the two bodies. 

The current upper bound on $\eta$ from ground-based experiments has been established in 2012 at $\eta\leq10^{-13}$ with a torsion balance experiment employing Beryllium-Titanium test-bodies~\cite{Wagner_2012}. In 2012 a similar bound was obtained 
employing the results from the Laser Lunar Ranging (LLR) experiment, which measured the differential acceleration between the Earth and the Moon with respect to the Sun~\cite{williams2012lunar}. However, the most stringent upper bound on the violations of the WEP to date comes from space, where one can exploit the possibility to compare the gravitational acceleration of two free falling test-bodies in orbit around the Earth for a very long time. The space mission MICROSCOPE~\cite{touboul2001microscope,touboul2020microscope} of the French space agency CNES, managed in two years of operation to push this upper bound to $\eta\leq1.3\times 10^{-14}$. This bound results from the use of Titanium-Platinum test-bodies, a reference pair of Platinum-Rhodium alloy masses, and the use of data recorded in 120 orbits with free-falling times of up to 8 days. However, this is still a preliminary analysis, which employs only $7\%$ of the data collected before the mission was decommissioned in late 2018. Further improvements are expected from the full analysis of the data~\cite{PhysRevLett.119.231101,Touboul_2019,touboul2020microscope}. 

Future endeavours to test of the WEP are expected to reach bounds which are two or three orders of magnitude better that what MICROSCOPE was designed to achieve, thus pushing the bound to around $\eta\leq10^{-17}\sim10^{-18}$~\cite{wolf2019exploring}. The Satellite Test of the Equivalence Principle (STEP)~\cite{mester2001step,kolodziejczak2007satellite,sumner2007step,overduin2009science} uses a similar concept to that of MICROSCOPE, with the target of bounding the WEP to $\eta\leq10^{-18}$ by employing four pairs of test masses instead of two. Similarly to STEP, the proposal for the small satellite Galileo-Galilei (GG)~\cite{nobili2012galileo,nobili2000galileo} and its proof-of-principle version on ground~\cite{nobili2000galileo} aim at bounds on the WEP at the  level of $\eta\leq10^{-17}$  using concentric cylindrical test masses~\cite{nobili2012galileo}.

\begin{figure}
\centering
    \includegraphics[width=0.6\linewidth]{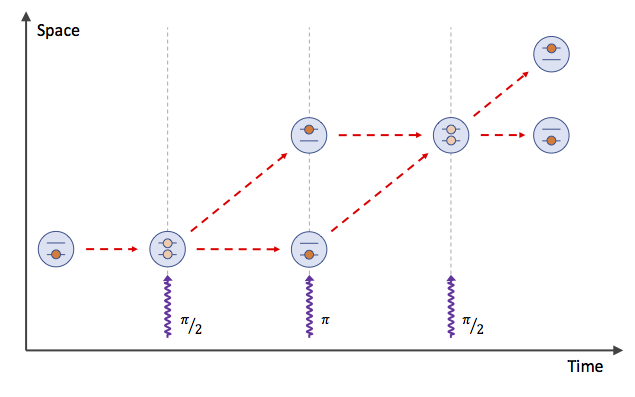}
    \caption{
   Simplified schematics of the typical atom interferometer, which mimics the functioning of an optical Mach-Zender setup. An initial $\pi/2$-pulse places the atom wave packet in a superposition of two different momenta by momentum transfer, and thus into a spatial superposition of two different trajectories. Successive additional $\pi$-pulses act like mirrors allowing to bring the trajectories back together. The trajectories are finally recombined by a second $\pi/2$-pulse acting as the final beam-splitter in the Mach-Zender interferometer. Figure adapted from \cite{hinton2017portable}.
    } \label{fig:inter}
\end{figure}

Dual species atom interferometry appears as the natural candidate for tests of the WEP employing quantum systems, and thus also possibly for exploring the interface  between quantum mechanics and gravitation. The typical atom interferometer mimics the functioning of an optical Mach-Zender interferometer by replacing the optical beam-splitters and mirrors with light pulses, as it is schematically represented in Fig.~\ref{fig:inter}. 

The populations at the two outputs of the interferometer encode the phase difference accumulated along the two trajectories. This allows to extract the free-fall acceleration of the  matter-waves with  respect to the laser source. Using different atomic species it is thus possible to compare the free-fall accelerations  and  constrain violations of the WEP via the E\"{o}tv\"{o}s parameter. This is the main working principle of several experiments performed on ground aiming at testing the WEP with quantum probes~\cite{fray2004atomic,PhysRevLett.98.111102,geiger2011detecting,schlippert2014quantum,PhysRevA.88.043615,PhysRevLett.115.013004,kuhn2014bose,PhysRevLett.120.183604,PhysRevLett.125.191101,tino2021testing} 
and of several space experiments in which we will now delve. More details on the specific experiments will be given in Sec.~\ref{ProofOfPrincipleImlementation}.

The panorama of space proposals that have the possibility to test the WEP (or other aspects of the EPP) with atom interferometry is currently vast.
Space experiments, both those in dedicated satellite missions and in missions exploiting the microgravity environment of existing satellites as the International Space Station (ISS), offer incredible perspectives for improving ground-based constraints due to the quadratic scaling of atom interferometry sensitivity to inertial accelerations with the free falling time~\cite{tino2013precision}. Increasing the free-falling time on ground-based experiments comes with the need for longer atomic paths and at the expense of the control of the experiment.
Such a problem is not present in space-based experiments where the atoms can free-fall inside a satellite which is also in free-fall. This is a common advantage of the space environment for free-falling experiments. 
It is thus not surprising that several efforts have been focused to make atom interferometry technology mature enough for the space environment.
The European Space Agency (ESA) founded the Space Atom Interferometer (SAI) project~\cite{sorrentino2010compact,sorrentino2011space}, which was aimed at demonstrating the possibility  of applying atom interferometry technologies in space missions and developed a compact, transportable atom interferometer prototype for space, based on Rubidium-87. The CNES-funded ICE~\cite{nyman2006ice}  
operated an atom interferometer for inertial sensing in reduced gravity on-board the NOVESPACE Zero-G plane~\cite{zerog}. The DLR-founded QUANTUS
developed preliminary studies for space-borne missions using the ZARM drop-tower in Bremen as microgravity environment from 2004 to 2013 (QUANTUS I)~\cite{QUANTUS} and from 2013 to 2018 (QUANTUS II)~\cite{rudolph2015high}. Later, QUANTUS moved to sounding rocket experiments with MAIUS (QUANTUS III-IV)~\cite{QUANTUS4} bridging the gap between drop-tower experiments and proper space missions with a first successful launch in 2017. The Atom Interferometry Test of the Weak Equivalence Principle in Space (Q-WEP) aims at testing the WEP with atom interferometry on board of the ISS, by exploiting the microgravity environment of the Columbus module~\cite{tino2013precision}. 
The  study, performed between 2012 and 2013, showed the prospect for tests  of the WEP at the level of $\eta\leq10^{-14}$ using Rubidium-85 and Rubidium-87 isotopes. Also the proposal Quantum Test of the Equivalence principle and Space Time (QTEST), to be implemented on the ISS with a similar concept to Q-WEP~\cite{williams2016quantum}, aimed at bounds of the WEP up to $\eta\leq10^{-15}$. This would  mean to arrive at the same level of the current MICROSCOPE data but by employing quantum systems and improving the current ground-based record for testing the WEP violation using atom interferometry~\cite{PhysRevLett.125.191101} by three orders of magnitude\footnote{It should be noted that, currently, also ground-based experiments are aiming at test beyond $\eta\sim10^{-15}$~\cite{tino2021testing}.}.
The first microgravity cold atom experiment on the ISS is the NASA-founded Cold Atom Laboratory (CAL)~\cite{elliott2018nasa}, which was successfully launched to the ISS in 2018, and complemented in 2020 by the science module SM3 which will allow for atom interferometry experiments. Furthermore, the NASA-DLR Bose-Einstein Condensate and Cold Atom Laboratory (BECCAL)~\cite{frye2021bose}, a new multi-user facility capable of performing experiments with Rubidium and Potassium cold atoms and Bose Einstein Condensates (BEC) aboard the ISS, is expected for launch in 2021~\cite{BECCAL}. 

In addition to experiments performed or proposed for drop-towers, sounding rockets and the ISS, there are also proposals for atom interferometry in dedicated satellite missions. The Space-Time Explorer and Quantum Equivalence Principle Space Test (STE-QUEST)~\cite{aguilera2014ste, altschul2015quantum, schuldt2015design} is a proposal 
to test different aspects of the EEP by employing 
dual atom interferometry with Rubidium-${85}$ and  Rubidium-${87}$ isotopes cooled down below the critical temperature for Bose-Einstein condensation (around few nK) and thus operating with  degenerate quantum gases. STE-QUEST forecasts the possibility to constraint the E\"{o}tv\"{o}s parameter to $\eta\leq10^{-15}$. The Space Atomic Gravity Explorer (SAGE)~\cite{tino2019sage} mission, which was proposed in 2016 to ESA in response to a call for ``New Ideas'', is based on the use of ultracold Strontium atoms for atomic clocks and atom interferometry and aims to use atomic interferometry for tests of the WEP. 
In this context we can mention also the GrAnd Unification and Gravity Explorer (GAUGE)~\cite{amelino2009gauge} proposal, which aims at combining experiments with macroscopic test-masses and atom interferometry for exploring a wide range of fundamental topics including measurements of the E\"{o}vt\"{o}s parameter at the  level of $\eta\sim 10^{-18}$. 
Finally, the Atomic Experiment for Dark Matter and Gravity Exploration (AEDGE)~\cite{abou2020aedge} and the Search for Anomalous Gravitation using Atomic Sensors (SAGAS)~\cite{wolf2009quantum} contemplate atom interferometers on-board with tests of the WEP as one of their aims.
  
\paragraph{Test of LPI} 
There are two main ways to test LPI: redshift experiments which test for the spatial dependence of non-gravitational experiments, and test of the time-constancy of fundamental constants, which test for the temporal dependencies~\cite{haugan2001principles,will2018theory,will2006confrontation}. 
In particular, tests of the constancy of fundamental constants -- like the fine structure constant, the weak interaction constant, and the electron-proton mass ratio -- have been performed by looking both at the present rate of variation and comparing the present value of these constants with that in the distant past. While the former strategy can be accomplished by comparing highly stable clocks of different kind, the latter requires to infer the values of fundamental constants by measuring relic signals from distant past processes and comparing them with current values~\cite{will2006confrontation,uzan2011varying}.  

Gravitational redshift experiments allow to test for spatial dependency violation of LPI. The gravitational redshift effect epitomises the relativity of time as described by GR. However,  this effect is not a peculiarity of GR, but stems from the EEP and thus is a feature of all metric theories of gravity. The paradigmatic gravitational redshift experiment considers the frequency shift $\Delta\nu$ between two identical frequency standards, namely clocks, positioned at different gravitational potentials. Assuming the EEP to be valid, it is not possible to locally distinguish between uniform acceleration and an external gravitational field. This simple observation is enough to derive the gravitational redshift expression $\Delta\nu/\nu=\Delta U/c^2$  from the standard Doppler effect formula in the weak-gravity limit of GR~\cite{1997gr.qc....12019C}.
Here,  $c$ is the speed of light and $\Delta U$ is the difference in the Newtonian potential between the two standards. If LPI is violated, and in particular the rate of the clocks is allowed to depend on their locations in the local freely-falling frames instantly at rest with them, then $\Delta\nu/\nu=(1+\alpha_\text{\tiny LPI})\Delta U/c^2$, where now $\alpha_\text{\tiny LPI}$ parametrises the violations of LPI. Several measurements of the gravitational redshift have been performed since the first series of experiments carried out by Pound, Rebka, and Snider in 1960-1965~\cite{PhysRevLett.3.439,PhysRevLett.4.337,PhysRevLett.13.539}. The Gravity Probe A (GPA) mission in 1976, and the successive analysis of the data, casts an upper bound $|\alpha_\text{\tiny LPI}|\leq1.4\times 10^{-4}$~\cite{vessot1979test,vessot1989clocks}. This experiment used a hydrogen maser on board a sounding rocket, which reached a height of 10,000\,km and was compared with one on the ground. In 2014, a typical case of {serendipity} happened when the Galileo satellites GSAT-0201 and GSAT-0202  of the European Global Navigation Satellite System (GNSS) Galileo, boarding an hydrogen maser, were unintentionally lunched into eccentric orbits. This allowed to improve over the LPI upper bound with respect to the GPA mission, furnishing the current strongest upper bound\footnote{Notice that there exist also ``null'' redshift experiments which cast bounds on the modulus of the difference between the parameters $\alpha_\text{\tiny LPI}$ of clocks with different compositions but evolving on the same trajectory~\cite{will2006confrontation}.} of $|\alpha_\text{\tiny LPI}|\leq2\times 10^{-5}$~\cite{delva2015test,herrmann2018galileo,PhysRevLett.121.231101}. 

The next leap in the test of LPI is expected from the proposed mission Atomic Clock Ensemble in Space (ACES)~\cite{cacciapuoti2009space,meynadier2018atomic}, which aims at improving the bound on redshift measurements to around $|\alpha_\text{\tiny LPI}|\sim3\times 10^{-6}$. ACES is an ESA mission scheduled to launch in 2021~\cite{ACES,pharao} and flown on the Columbus module of the ISS where it will operate highly stable atomic clocks in microgravity environment~\cite{cacciapuoti2006aces,cacciapuoti2009space,meynadier2018atomic}. The  ACES payload will include the space Hydrogen maser (SHM) and the cold Cesium atom clock PHARAO (Projet d'Horloge Atomique par Refroidissement) achieving fractional frequency stability of $\sim10^{-16}$, as well as a microwave link. The latter will ensure accurate time and frequency transfer for direct clock comparison, both space-to-ground and ground-to-ground. This is a crucial step to overcome the current limitations in the comparison of ground-based atomic clocks at intercontinental distances, currently exploiting navigation and communication satellites. 
ACES will test both LPI and LLI. In particular, it will measure the constancy in time of the fine structure constant $\alpha$ with a target of $\alpha^{-1} \exd \alpha/\exd t<10^{-16}/\text{year}$, furnishing a direct test of LPI. Another test of LPI will be achieved with measurements of gravitational redshift, comparing ACES ultrastable atomic clock with atomic clocks on ground, with the target of constraining the violation parameter to $|\alpha_\text{\tiny LPI}|\leq3\times 10^{-6}$.

Ultrastable optical atomic clocks, combined with high-performance time/frequency links, open the way to accurate tests of LPI and 
the panorama of proposals employing them is vast~\cite{gill2008optical}.  The proposals for the Space  Optical  Clock SOC and I-SOC were submitted in 2005 in response to a call for scientific experiments on the ISS. The former was funded  by ESA in 2006-2009 and followed by the EU-FP7 project SOC2 in 2010-2015. The I-SOC mission~\cite{isoc} is the natural follow up to ACES and aims at operating an optical clock on the ISS by 2023 with performance improved by a factor of at least one order of magnitude in both clock and frequency link with respect to ACES~\cite{bongs2015development}. Among the primary scientific objectives, I-SOC aims at measurements of the gravitational redshift at the level of $\Delta\nu/\nu\sim2\times 10^{-7}$, $ 10^{-6}$ and $2\times 10^{-4}$ in  the Earth, Sun and Moon fields respectively. 

The Einstein  Gravity  Explorer  (EGE)  mission~\cite{schiller2009einstein} proposal was submitted in 2007 to ESA in the framework of ESA's Cosmic Vision program. It aims at a measurement of the gravitational redshift at the  level of $\Delta\nu/\nu\sim2.5\times 10^{-8}$ in the Earth field and  $10^{-6}$ in that of the Sun, and a measurement of the  space and time variability of fundamental constants with $3\times 10^{-8}$ accuracy. Another proposal submitted in 2007 to ESA in the framework of ESA's Cosmic Vision program is the SAGAS~\cite{wolf2009quantum}, which aims at flying  highly sensitive atomic sensors on a Solar System escape trajectory. SAGAS envisages employing a Strontium$^+$ ion atomic clock with a fractional frequency stability of $10^{-17}$ and aims at measuring  solar gravitational redshift at the $|\alpha_\text{\tiny LPI}|\sim 10^{-9}$ level. This mission will also constrain the spatial and temporal variations of the fine structure constant. 

\begin{figure*}[t!]
\centering
\includegraphics[width=\textwidth]{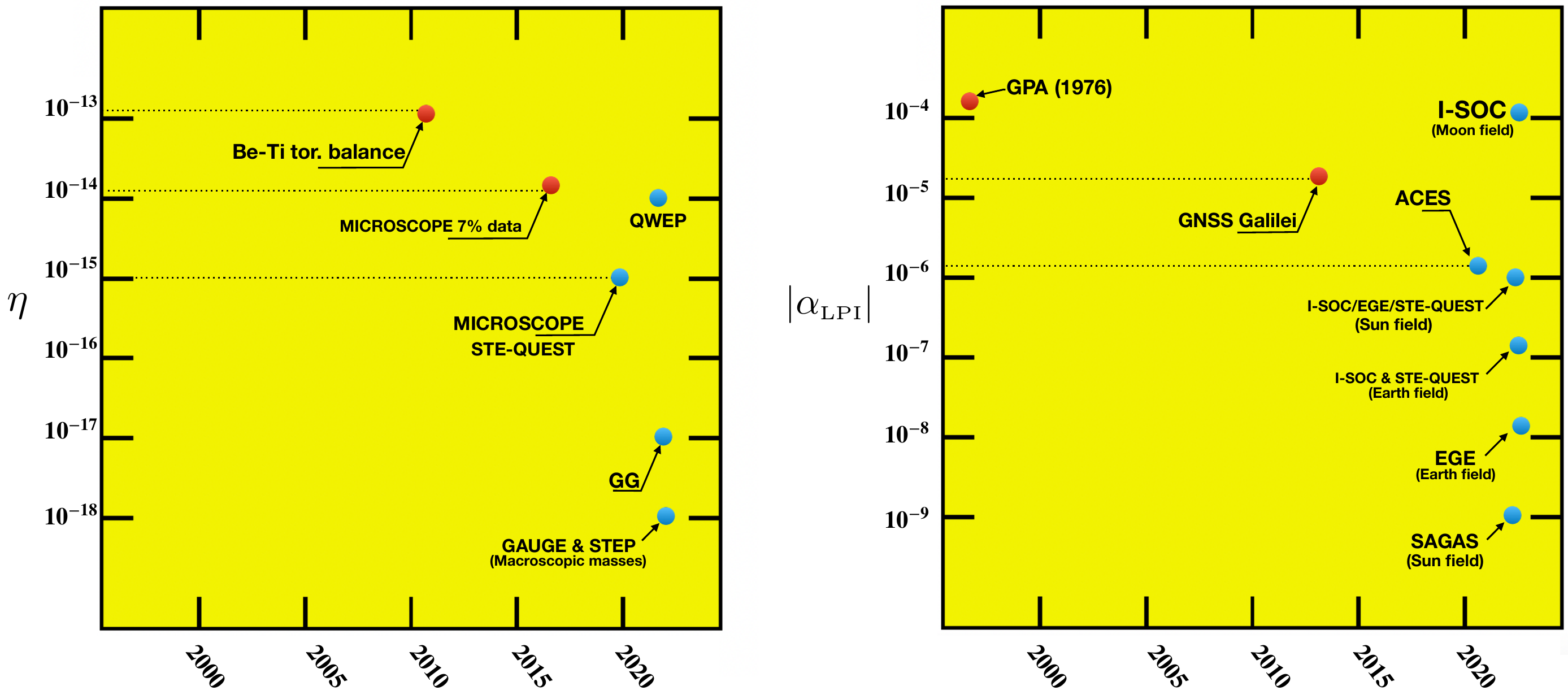}
\caption{Summary of bounds on the WEP and LPI reported in the text. The red dots refer to test of the WEP (left panel) and LPI (right panel) via gravitational redshift which has been performed. Conversely, the blue dots represent the prospects of the various space proposals considered in the text.}\label{Sec2:FigTestEEP}
\end{figure*}

The STE-QUEST and SAGE proposals will possibly include atomic clocks (the successor of PHARAO in  the case of STE-QUEST) and microwave links for comparison between ground-based clocks in their payloads. STE-QUEST aims to constraints of $\alpha_\text{\tiny LPI}\leq 10^{-6}$ via redshift measurements in the Sun gravitational field by direct comparison of ground-based clocks. Moreover, the possible inclusion of the on-board atomic clock would allow STE-QUEST to improve on ACES prospected constraints by one order of magnitude, arriving at around $|\alpha_\text{\tiny LPI}|\sim2\times 10^{-7}$ for the redshift measurements in the Earth's field. SAGE will also include Strontium atomic clocks, for which a fractional frequency stability of $10^{-18}$ has been demonstrated~\cite{nicholson2015systematic}. This should allow improving on the results achievable with ACES for the measurement of the gravitational redshift by several orders of magnitude. 
An overview of the bounds on $\alpha_\text{LPI}$ is given in Fig.~\ref{Sec2:FigTestEEP}.

\paragraph{Tests of LLI} 
Tests of LLI 
are effectively fundamental tests of the modern apparatus of theoretical physics, from special relativity up to the standard model of particle physics, which are both based on Lorentz symmetry. Quite recently, an enormous theoretical and experimental effort has been poured in the search for violations of Lorentz Invariance (see~\cite{liberati2013tests,mattingly2005modern,RevModPhys.83.11} and references therein). 
Quantum gravity is quite relevant here, since several models~
\cite{ellis2004synchrotron,RevModPhys.73.977,PhysRevD.59.124021,urrutia2006corrections,mavromatos2007lorentz,amelino2001testable,PhysRevD.84.084010} directly or indirectly imply high-energy violations of Lorentz Invariance, often linked to the assumed discreetness of spacetime\footnote{In some quantum gravity models, fundamental spacetime discreteness can coexist with Lorentz symmetry to a certain extent~\cite{bombelli2009discreteness,bombelli1987space,PhysRevD.83.104029}, which was also studied from a phenomenological perspective~\cite{sorkin2009does,belenchia2015nonlocal,PhysRevLett.116.161303,PhysRevD.94.061902}.}. 

The easiest way to characterise Lorentz Invariance Violations (LIV) is via a kinematical framework, namely the Standard Model Extension (SME) \cite{colladay1998lorentz},  in which Lorentz violating corrections are added to the relativistic particles dispersion relation formula
\begin{equation}
E^2=c^4m^2+c^2p^2+c\,E_\text{\tiny{P}}f^{(1)}_ip^i+c^2f^{(2)}_{i,j}p^ip^j+c^3\frac{f^{(3)}_{i,j,k}}{ E_\text{\tiny{P}}}p^ip^jp^k,
\end{equation}
where $c$ is the speed of light, $m$ is the mass of the particle, $p^i$ are the components of its momentum, the Planck energy $E_\text{\tiny P}$ has been factor out,  the dimensionless coefficients $f^{(i)}$ parametrise the LIV, and we used the Einstein's summation convention. A dynamical framework for the characterisation of LIV is also present. In the SME~\cite{colladay1998lorentz} an effective field theory approach is employed in order to parametrise all possible Lorentz violating operators emerging in the various sectors of the SM when Lorentz symmetry is broken.   
At present a variety of ground-based, astrophysical, and cosmological observations~\cite{jacobson2003strong,mattingly2005modern,RevModPhys.83.11}, as well as theoretical arguments~\cite{PhysRevLett.93.191301,belenchia2016lorentz}, have cast stringent constraints on LIV in the matter sector of the SM. Less stringent constraints are available in the gravitational sector, where a variety of models introduced violations yet to be proved or disproved. It should be noted that, in this case, LLI is extended to gravitational experiments -- so that the strong equivalence principle is probed -- and it is thus intimately related to the validity of Local Lorentz symmetry~\cite{mattingly2005modern,liberati2013tests}.

Also for tests of LLI space offers interesting possibilities. The data obtained by the  MICROSCOPE experiment have been already used to cast improved constraints on some SME parameters \cite{2017arXiv170511015P} and atomic sensors have been proposed as an effective way to test LLI in the framework of the SME~\cite{PhysRevD.80.016002,muller2008atom,gill2008optical,wolf2009space}, thus offering the chance to improve current bounds. Different type of LLI tests can be performed in this context. Optical clocks can be used to test the independence of the speed of light $c$ on the velocity of the source and orientation of the light path. Following the notation in~\cite{gurlebeck2018boost}, these violations can be easily parametrized using the Robertson-Mansouri-Sexl test theory~\cite{RevModPhys.21.378,mansouri1977test,mansouri1977test2,mansouri1977test3} as
\begin{equation}\label{RMS}
    \frac{c(\theta,\vec{v})}{c_0}=1+(\beta-\alpha-1)\frac{\vec{v}^2}{c_0^2}+\left(\frac{1}{2}-\beta+\gamma\right)\frac{\vec{v}^2}{c_0^2}\sin\theta^2+\mathcal{O}\left(\left|\vec{v}^3/c_0^3\right|\right),
\end{equation}
where $\vec{v}$ is the velocity of the experiment's rest frame and $\theta$ the orientation of the light's path both with respect to a preferred frame where the speed of light is isotropic and equal to $c_0$. Special relativity predicts that the parameters $\alpha$, $\beta$ and $\gamma$ take the following values: $\alpha=-\beta=-1/2$ and $\gamma=0$. Constraints on these coefficients can come from different experiments, which can access different combinations of these violation coefficients, and can also be reinterpreted in the modern language of the SME. 

The BOOst Symmetry Test (BOOST) mission~\cite{gurlebeck2018boost}, which was proposed as  a future DLR small satellite mission, will perform Kennedy-Thorndike experiment~\cite{hils1990improved} type measurements, among others. It will cast constraints on the independence of the speed of light on the velocity of the source as encoded in $\alpha_\text{\tiny KT}=\beta-\alpha-1$. This will be possible by comparing two highly stable frequency references -- an optical cavity and an Iodine frequency reference -- and will improve bounds on the LIV parameters of the electron sector of the SME. Also the EGE and the STE-QUEST proposals envisage this kind of measurements by comparing a highly  stable  local optical  cavity frequency and a ground clock frequency. These space experiments offer the advantages of an high orbital velocity $\vec{v}$ and strongly reduced cavity deformations due to microgravity~\cite{lammerzahl2001optis,scheithauer2006analytical}. Both proposals also include a measure of the independence of the Zeeman splitting frequency between two levels of the atomic species in the clock, in the direction of an applied  static magnetic field which is applied continuously in the clock, namely a Hughes-Drever type experiment~\cite{gurlebeck2018boost}. These tests promise to improve constraints on  LLI in the matter sector of the SME. The ACES mission will also measure the propagation delays of electromagnetic signals between the ISS and clocks on the ground, similarly to what done in~\cite{wolf1997satellite}, thus casting constraints on deviations from special relativity of the order of $\delta c/c\leq10^{-10}$~\cite{cacciapuoti2006aces}. 
Finally, SAGAS envisions a Ives-Stilwell experiment~\cite{gurlebeck2018boost}, which is a direct test of time dilation,
by measuring the frequency difference between space and ground-based clocks. 
This kind of measurement casts  constraints on the photon sector of the SME~\cite{wolf2009space}, thus testing special relativity and LLI, and it corresponds to a bound on $\alpha_\text{\tiny IS}=\alpha+1/2$ in Eq.~\eqref{RMS}.

\subsection{Searches for Dark Matter and Dark Energy}
Dark matter (DM)~\cite{RevModPhys.90.045002} and dark energy (DE)~\cite{RevModPhys.75.559,huterer2017dark} are arguably among the  greatest  open problems in cosmology. Cosmological and astrophysical observations point towards the fact that the SM's matter fields constitute only around the $5\%$  of the energy-matter content of the Universe, with the $68\%$ being accounted for by dark energy and the remaining $27\%$ by the so called dark matter. While the existence of dark matter can be currently inferred only via its gravitational effects, several observations point towards the fact that it could actually be some new form of matter~\cite{bertone2005particle}, constituting thus the $84\%$ of the matter content of the Universe\footnote{It should be mentioned that alternatives to the particle interpretation of dark matter exist. The major one is the  Modified Newtonian Dynamics~\cite{milgrom1983modification,famaey2012modified}, which was proposed in 1983~\cite{milgrom1983modification} and rests on the assumption that gravity is modified at large scales to justify the astrophysical and cosmological observations attributed to dark matter.}. These observations make the research for DM and DE an interdisciplinary field lying between cosmology and high-energy physics, and constitute an important share of the research for physics beyond the SM. 

To date, the nature of DM and DE remains elusive. Extensive searches~\cite{strigari2013galactic} for massive dark matter particles, with a combination of direct and indirect detection strategies and from astrophysical, cosmological, and high-energy physics experiments,
have not led to the detection of possible candidates in the family of the so-called weakly interacting massive particles (WIMPs) with masses in the GeV-TeV range. Nonetheless, given the unknown nature of dark matter, many theoretical proposals for DM candidates have been advanced over the years~\cite{RevModPhys.90.045002}. In particular, a growing interest is mounting for light DM candidates with masses smaller than an eV down to less than $10^{-22}$\,eV~\cite{PhysRevD.81.123530}, like axion-like particles, the Quantum Chromodynamics axion pseudoscalar, or ultra-light bosons and scalar particles suggested by high-energy physics models~\cite{battaglieri2017us}.
\begin{figure*}[ht]
   \centering
    \includegraphics[width=0.8\linewidth]{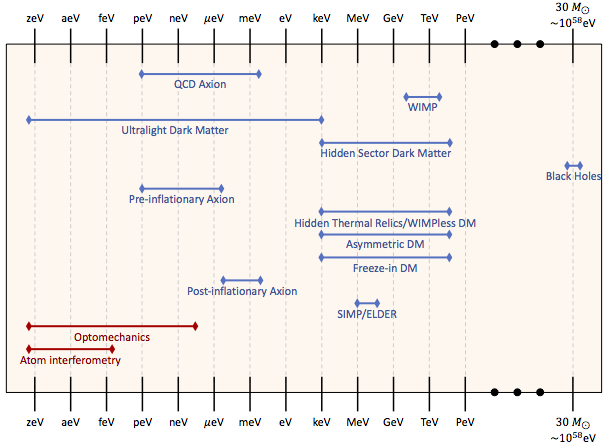}
 \caption{
 Summary of some among the dark matter candidates and their mass (ranges highlighted in blue) compared to the range of parameters covered by 
 atom interferometry~\cite{tino2021testing} and optomechanical set-ups~\cite{manley2021searching,carney2021mechanical}, which are highlighted in red and have been proposed to test ligh DM candidates. See~\cite{battaglieri2017us} and references therein for additional details. Figure adapted from~\cite{battaglieri2017us}.}
 \end{figure*}
At present, DM searches with mechanical quantum sensors are attracting growing interest~\cite{PhysRevD.88.116005,PhysRevD.102.072003,PhysRevD.99.023005,bateman2015existence,cheng2020dark,PhysRevLett.125.181102,manley2021searching}. The impressive sensitivity reached by optomechanical systems in measuring small displacements and forces make these devices very appealing to test DM by searching for the tiny accelerations produced by its interaction with ordinary matter. Recent studies have shown the potential of optomechanical accelerometers to test WIMPs and ultralight DM~\cite{manley2021searching} (see also the recent white paper~\cite{carney2021mechanical} and references therein).  Moreover, possible applications of quantum optomechanical experiments for DM searches have been suggested to be performed in space~\cite{kaltenbaek2015testing}.
In addition, Gravitational Wave (GW) detectors, such as the Laser Interferometer Space Antenna (LISA)~\cite{baker2019laser}, can be exploited to provide insight on DM and DE \cite{barausse2020prospects}.

Together with the novel avenues offered by optomechanical systems in the search of DM candidates, atomic sensors -- atom interferometry and atomic clocks -- have shown the potential to be key players for the search light candidates~\cite{PhysRevLett.117.261301,PhysRevD.93.075029,hamilton2015atom,bongs2019taking}. This is related to the fact that investigations with atomic sensors connect naturally with the search for EEP violations discussed in the previous section. Ultra-light DM candidate phenomenology is, in fact, associated with temporal variation of fundamental constants~\cite{arvanitaki2016sound,PhysRevD.98.064051,PhysRevD.91.015015,derevianko2014hunting,wcislo2018new} -- in particular of the fine structure constant and the electron mass -- and with violations of the WEP~\cite{PhysRevLett.103.011301,PhysRevD.98.064051}. 

Moreover, atom interferometry has been also investigated as a possible probe for dark energy. In several models, DE is described by modifications of gravity and the introduction of a dynamical ultra-light field -- a ``fifth force'' -- that, when coupled to the SM fields, affect the constancy of fundamental constants. This relates, once again, to searches for violations of the EEP~\cite{hui2009equivalence}. In particular, specific theoretical models argue that DE candidates could present a screening mechanism, suppressing the effects of the fifth-force in high matter-density regions and thus evading tests of the EEP with macroscopic masses. Among these, the chameleon field~\cite{PhysRevLett.93.171104} and the symmetron~\cite{PhysRevLett.104.231301} have attracted much interest. The potential of atom interferometry, in this context, has been widely investigated~\cite{hamilton2015atom,PhysRevD.94.044051,PhysRevD.101.083501,PhysRevLett.123.061102,burrage2015probing,PhysRevLett.123.061102,jaffe2017testing,bongs2019taking}. In~\cite{hamilton2015atom}, first experimental results have been reported where a high-vacuum chamber was used to reproduce conditions in which the fifth-force field should be long-ranged, and thus detectable with the use of microscopic probe masses\footnote{For further details on DM and DE searches with atom interferometry, we refer the interested reader to~\cite{tino2021testing} and references therein.}.

In the searches for DM and DE, space-based experiments can offer all the advantages previously discussed when dealing with the EEP and atom inteferometry. The QTEST~\cite{williams2016quantum}, AEDGE~\cite{abou2020aedge}
 and SAGE~\cite{PhysRevD.93.075029,PhysRevLett.117.261301,tino2021testing} proposals fall in this category and, for the last two, the detection of light DM is one of the primary scientific objectives. AEDGE envisages using atom interferometry also for probing DE, and the possibility for experiments on the ISS with the NASA-founded experiment CAL were outlined~\cite{chiow2018multiloop}.

\subsection{Interface between quantum physics and relativity}

General relativity and quantum mechanics are  two of the main pillars on which we base our understanding of the physical world. Quantum mechanics predicts with great accuracy the behaviour of the microscopic world, while general relativity provides an accurate description of gravity and of the Universe at large length-scales. However, we do not fully understand what happens when these two theories are combined together despite almost a century of investigation.

Combining quantum mechanics with special relativity was one of the hallmark of 20th century physics with the development of  relativistic Quantum Field Theory (QFT) in flat spacetime. QFT is at present the physical theory with the most stringently tested predictions in physics, despite being plagued by divergences. While extensions of QFT to curved background are possible and have been extensively investigated~\cite{birrell1984quantum}, the gravitodynamics characteristic of GR is essentially ignored by these attempts. 
The only known and controllable way to include gravitational effects in QFT is through a non-renormalizable effective field theory~\cite{shomer2007pedagogical,doughty2018lagrangian}, and beyond that lies the yet not fully charted territory of quantum gravity (QG). In this context, due to the century-long attempts to quantize gravity and the lack of a consistent way to do so, part of the scientific community has also considered the possibility that it is not GR that needs to bend to the rules of quantum physics, but it is quantum mechanics that must be modified to accommodate GR as a fundamentally classical theory~\cite{penrose1996gravity}. However,  these attempts have not yet succesfully provided a fully consistent theory.

One of the major obstacles in this field of research is the absence of experimental guidance.
We know from the famous COW (Colella, Overhauser and Werner) experiment performed in 1975 that the interaction of a quantum system with a weak gravitational field, such the one generated by the Earth, produces a phase shift in the wave function as any other external potential would do~\cite{colella1975observation}. Apart from that, we still do not have experimental evidence on how the curvature of space–time affects quantum systems or on the gravitational field generated by a system in a quantum superposition, although several proposals exist~\cite{pikovski2012probing,bose2017spin,marletto2017gravitationally,carlesso2019testing,carney2019tabletop,krisnanda2020observable}. Two are the major problems faced in order to deliver this kind of  experiments. On the one hand,  gravitational forces in the microscopic regime are very small and thus easily spoiled by any other residual force or source of noise present in the experimental setup~\cite{ritter1990experimental,schmole2016micromechanical,rijavec2021decoherence}. On the other hand, velocities and distances of systems in ground-based experiments are strongly limited. In this context,  space  can offer a viable option to mitigate both these problems. Indeed, long free-fall times and the large distances and velocities available in space could provide the right setting for tests at the interface between quantum physics and relativity. 

\subsubsection{Relativistic quantum information}
The incredible success of quantum information in the second half of the 20th century has led to incredible advancements in both  the fundamental understanding of quantum theory and  the development of quantum technologies. Over the last couple of decades, the novel field of relativistic quantum information (RQI)~\cite{peres2004quantum,mann2012relativistic} has started to investigate what happens when quantum information results and protocols are extended into the relativistic domain.
In particular, the study of curved spacetime effects on a quantum systems from an informational perspective have produced several interesting results: from considerations on entanglement being an observer-dependent concept~\cite{alsing2012observer}, which can be degraded by non-inertial motion~\cite{bruschi2014testing}, to those proving that special relativistic effects are relevant for quantum teleportation protocols~\cite{alsing2004teleportation}.
Testing these effects would prove the applicability of quantum information concepts at the intersection of quantum and relativistic regimes~\cite{rideout2012fundamental}. In this context, experimental setups with entangled photons between ground and satellites~\cite{vedovato2017extending,yin2017satellitetoground,liao2017long} have proven the possibility to test entanglement at long distances and could be used to study the effects of the Earth gravitational field on photons. Proposals to use low Earth orbit satellite such as the CanX4 and CanX5 CubeSats~\cite{bonin2015canx}  
to realize a refined version of the COW experiment on large distances and to test post Newtonian effects of gravity have also been presented~\cite{zych2012general,bruschi2014testing, pallister2017blueprint}.
Furthermore, planned space-based missions to deliver quantum communication protocols at large distances such as the Canadian Quantum Encryption and Science Satellite (QEYSSat) micro-satellite mission~\cite{jennewein2014qeyssat}, the NASA-founded Deep Space Quantum Link~\cite{mohageg2018deep} or the SAGE mission~\cite{tino2019sage} will allow to analyse the effects of the spacetime curvature around the Earth on propagating photons.

\subsubsection{Gravitational Decoherence}
Gravitational decoherence is the loss of coherence of a quantum superposition caused by gravity related effects, and it is predicted in several theoretical models~\cite{bassi2017gravitational,plato2016gravitational}. It can be regarded also as a low-energy effect which does not require to enter into the domain of quantum gravity,  but, at the same time, opens a window on the interface between quantum physics and gravitation. 
The term gravitational decoherence, as intended here, actually encompasses several different mechanisms leading to the loss of coherence in quantum systems. The most straightforward way gravitational decoherence can ensue is when a quantum system is exposed to the stochastic fluctuations of a classical gravitational field, namely a stochastic gravitational wave background~\cite{power2000decoherence,lamine2002gravitational,reynaud2002decoherence,reynaud2004hyper,lamine2006ultimate,breuer2009metric}.    
At the quantum level, all the models, which perturbatively quantize general relativity, predict the existence of the graviton as a quantum fluctuation of the gravitational field, which can serve as an additional decoherence channel~\cite{blencowe2013effective,anastopoulos2013master,lagouvardos2020gravitational}.
Gravitational decoherence also arises in models that try to ``gravitise'' quantum mechanics, modifying it in order to make it compatible with GR, and that need to introduce ``ad hoc'' spacetime stochastic fluctuations to guarantee the consistency of the modified theory with GR principles~\cite{karolyhazy1966gravitation,karolyhazy1986b,milburn1991intrinsic,altamirano2018gravity,khosla2018classical,kafri2014classical,kafri2015bounds,kafri2013noise,gaona2021gravitational}. 
Even static gravitational potentials can be responsible for quantum decoherence through dephasing as discussed 
in~\cite{pikovski2015universal,carlesso2016decoherence}, or as described by the attempts to make quantum field theory compatible with closed time-like curves like in the Event Formalism where decoherence induced by the curvature of spacetime is present 
~\cite{ralph2014entanglement,ralph2009quantum}.

The study of gravitational decoherence  can be of great value for understanding the interplay between quantum mechanics and GR, and a possible observation would be a groundbreaking result. The advent of new technologies, from atomic interferometry to optomechanical systems, has the potential to make the observation of gravitational decoherence feasible in the near future. In this context, 
the exquisite sensitivity of quantum interferometric experiments to decoherence sources can be used to infer the small phase-shifts induced by gravitational decoherence. Moreover, space-based experiments, as already discussed, offer many advantages for performing such experiments.
In 2019, the Quantum Experiments at Space Scale (QUESS) space mission was able to deliver a first test on the decoherence predicted by the Event Formalism on a photon travelling in the Earth's gravitational field by performing an interferometric experiment between the ground station at Ngari in Tibet and the Micius satellite~\cite{xu2019satellite}. Current and past proposals for interferometric experiments in space using massive nanoarticles or cold atoms, like  MAQRO~\cite{CDFMAQRO2019}, STE-QUEST~\cite{joshi2018space} and GAUGE \cite{sumner2007step}, also include in their scientific objectives the investigation of different models entailing gravitational decoherence effects.

\subsection{Space-based Gravitational Wave detectors}
Gravitational waves (GWs) are curvature deformations of space-time generated by the accelerated motion of massive objects, as predicted by general relativity~\cite{weinberg1972gravitation}.
Their direct detection 
by the Laser Interferometer Gravitational-Wave Observatory (LIGO) and Virgo collaboration~\cite{abbott2016observation} in 2016  marked the beginning of a new and exciting era for observations of our Universe. This detection was a striking confirmation of Einstein's theory and, more importantly, provides an entirely new window of opportunities, from investigations into the physics of black hole and neutron star to new tests of GR and the observation of low mass sources at low redshift~\cite{amaro2017laser,caldwell2019astro2020}. 
At present, several observations have been confirmed with sources being the merges of two black holes and binary neutron star inspirals~\cite{abbott2016observation,abbott2016gw151226,abbott2016gw150914,abbott2016binary,scientific2017gw170104,abbott2017gw170814,abbott2017gw170608,abbott2017gw170817,abbott2019gwtc}. Several other potential cosmological sources have been suggested. They include cosmic strings~\cite{siemens2007gravitational}, inflationary scenarios~\cite{turner1997detectability,ungarelli2005gravitational,boyle2008probing,bartolo2016science}, warped extra dimensions~\cite{randall2007gravitational}, and various first-order cosmological phase transitions~\cite{witten1984cosmic,hogan1986gravitational,turner1990relic,maggiore2000gravitational}.

Processes occurring in early stages of the Universe can produce high-frequency GW, which will be eventually redshifted as the Universe expands, and are thus expected to be detectable in the sub-Hertz band. This is the frequency interval where we expect to observe the heaviest and most diverse GW sources~\cite{amaro2017laser}. However, this band is outside the capability of ground-based detectors, which are currently limited to the $10\sim10^3$\,Hz~\cite{abbott2019gwtc}. The low-frequency window below 1\,Hz is not accessible to ground-based interferometers due to unavoidable noises, like those caused by seismic density fluctuations that are increasingly severe at low frequencies~\cite{baker2019laser}. 
In this context, space-based detectors hold the promise to be able to cover the sub-Hertz band opening an observational window on the early Universe. In addition, also the sensitivity to stochastic GW backgrounds should improve at low frequency \cite{larson2000sensitivity,binns2009fundamental,jinno2014studying,cornish2019discovery,schmitz2020new}.
The observation of such a background could provide insight into the theory of inflation and the physics beyond the standard model~\cite{weir2018gravitational,huang2018exploring,baldes2018high,chiang2019revisiting,baldes2019strong,ellis2019maximal,madge2019leptophilic,ahriche2019gravitational,beniwal2019gravitational,angelescu2019multistep,abou2020aedge}.

\begin{figure*}[ht]
   \centering
    \includegraphics[width=0.8\linewidth]{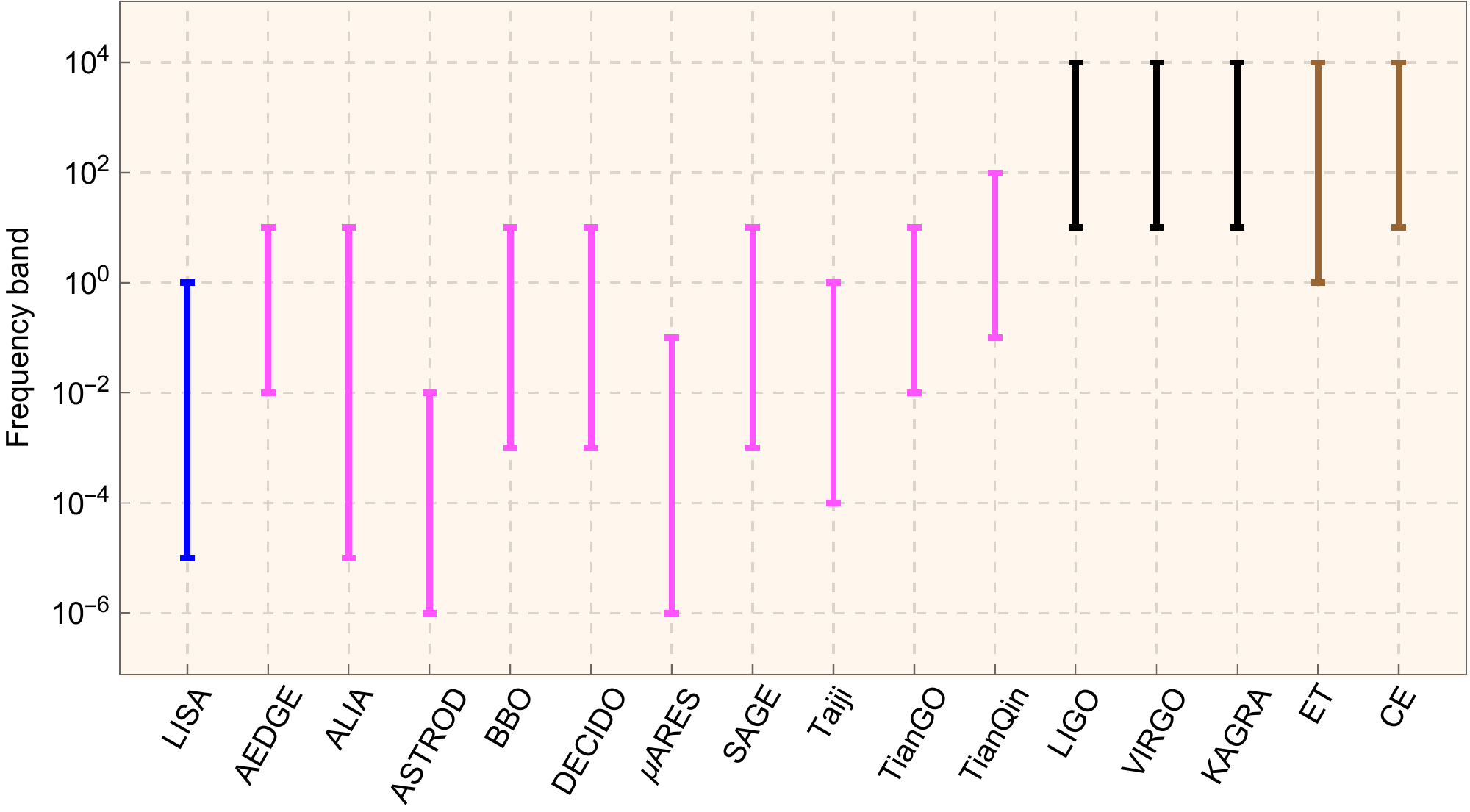}
 \caption{\label{fig.GWdetectors} Comparison of the frequency range which will be covered by different proposals for space-based GW detectors. LISA, which is highlighted with the blue line, is the only currently scheduled space-based GW detector. AEDGE, ALIA, ASTROD, BBO, DECIDO, $\mu$Ares, SAGE, Taiji, TianGO and TianQin, which are represented with magenta lines, are still at the level of proposals. For comparison we report also the values of the frequency bands for LIGO, VIRGO and Kamioka Gravitational Wave Detector (KAGRA) \cite{abbott2018prospects} with black lines, and for the proposals for ground-based GW detectors  Einstein Telescope (ET) \cite{punturo2010einstein} and Cosmic Explorer (CE) \cite{reitze2019cosmic} with brown lines.}
 \end{figure*}

\begin{figure*}[t]
   \centering
 \includegraphics[width=0.8\linewidth]{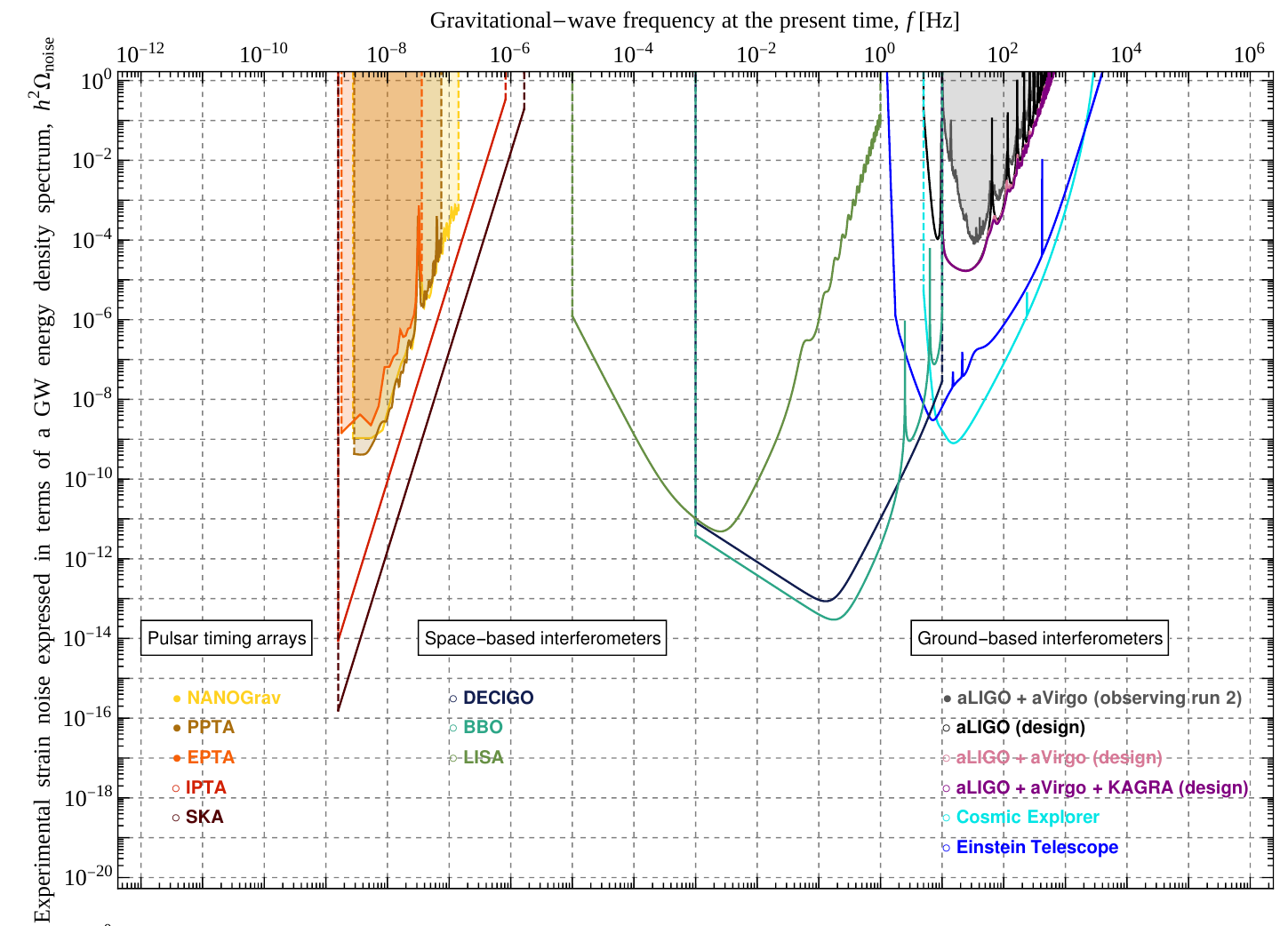}
 \caption{\label{fig.GW.space}  Comparison of the potential sensitivity of different ground-based and space-based GW detectors: Square Kilometre Array (SKA) \cite{hobbs2010international}, International Pulsar Timing Array (IPTA) \cite{hobbs2010international}, European Pulsar Timing Array (EPTA) \cite{kramer2013european}, Parkes Pulsar Timing Array (PPTA) \cite{manchester2013parkes}, North American Nanohertz Observatory for Gravitational Waves (NANOGrav) \cite{mclaughlin2013north}, Einstein Telescope (ET) \cite{punturo2010einstein}, Cosmic Explorer (CE) \cite{reitze2019cosmic}, Hanford-Livingston-Virgo-KAGRA (HLVK), Hanford-Livingston-Virgo (HLV), Hanford-Livingston (HL), Hanford-Livingston-Virgo observing run 2 (HLV (O2)) \cite{schmitz2020new}. Figure  from~\cite{schmitz2020new}.  }
 \end{figure*}

At present, there are several proposals for space-based GW detectors, such as the
Atomic Experiment for Dark Matter and Gravity Exploration in Space (AEDGE)~\cite{abou2019aedge,abou2020aedge}, other proposals based on atom interferometry~\cite{hogan2016atom,Loriani2019graviwave}, the Advanced Laser Interferometer Antenna (ALIA)~\cite{gong2015descope,bender2004additional}, the Astrodynamical Space Test of Relativity using Optical Devices (ASTROD)~\cite{ni2013astrod},
the Big Bang Observer (BBO)~\cite{harry2006laser,seto2006correlation,smith2006direct}, the Deci-hertz Interferometer Gravitational wave Observatory (DECIGO)~\cite{seto2001possibility,smith2006direct,kawamura2006japanese,kawamura2011japanese,kawamura2020current}, $\mu$Ares~\cite{sesana2019unveiling}, the Space Atomic Gravity Explorer (SAGE)~\cite{tino2019sage}, Taiji~\cite{hu2017taiji,guo2018taiji,luo2020brief} and its extension forming the LISA-Taiji network~\cite{ruan2020lisa}, TianQin~\cite{luo2016tianqin}, TianGo~\cite{kuns2020astrophysics}, and finally the Laser Interferometer Space Antenna (LISA)~\cite{amaro2017laser}, which is currently scheduled for launch in 2030's; we will briefly delve into its proposal in the next subsection. In Fig.~\ref{fig.GWdetectors},  we compare the frequency range of all these proposals while in Fig.~\ref{fig.GW.space} we report their sensitivities. 

\subsubsection{Laser Interferometer Space Antenna (LISA)}
 
LISA is envisaged as a space-based GW detector composed of three independent spacecrafts each placed at one vertex of an approximately equilateral triangle on an orbit far from Earth where the thermal, magnetic, and gravitational environment are sufficiently stable to eventually detect GW. Each spacecraft contains two test masses and it is forced to follow them in their geodesic trajectories with sub-g/$\sqrt{\text{Hz}}$ spurious accelerations~\cite{amaro2017laser}.  
The proper distance between the masses in the different spacecrafts will be monitored using optical interferometry over a 2.5 Mkm baseline, forming a configuration which is expected to be passively stable over the lifetime of the LISA mission, which should comprise 1~yr of commissioning and calibration, 4~yrs of science operations and 6~yrs of potential extended operations~\cite{folkner1997lisa}.

LISA will conduct the first survey of the millihertz GW sky, with possible sources ranging from white-dwarf binaries to mergers of massive black holes. Furthermore, among the scientific objectives of LISA there is the direct detection of a stochastic GW background of cosmological origin and stochastic foregrounds. As mentioned before, measuring such a background, or placing upper limits on it, would constrain models of the early Universe and particle physics beyond the standard model~\cite{caprini2016science,bartolo2016science,amaro2017laser}.

LISA technical readiness is strongly supported by two previous flight demonstrations: LISA Pathfinder and the Gravity Recovery And Climate Explorer Follow-On (GRACE-FO) missions~\cite{baker2019laser}. 
The ESA-led LISA Pathfinder mission (2015-2017) provides the basis for the predicted acceleration performance of the LISA mission~\cite{baker2019laser}. 
LISA Pathfinder included a pair of representative LISA test masses. A first one was used to determine the drag-free geodesic trajectory, while the second was used to measure the residual acceleration of the primary test mass. GRACE-FO instead includes a laser ranging instrument, which will be used to determine the inter-satellite distance together with the primary microwave ranging instrument. First results already show a nanometer accuracy level over a 210~km link~\cite{abich2019orbit}.
 
\subsection{Quantum Foundations}
Despite being almost one century old, quantum mechanics continues to puzzle and amaze  physicists and philosophers of science. While the predictive power of the theory is having a huge technological and theoretical impact, 
there are some profound interpretational issues with fundamental implications which are at the center of a heated debate. Among them, the measurement problem~\cite{leggett2005quantum}, and the related emergence of the classical world we experience from the microscopic quantum laws of physics~\cite{leggett1980macroscopic,zurek1991quantum,penrose1996gravity,bell2004speakable,weinberg2014precision}, is among the most puzzling ones. As  is the case with quantum gravity~\cite{liberati2011quantum},  in the field of quantum foundations a drastic improvement of the discussion is linked to the possibility of empirically falsifying models and alternatives to 
quantum mechanics. Space offers a promising stage for this and other investigations of fundamental 
aspects of the quantum world. In this section we detail how space-based experiments can provide the possibility of testing the predictions of, and possible deviations from, quantum mechanics. 

\subsubsection{Predictions of Quantum Mechanics}
\paragraph{Entanglement and Non-locality}
Quantum  entanglement~\cite{schrodinger1935discussion,horodecki2009quantum} is one of the most peculiar traits of quantum mechanics. In turn, one of the most intriguing aspects of entanglement is quantum non-locality~\cite{horodecki2009quantum}, i.e. the fact that quantum correlations can violate classical probability's inequalities respected by all models satisfying basic locality and causality assumptions. This aspect of the quantum formalism has had a tremendous impact on the entirety of quantum physics. On the one hand, it helped to reshape the understanding of the quantum formalism. On the other hand, it had a great impact at the technological level, where it opened the way to the second quantum revolution. While the two concepts are not equivalent -- non-locality requires entanglement but not the other way around~\cite{brunner2014bell} -- they both represent the basic building  blocks  of important quantum technologies, from quantum key distribution (QKD) to quantum computation. Some of these technological applications will be discussed in Sec.~\ref{Applications} when relevant for space;  here we focus on the fundamental aspects that can be investigated with quantum technologies in space. In the following discussion we will focus on the simple case of bipartite quantum systems for which  the theory of entanglement and non-locality is fairly well understood and many results have been formulated. For more details, we refer the interested reader to specialized reviews~\cite{horodecki2009quantum,brunner2014bell,adesso2016measures,plenio2014introduction} and references therein.

\begin{figure}[t!]
 \centering
 \includegraphics[width=0.8\linewidth]{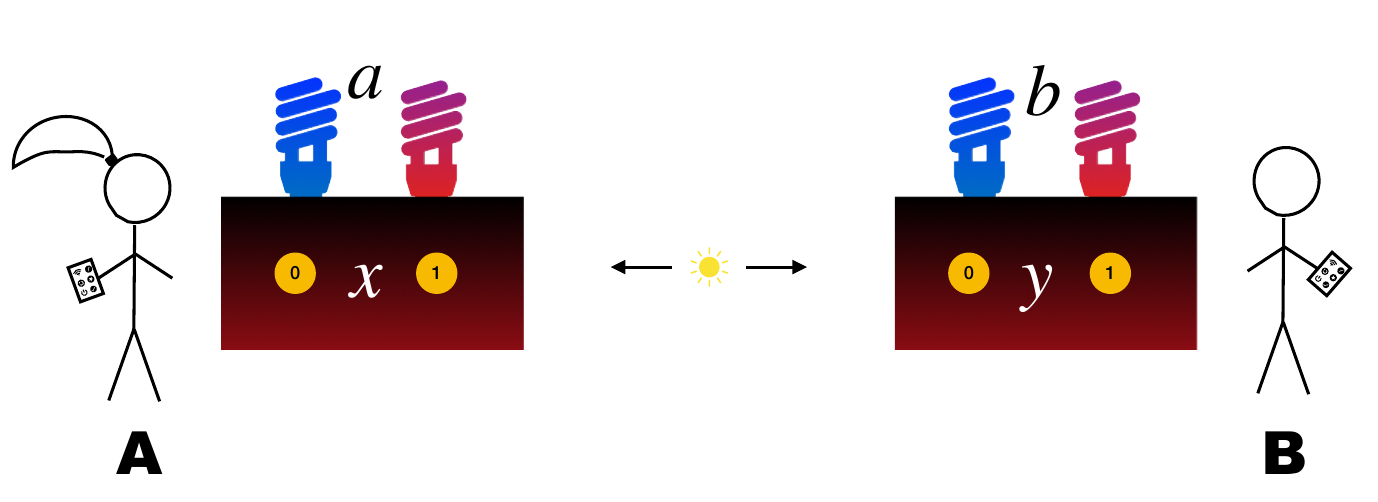}
 \caption{\label{fig.Entanglement} Quantum non-locality: In the paradigmatic example of a bipartite system subject to dicotomic measurements. Two  observers, Alice and Bob, are given two black-box devices with dicotomic measurement choices $x,y\in\{0,1\}$  and dicotomic outcomes of the measurements $a,b\in\{\text{blue}, \text{red}\}$. In any local model, i.e. under the assumptions of Bell's local causality and measurement outcomes that depend on hidden variables, the probability distribution of the outcomes  assumes the form in Eq.~\eqref{loca_caus}. It is then trivial to show that such form of the probability density function cannot  violate the celebrated CHSH inequality~\cite{PhysRevLett.23.880,brunner2014bell}. On the other hand, quantum mechanics predicts the violation of such inequality with entangled states. These violations have been experimentally observed~\cite{brunner2014bell,aspect2015closing}.}
 \end{figure}

In order to grasp the significance of non-locality, let us consider the paradigmatic set-up that sees two spacelike separated observers, Alice and Bob, each with access to one part of a bipartite quantum system. Alice and Bob can locally measure their subsystem by choosing a measurement set-up,  $x$ for Alice and $y$ for Bob, and obtain an outcome $a$ and $b$ respectively, see Fig.~\ref{fig.Entanglement}. Any local hidden-variable model essentially assumes the existence of a set of past factors $\lambda$ with joint causal influence on the two outcomes $a$ and $b$ and that completely takes into account their correlations. This is tantamount to assuming that the  outcome obtained by Alice cannot have any direct influence on that of Bob, in accordance with Einstein's causality. This is known as Bell locality. It implies that the joint conditional probability $p(a,b|x,y)$ of the outcomes given the choice of measurements takes the form~\cite{brunner2014bell}
\begin{equation}\label{loca_caus}
    p(a,b|x,y)=\int d\lambda\, q(\lambda)p(a|\lambda,x)p(b|\lambda,y),
\end{equation}
where $q(\lambda)$ is the probability distribution characterising the hidden variables, which can be in principle stochastic, while $p(a|\lambda,x)$ and $p(b|\lambda,y)$  are the conditional probabilities of the outcomes given the measurement and the hidden variables configuration.
It is possible to derive inequalities that need to be satisfied by any local hidden-variable model. These are known as Bell's inequalities and can be violated in quantum mechanics. These violations -- often referred to as quantum non-locality -- have been observed in an impressive series of works, see for instance~\cite{aspect2015closing} and references therein for an overview on the subject. 
 
From the first verification in the '70s~\cite{PhysRevLett.28.938,pipkin1979atomic,PhysRevLett.36.1223,clauser1978bell}, the history of Bell's inequalities violations (BIV) is one of theoretical investigations, technological applications, and subsequent experiments aimed at solving various loopholes in their detection and interpretation~\cite{brunner2014bell,acin2018quantum,aspect2015closing}. In particular, the {locality loophole}~\cite{bell2004speakable}, i.e. the fact that Alice and Bob should be spacelike separated so that the choice of measurements cannot influence one another, was first closed after the '80s~\cite{PhysRevLett.49.1804,PhysRevLett.81.5039}. On the other hand,  the {detection loophole}~\cite{PhysRevD.2.1418,PhysRevD.10.526}, which is related to the inefficiencies of detectors, was closed in both photon and atomic experiments in~\cite{giustina2013bell,PhysRevLett.111.130406,rowe2001experimental,PhysRevLett.100.150404}. In 2015, three different experiments~\cite{hensen2015loophole,PhysRevLett.115.250401,PhysRevLett.115.250402} were finally able to close the previous two-loopholes together, a result with important fundamental as  well as applied implications\footnote{Let us remark that there are also other loopholes considered in the literature of BIV, see for example~\cite{aspect2015closing,brunner2014bell}.}.   

Several possibilities of employing satellite configurations for testing quantum nonlocality and entanglement at long distance have been considered in the past~\cite{rideout2012fundamental}. Current technological advancement in this direction is progressing at a fast pace~\cite{yin2017satellite,liao2017satellite,jennewein2014qeyssat,mohageg2018deep} and the impact on the future of quantum communication cannot be overstated. Quantum mechanics does not predict any bound on the distance between two entangled systems so that BIV are expected also when Alice and Bob are far away. In this context, space offers the possibility to probe these predictions of quantum mechanics at larger and larger distances and in a regime in which special and general relativistic effects may come into play. Indeed, in ground-based experiments the distance between Alice and Bob cannot be greater than few hundred kilometers\footnote{ In~\cite{scheidl2010violation} BIV were observed at distance of 144~km closing the locality loophole together with the so-called freedom-of-choice one} due to either losses in fibre experiments or the Earth curvature for free-space propagation experiments. Both these issues can be overcome by moving experiments to space. In 2017, the QUESS mission of the Chinese Academy of Sciences used the satellite Micius for demonstrating satellite-mediated entanglement distribution and BIV between locations on Earth more than 1200\,km apart~\cite{yin2017satellite,liao2017satellite}. 

Several other projects allowing to test BIV and quantum mechanical entanglement at large scales are been planned at the national and international level, driven mostly by the alluring possibility of a global quantum internet and long distance quantum communication and cryptography, which are topics we will cover in Sec.~\ref{Applications}. Examples are the Canadian QEYSSat micro-satellite mission expected for 2022~\cite{jennewein2014qeyssat} and the NASA's Deep Space Quantum Link (DSQL), which will employ the Lunar Orbital Platform-Gateway -- a space station orbiting the moon -- to establish a quantum link with ground stations which will allow testing BIV~\cite{mohageg2018deep} in conditions in which the spacetime curvature is expected to play a role.  The SAGE mission proposal includes among its main scientific objectives the test of BIV with two satellites at a distance between 5000 and 30000\,km playing the role of Alice and Bob. At these distances, special and general relativistic effects are expected to be relevant. 

Finally, it should be mentioned that the experimental investigation of BIV represents a very good example of a fundamental problem in theoretical physics, born out of an academic debate on the meaning of quantum theory, which had profound repercussions on applied physics and technologies. This problem has been tackled by the physics community from several different angles employing photonic setups (as already discussed), atomic~\cite{PhysRevLett.119.010402} and, more recently, optomechanical setups~\cite{PhysRevLett.121.220404}. In this respect, BIV are an excellent example of the success of a multidisciplinary approach to a fundamental problem and of the far-reaching implications that fundamental studies can entail. 

\paragraph{Delayed choice}
One of the most intriguing feature of quantum mechanics is that  quantum objects may possess properties that are equally real, but mutually exclusive, as illustrated by Bohr's complementary principle~\cite{bohr1928quantenpostulat}.
 
\begin{figure}[t!]
 \centering
 \includegraphics[width=0.8\linewidth]{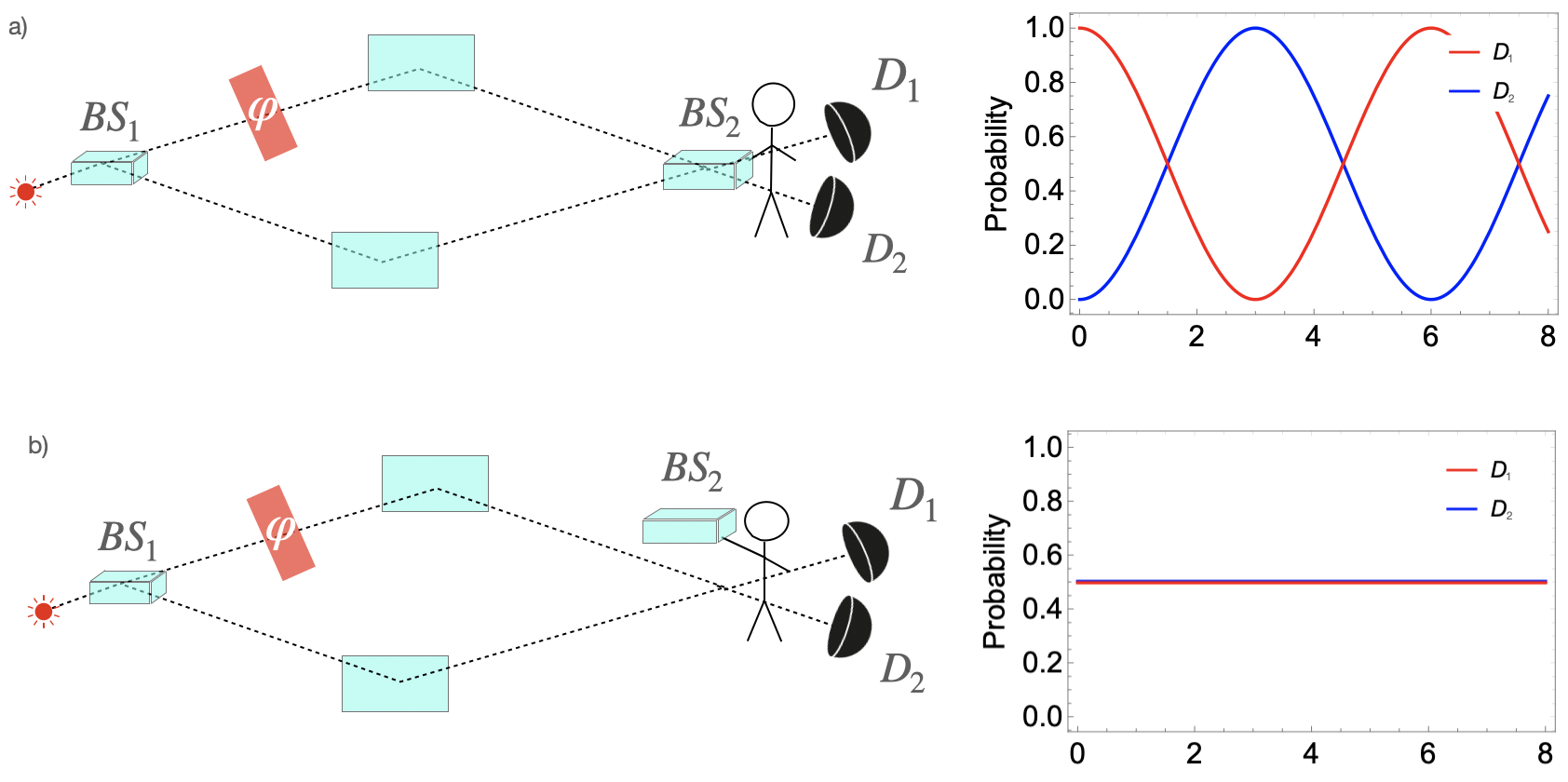}
 \caption{ \label{fig.wheeler}
   Idealized Mach-Zhender interferometer.
  Upper panel: standard configurations where the two laser path are recombined in a beam-splitter BS$_{2}$ before being detected. Lower panel: the delayed-choice version of the experiment where the second beam-splitter is removed after the photon has passed through the first one. On the right, the corresponding counting probabilities on the two detectors D$_1$ and D$_2$, as a function of the phase-shift $\varphi$. The difference in the counting of the detectors in the two cases leads to the conclusion that the photons behave as a particle or wave depending on the presence of BS$_{2}$.
}
\end{figure}
 
The most iconic example is represented by the wave-particle duality, which can be exhibited by a photon in a Mach-Zehnder interferometer, as schematically depected in Fig.~\ref{fig.wheeler}. If the paths through the interferometer are recombined in a beam-splitter, the probability for a photon to exit at one of the detectors only depends on the phase difference between the two arms. If no information about which path the photon has followed is acquired, a wave-like interference pattern is observed after recording many particle detection events.
On the other hand, if we remove the second beam-splitter, the probability of a photon to exit at one of the detectors depends on which path the photon took, and no interference will occur. In this case, the particle behaviour of each photon becomes apparent.

Following classical common sense, we would be forced to think that in the first case the photon must have entered the Mach-Zehnder as a wave, while in the second case as a particle, leading to the  conclusion that the act of the measurement at the end of the protocol is changing the behaviour of the photon at the entrance of the device.
This was a central aspect on the debate between Bohr and Einstein on the interpretation of mathematical entities in quantum theory, which culminated in the Einstein-Podolsky-Rosen paradox in 1935~\cite{einstein1935can} and later guided Bell to his famous theorem on local causality in 1964~\cite{bell1964einstein}. 

In the 1970s, Wheeler wanted to push the idea of complementarity further~\cite{wheeler1978past,wheeler2014quantum}, posing the question of what happens if in a double slit experiment the choice of including or not a second beam splitter is taken after the photon passed through the first beam splitter. Could it be possible for the photon to change from a wave to a particle if the second beam splitter is removed?
In a modern language, this is equivalent to asking whether the choice of measurement can alter states that are spatially-separated. Obviously if this were possible, as predicted by quantum mechanics, and the change of the states constitutes a real physical change carrying information, we would have an apparent violation of relativistic causality. {This line of reasoning led Wheeler to embrace Einstein's view that the wave function cannot be a  physical entity, but only a mathematical object representing the knowledge about a physical system}. After the successful performance of the experiment proposed by Wheeler~\cite{hellmuth1987delayed}, which confirmed the correctness of quantum mechanical predictions in this context, 
the interest in delayed-choice type experiments has never vanished and several realizations of such experiments have been developed during the years~\cite{jacques2007experimental,ma2012experimental,tang2012realization,manning2015wheeler}. The culmination of these experiments was reached with the impressive result by Vedovato \textit{at al.}~\cite{vedovato2017extending},
which consisted in a Wheeler delayed-choice experiment over a distance of 3500\,km by exploiting the temporal degree of freedom of photons reflected by a rapidly moving satellite in orbit. This confirmed once more the wave-particle duality at unprecedented distances, opening the way for novel applications of quantum mechanics in space. 

\subsubsection{Deviations from Quantum Mechanics: collapse models}

The measurement problem is one of the main open questions in quantum theory. It stems directly from the contrast between the linear and deterministic evolution of the Schr\"odinger equation and the necessity of a non-linear and stochastic description of the measurement process. Furthermore, quantum theory allows for superpositions of macroscopically distinct states even if that seems to be in distinct disagreement with our everyday experience of the macroscopic world. Although it is commonly believed that environmental decoherence can account for the apparent classical behaviour at the macroscopic level, it should be pointed out that decoherence does not solve the measurement problem~\cite{weinberg2012collapse}. Other models tackle this problem by requiring a failure of the quantum superposition principle in the macroscopic regime and the collapse of the wavefunction during a measurement. Among them, the most widely studied are the so called collapse models~\cite{bassi2003dynamical,bassi2013models}, which have the unique feature of being experimentally testable. Examples of such models are the Ghirardi-Rimini-Weber (GRW) model \cite{ghirardi1986unified}, the Continuous Spontaneous Localization (CSL) model \cite{pearle1989combining,ghirardi1990markov}, and the gravity-induced collapse models such as the  Diosi-Penrose (DP) model \cite{diosi1987universal,diosi1989models,penrose1996gravity}.

 \paragraph{Ghirardi-Rimini-Weber model}
The Ghirardi-Rimini-Weber (GRW) model was the first proposed consistent collapse model \cite{ghirardi1986unified}. Its underlying idea is that each particle of a composite system independently undergoes a spontaneous collapse process that localises the  wave function in space. This process is stochastic in time and space independently. It follows a Poissonian distribution in time with a rate $\lambda$, and between two consequent collapses the dynamics of the system is described by the standard Schr\"odinger equation. The probability density in space for having a collapse in $\x_0$ is given by $P_i(\x)=\braket{\psi_{\x_0}^{i}|\psi_{\x_0}^{i}}$, where
$\ket{\psi_{\x_0}^{i}}=\hat L_{\x_0}^{i}\ket \psi$, with $\ket \psi$ denoting the multi-particle wave function and
\begin{equation}
    \hat L_{\x_0}^{i}=(\pi \rC^2)^{-3/4}e^{-(\hat \q_i-\x_0)^2/2\rC^2},
\end{equation}
is the localization, or collapse, operator at the position $\x_0$. Here, $\hat \q_i$ is the position operator of the $i$-th particle and $\rC$ is the characteristic length of the localization.
The collapse process is then described through the following expression
\begin{equation}
    \ket\psi\to\frac{\ket{\psi_{\x_0}^{i}}}{\sqrt{\braket{\psi_{\x_0}^{i}|\psi_{\x_0}^{i}}}},
\end{equation}
whose form underlines the non-linearity of the collapse process.
The action of the GRW mechanism depends on the two parameters $\lambda$ and $\rC$ just introduced. These are phenomenological parameters which should be eventually fixed by experiments. Originally, the value of $\lambda=10^{-16}\,$s$^{-1}$ was theoretically proposed \cite{ghirardi1986unified}; later, Adler considered a stronger value of $\lambda=10^{-8\pm2}\,$Hz as more adequate \cite{adler2007lower}.
The value  $\rC=10^{-7}\,$m has instead met a stronger consensus, since it denotes a length-scale dividing the quantum microscopic world from the classical macroscopic one. Although being a very simple model, GRW satisfies the requirements for a well-defined collapse model. In particular, it embodies an amplification mechanism for which the larger the system is the stronger is the collapse becomes. Such a mechanism is fundamental for describing the collapse of more macroscopic systems. Indeed, by averaging over the relative degrees of freedom, it turns out that the effective collapse rate $\lambda_\text{\tiny eff}$ for the center-of-mass wavefunction is given by  $\lambda_\text{\tiny eff}=\mathcal N \lambda$, where $\mathcal N$ is the number of particles of the system. Thus, larger macroscopic systems collapse faster and more efficiently, while the collapse  on microscopic system is essentially negligible \cite{bassi2003dynamical,bassi2013models}.\\
The GRW model predicts a violation of the energy conservation, with a rate of $\lambda \hbar^2/4 m_0 \rC^2$ for a free particle, where $m_0$ is the mass of a nucleon. Such tiny violations can be be avoided by considering an energy conserving extension of the model \cite{smirne2014dissipative}. Moreover, while the GRW model does not apply to identical particles, a suitable extension to comprise them was proposed in~\cite{tumulka2006spontaneous}. Finally, we recall that the model is constructed in a non-relativistic setting, while the possibilities to extend it to the relativistic case were investigated in~\cite{tumulka2006relativistic,jones2019possibility}.

 \paragraph{Continuous Spontaneous Localization model}
The Continuous Spontaneous Localization (CSL) model  \cite{pearle1989combining,ghirardi1990markov} can be considered the natural extension of the GRW model when the collapse process occurs continuously in time and, since it is formulated in the second quantized formalism, it can be straightforwardly applied to identical particles. The modified Sch\"odinger equation reads \cite{bassi2003dynamical}
  \begin{equation}\label{eq.CSL}
 \exd\ket{\psi_t}=
 \left[
 -\frac i\hbar \hat H\exd t+\sqrt{\lambda}\int\exd \x \,\hat N_t(\x)\exd W_t(\x)
-\frac{\lambda}{2m_0^2}\int \exd\x \int\exd\y\,g(\x-\y)\hat N_t(\x)\hat N_t(\y)\exd t
 	\right]\ket{\psi_t},
 \end{equation}
 with $\hat N(\x)=\hat M(\x)-\braket{\psi_t|\hat M(\x)|\psi_t}$ denoting a non-linear operator where
 \begin{equation}
\hat M(\x)=\sum_j m_j\sum_s\hat a^\dag_j(\x,s)\hat a_j(\x,s),
 \end{equation}
is the mass density operator with $\hat a_j(\y,s)$ denoting the annihilation operator of a particle of type $j$, mass $m_j$, and spin $s$ at the position $\y$. Here, $m_0$ is the mass of a nucleon, $\hat H$ is the Hamiltonian generating the standard Schr\"odinger evolution and $ W_t(\x)$ is a Wiener process defined by zero average and $\mathbb E[\exd W_t(\x)\exd W_t(\y)]=g(\x-\y)\exd t$, where $\mathbb E[\,\cdot\,]$ denotes the average over the stochastic process. The spatial correlation of the noise is chosen as $g(\x-\y)=e^{-( \y-\x)^2/4\rC^2}$. The parameters $\lambda$ and $\rC$ have the same meaning as in the GRW model and, since the CSL model reduces to GRW for a single particle, also the theoretical proposals for their values are the same. The choice $\hat M(\x)$ of the collapse operator ensures the localization in position basis as well as the amplification mechanism, which is underlined by the proportionality   of $\hat M(\x)$ on the mass of the particles undergoing the collapse.
Similarly to the GRW model, a dissipative extension of the CSL model was developed \cite{smirne2015dissipative,nobakht2018unitary}, while there is no satisfactory extension of the CSL model to the relativistic regime \cite{jones2020mass}. Moreover, several studies \cite{adler2008collapse,bassi2010breaking,ferialdi2012dissipative,torovs2017colored,carlesso2018colored,adler2019testing} were performed when the white noise $ W_t(\x)$ is substituted by a colored one, which is characterized by a suitable time-correlation function.

\paragraph{Connection to gravity}
A possible connection of collapse models with gravity has been often discussed  \cite{diosi1987universal,diosi1989models,penrose1996gravity,bahrami2014role,bell2016quantum,gasbarri2017gravity,bassi2017gravitational}. There are two main motivations for this connection. First, all the other fundamental forces apart from gravity have been quantized and thus cannot provide the correct coupling needed for the non-linear collapse. Second, the requirement of an amplification mechanism suggests that the collapse should scale with the mass of the system, as in the CSL model. It is then natural to chose the mass density as the collapse operator and gravity as a well-motivated source of quantum collapse. 

Different models were proposed to implement this idea. Diosi formulated a collapse model with the same structure as in Eq.~\eqref{eq.CSL} where the spatial correlation of the noise is related to the Newtonian potential $g(\x-\y)=1/|\x-\y|$ and the CSL coupling $\lambda/m_0^2$ is substituted by $G/\hbar$, where $G$ is the universal gravitational constant~\cite{diosi1987universal,diosi1989models}. For Penrose \cite{penrose1996gravity}, a spatial superposition of a massive system implies a superposition of spacetimes with a consequent ill-defined time-translation operator which reflects an uncertainty in energy~\cite{bahrami2014role}. As a consequence, these superpositions are suppressed with a characteristic time coinciding with that predicted by the Diosi model. These two proposals are commonly known as forming the Diosi-Penrose (DP) model which, in turn, can be considered to be a particular instance of a broader class of models~\cite{kafri2014classical,kafri2015bounds,tilloy2017principle,altamirano2018gravity,khosla2018classical,gaona2021gravitational}. Adler suggested that the collapse mechanism could be driven by complex stochastic fluctuations of the spacetime metric $g_{\mu\nu}$ \cite{bell2016quantum}. Based on this idea, a fully consistent non-Markovian collapse model has been then developed in~\cite{gasbarri2017gravity}. In particular, in the Markovian limit, such a model can be reduced to the CSL one with a correlation length $r_\text{\tiny C}$ and a collapse strength parametrized by $\xi$. 
 
  \begin{figure}[ht]
  \centering
 \includegraphics[width=0.5\linewidth]{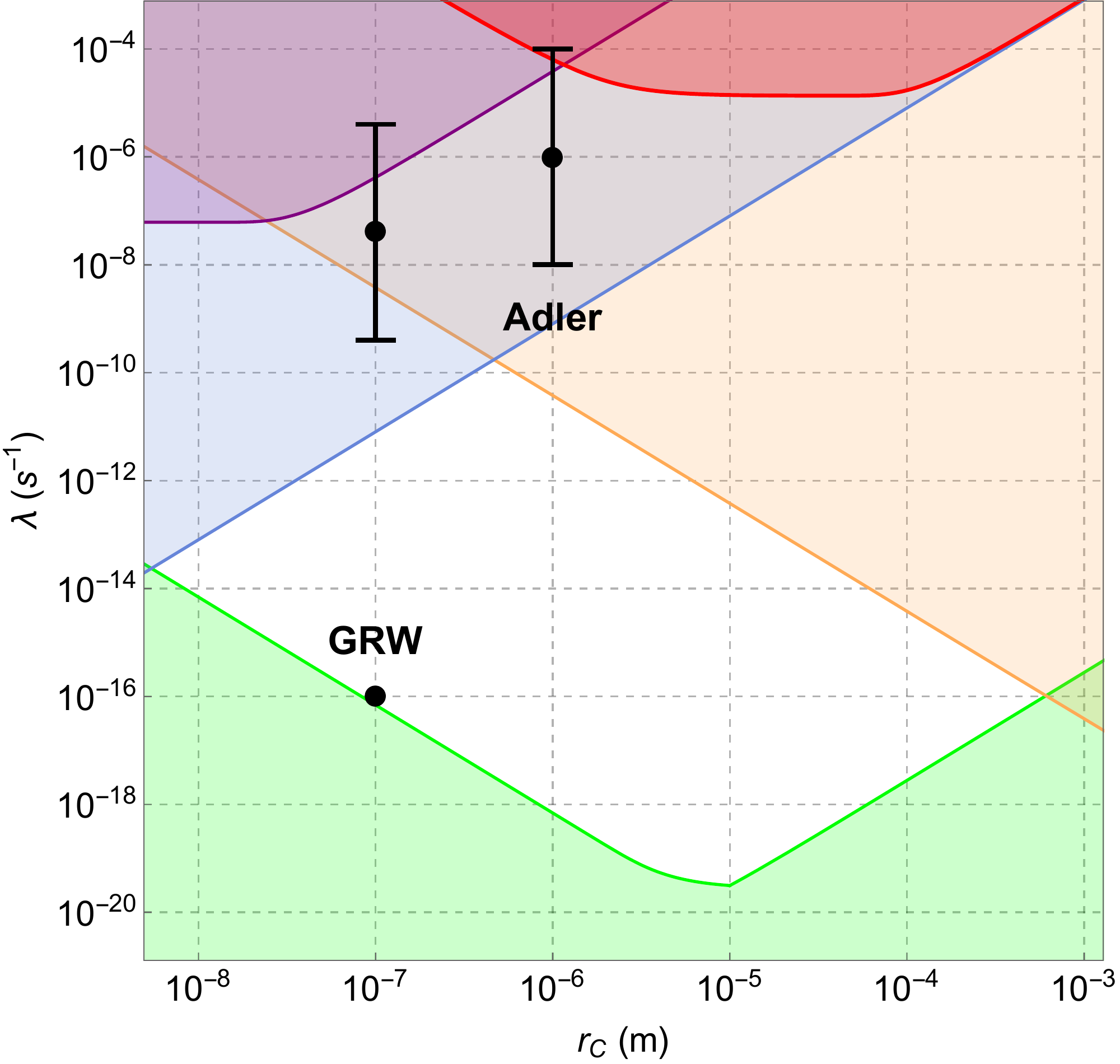}
 \caption{\label{fig.CSL.bounds} Exclusion plot for CSL model, which is characterized by the collapse rate $\lambda$ and correlation length $r_\text{\tiny C}$. Experimental bounds: orange area from LISA Pathfinder analysis \cite{carlesso2016experimental, carlesso2018non}, blue area from X-rays measurements \cite{piscicchia2017csl}, purple area from interferometric experiments with macromolecules \cite{torovs2017colored,torovs2018bounds,fein2019quantum}, red area from experiments with entangled diamonds. Theoretical lower bound: green area from localization of macroscopic objects \cite{torovs2017colored,torovs2018bounds}. }
 \end{figure}

   \begin{figure}[t]
   \centering
 \includegraphics[width=\linewidth]{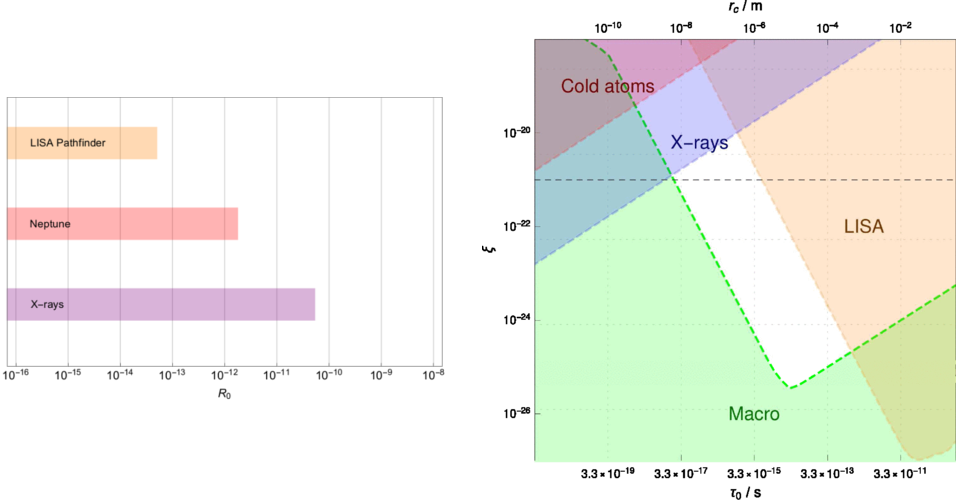} 
 \caption{\label{fig.DP.bounds} (\textbf{Left panel}) Exclusion plot for the DP model, which is characterized by the correlation length $R_0$. The colored lines identify the vaues of $R_0$ that have been excluded. Experimental bounds: Blue line from LISA Pathfinder data \cite{helou2017lisa, vinante2019testing}, 
 red line from the blackbody emission of Neptune \cite{adler2019testing}, blue line from X-rays measurements \cite{donadi2020underground}.  (\textbf{Right panel}) Exclusion plot for the gravity induced collapse model proposed by Adler \cite{bell2016quantum} and developed in Ref.~\cite{gasbarri2017gravity}. The model is characterized by the collapse strength $\xi$ and correlation length $r_\text{\tiny C}$ or equivalently time $\tau_0=r_\text{\tiny C}/c$. Experimental bounds: orange area (LISA) from LISA Pathfinder data \cite{carlesso2016experimental}, blue area (X-rays) from X-rays measurements \cite{curceanu2015x}, purple area (Cold atoms) from cold atom analysis \cite{bilardello2016bounds}. Theoretical lower bound: green area (Macro) from localization of macroscopic objects \cite{torovs2017colored,torovs2018bounds}. As a reference, the dashed grey line quantifies the magnitude of the first observed  gravitational wave in LIGO \cite{abbott2016observation}. Figure in the right panel from~\cite{gasbarri2017gravity}.
 }
 \end{figure}
  
\paragraph{Experimental tests}
The most direct and natural tests of collapse models are those based on the generation and probe of a massive superposition. Among them, experiments generating superpositions of electrons~\cite{davisson1927scattering,thomson1927diffraction}, atoms~\cite{kovachy2015quantum} and molecules~\cite{eibenberger2013matter,torovs2017colored,torovs2018bounds,fein2019quantum} by now confirmed the validity of quantum mechanics up to masses of $2.5\times10^4$\,amu and thus impose first important bounds on the parameters of collapse models. Yet, these results are orders of magnitude weaker than those obtained by the non-interferometric class of experiments~\cite{carlesso2019collapse}. With reference to optomechanical systems~\cite{vinante2016upper,carlesso2016experimental,vinante2017improved,zheng2020room,vinante2020challenging,helou2017lisa} (see also discussion in Sec.~\ref{Applications} and Sec.~\ref{ProofOfPrincipleImlementation}) the collapse acts on the mechanical system as a noise, whose strength is determined by the specific parameters of the model. An alternative way to determine the character of the noise is to search for a possible collapse induced heating of the internal degrees of freedom of bulk materials~\cite{adler2018bulk,tilloy2019neutron,adler2019testing}, or an enhanced expansion of a cold atom cloud~\cite{bilardello2016bounds}. Additionally, if a noise acts on an electrically charged particle, there will be a spontaneous radiation emission which can be monitored to further test  collapse models~\cite{piscicchia2017csl, donadi2020underground}. The state-of-art exclusion plots for the parameters of CSL are reported in Fig.~\ref{fig.CSL.bounds}, while those for the DP and Adler's model are shown in Fig.~\ref{fig.DP.bounds}.
It should be noted that the results of interferometric experiments are robust against possible modifications of the  collapse process, like in colored and dissipative~\cite{smirne2015dissipative,bahrami2014role} extensions. For these extensions, some of the bounds from non-interferometric experiments can in fact be evaded~\cite{carlesso2018colored,nobakht2018unitary}.

Implementing tests of the validity of quantum mechanics at macroscopic scales in space would avoid some of the limitations plaguing current ground-based experiments~\cite{gasbarri2021testing}. Experiments performed in space could benefit from a microgravity environment and long observation times that would allow to achieve high levels of precision. A microgravity environment is not strictly required for some types of non-interferometric experiments, but it would be beneficial in order to avoid vibrations from masking potential effects resulting from collapse mechanisms. On the other hand, a free-falling system is a fundamental aspect of interferometric experiments; using near-field interferometry~\cite{RevModPhys.84.157,Bateman2014,PhysRevA.100.033813}, as in state-of-art experiments, tests with masses of $10^6$\,amu would require a free-fall time of around $1\,$s to form an interference pattern. Such free-fall times can, in principle, be achieved in ground-based experiments~\cite{Bateman2014,fein2019quantum,gasbarri2021prospects}. However, for larger masses significantly longer free-fall times are necessary, which makes space particularly attractive for experiments employing massive objects.
MAcroscopic Quantum ResOnator (MAQRO)~\cite{kaltenbaek2012macroscopic,kaltenbaek2013maqro,hechenblaikner2014cold,zanoni2016thermal,kaltenbaek2014maqro,kaltenbaek2016macroscopic} is a  proposal for a scientific, medium-size space mission that is fully dedicated to testing quantum physics of massive nanospheres, with a precision that allows testing for possible deviations from the predictions of quantum mechanics in a yet uncharted parameter regime. MAQRO aims at performing interferometric experiments with masses up to around $10^{10}\,$amu by exploiting the favourable conditions provided by space. As a result, it could provide strong experimental bounds for theoretical modifications of quantum theory. In particular, MAQRO plans to provide a platform for quantum experiments at cryogenic temperatures lower than $20\,$K, extremely low vacuum pressures $\leq 10^{-13}\,$Pa, and a  microgravity environment $\leq 10^{-9}\,g$. These conditions would allow matter-wave interferometry with free-fall times on the order of $100\,$s and test-masses ranging from $10^8$\,amu ($30$~nm radius for fused silica spheres) to $10^{11}$\,amu ($150\,$nm radius with Hafnium dioxide spheres)~\cite{kaltenbaek2012macroscopic,kaltenbaek2016macroscopic}. The ESA's Concurrent Design Facility (CDF) study of the mission~\cite{CDFMAQRO2019}, which subsequently identified the possible space mission as the Quantum Physics Payload platForm (QPPF), has highlighted that such experiments are feasible in principle, but that existing technology limits the achievable vacuum to $\sim 10^{-11}\,$Pa. This limits the maximum mass of test particles to about $5\times 10^9$\,amu.  MAQRO and QPPF will be further discussed in Sec.~\ref{ProofOfPrincipleImlementation}.

\section{Applications of Quantum Technologies in Space}\label{Applications} 
\begin{figure}
    \centering
    \includegraphics[width=0.9\linewidth]{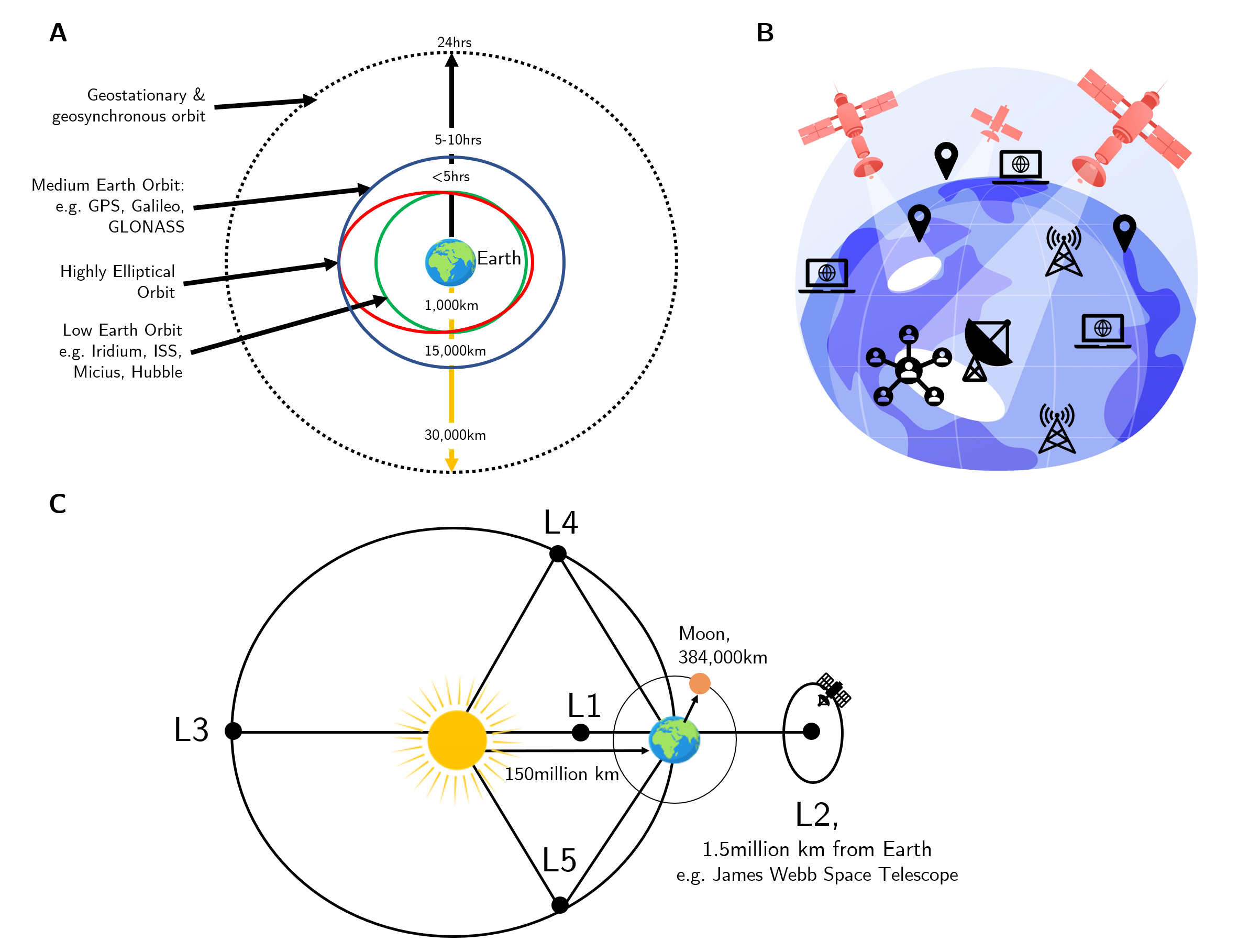}
    \caption{Figure A shows typical orbits around Earth, defined by the distance from Earth and the orbit cycle time. The use of quantum technologies in space can enhance ground based applications such as navigation, positioning, or providing greater coverage of communications including encrypted data transfer. Figure C shows orbits much further away from Earth than shown in Figure A; the Lagrange points are labelled (not to scale). These are  positions in space where objects will stay relatively fixed because the gravitational pull of two large masses precisely equals the centripetal force required for a small object to move with them. These points in space can be used by spacecraft to reduce fuel consumption needed to remain in position. The L2 Lagrange point is recommended for the MAQRO mission because of the high thermal stability, low gravitational gradients, reasonable communication bandwidth, and the feasibility of excellent microgravity conditions.
    }
    \label{fig:orbits}
\end{figure}

The same advantages space offers for fundamental tests also make it attractive for applied purposes. The long lines of sight and the low losses of free-space optical transmission in space, compared with the losses in fibre or the free-space transmission through the atmosphere, are the driving motivations for satellite quantum key distribution (QDK) 
and long-term applications beyond QKD, like entanglement distribution and the quantum internet~\cite{sidhu2021advances}. In this quest, the vantage point from low Earth orbits (LEOs) -- generally at an altitude between 160 and 1000 km -- is valuable for remote sensing and remote observation, as for example 
for Earth observation. Other orbits (see Fig.~\ref{fig:orbits}) are useful in alternative applications. For example, geostationary orbits (GEOs) -- at an altitude of 35,786 km -- are useful for meteorological analysis over specific areas and telecommunications where ground stations can establish continuous links with the satellite. In particular, as little as three equally-spaced satellites in GEO are required to establish near global coverage. Satellites in medium Earth orbit (MEO) are anywhere between LEO and GEO, and a constellation of such satellites is generally used to provide global navigation communications. Finally, a Sun-synchronous orbit (SSO) -- generally at an altitude of between 500 to 800 km -- ensures the same region on Earth is observed at the same local time, which is valuable for monitoring changes at ground or atmospheric level.

The development of quantum sensors may bring many benefits to existing markets, and these developments may open up new avenues of exploration. As a relevant example, space-based quantum clocks allow for the distribution and synchronisation of timing information and promise performance upgrades of existing global navigation satellite systems (GNSS). These clocks  
also enable distributed quantum information processing such as faster algorithmic processing of data through distributed quantum computing~\cite{fitzsimons2017private}. 

In this section, we review the applications of space quantum technology and the benefits and features that quantum technologies may bring.
 
\subsection{Quantum Communication}
{Satellite-based quantum communications was proposed in the late 1990s as a solution to long-distance quantum secured links~\cite{Hughes1999_SPIE}. Indeed, exponential losses in optical fibres constrain the range of quantum channels. Quantum repeaters can extend this range by segmenting the channel into separate links~\cite{Briegel1998_PRL,Jiang2009_PRA,Munro2010_NP,Zwerger2018_PRL}, which can be generalised to arbitrary quantum networks providing efficient routing of entanglement~\cite{Pant2019_NPJQI}. The use of quantum memories at each node can then also improve entanglement distribution rates to sub-exponential scaling with total distance~\cite{Gundogan2020_arxiv}, but remains sub-optimal for global networking. However, deterministic correction of errors requires small inter-node distances, which places deployment difficulties~\cite{Muralidharan2014}. Finally, terrestrial free-space transmission can extend the range of quantum links, but it suffers from losses due to long atmospheric path length and is restricted by the curvature of the Earth\footnote{So far, the longest terrestrial free-space quantum link is 144 km between two mountains in the Canary Islands, which were chosen to establish line of sight and to reduce the optical depth of the atmosphere~\cite{ursin2007entanglement}.}.

In this context, the use of space-based segments provides the most promising route to global scale-up of quantum networks~\cite{chen2021integrated}. Optical signals can propagate almost indefinitely in space, suffering negligible absorption in high vacuum. The main reduction in transmission is due to diffractive spreading of a beam as it freely propagates. Such a reduction in intensity follows an inverse square law beyond the near field regime, which is exponentially better to what can be achieved in fibres~\cite{feynman1965flp}. 

Satellite-based quantum communications also have their own unique challenges. Integrating space segments with terrestrial networks may generate further losses. For an Earth-space link, the beam needs to traverse the atmosphere and hence suffers losses due to absorption, scattering and turbulence-induced beam wander, but these losses are independent of the total range being confined mostly to the lower few kilometres of the atmosphere~\cite{vasylyev2019satellite}.}

\subsubsection{Quantum Key Distribution} 
\label{subsec:QKD}
The goal of a QKD protocols is to privately share a secure random encryption key between two remote trusted parties \cite{gisin2002quantum, duvsek2006quantum, diamanti2016practical, pirandola2019advances,sidhu2021advances}. Upon successful realization of the protocol, the security of the shared key is guaranteed by the fundamental principles of quantum physics. Access to a link granting such level of security and integrity is an important asset to critical infrastructure providers, governmental, military and corporate sectors \cite{bedington2017progress, laenger2010etsi}. A QKD link can be used for protection of data backup, continuity processes, transactions, as well as for securing network infrastructure, systems of supervision and control. 
	
While sharing a common goal, there is a variety of approaches to how exactly the QKD protocol can be implemented, differing in preparation techniques, encoding, measurement types and assumptions made regarding used equipment, to name a few. The QKD protocols can be divided into two main families, namely discrete-, and continuous-variable, DV and CV respectively. The former make use of quantum systems defined on finite-dimensional Hilbert spaces and encode key bits onto discrete degrees of freedom of a carrier system~\cite{duvsek2006quantum}, ideally a single photon, and employ single-photon avalanche diode (SPAD) or superconducting nanowire single-photon detector (SNSPD) for measurements. The well-known representative of the DV family is the seminal BB84 protocol that operates with polarization qubits~\cite{bennett2014gb}. On the other hand, CV QKD protocols use quantum systems described on infinite-dimensional Hilbert spaces, and encode the key bits onto continuous observables of the light field, such as the quadratures of generally multiphoton Gaussian coherent or squeezed states~\cite{braunstein2005quantum, weedbrook2012gaussian}, measured on homodyne or heterodyne detector using positive-intrinsic-negative (PIN) diodes.   
	
\begin{figure}[t]
        \centering
        \includegraphics[width=.75\linewidth]{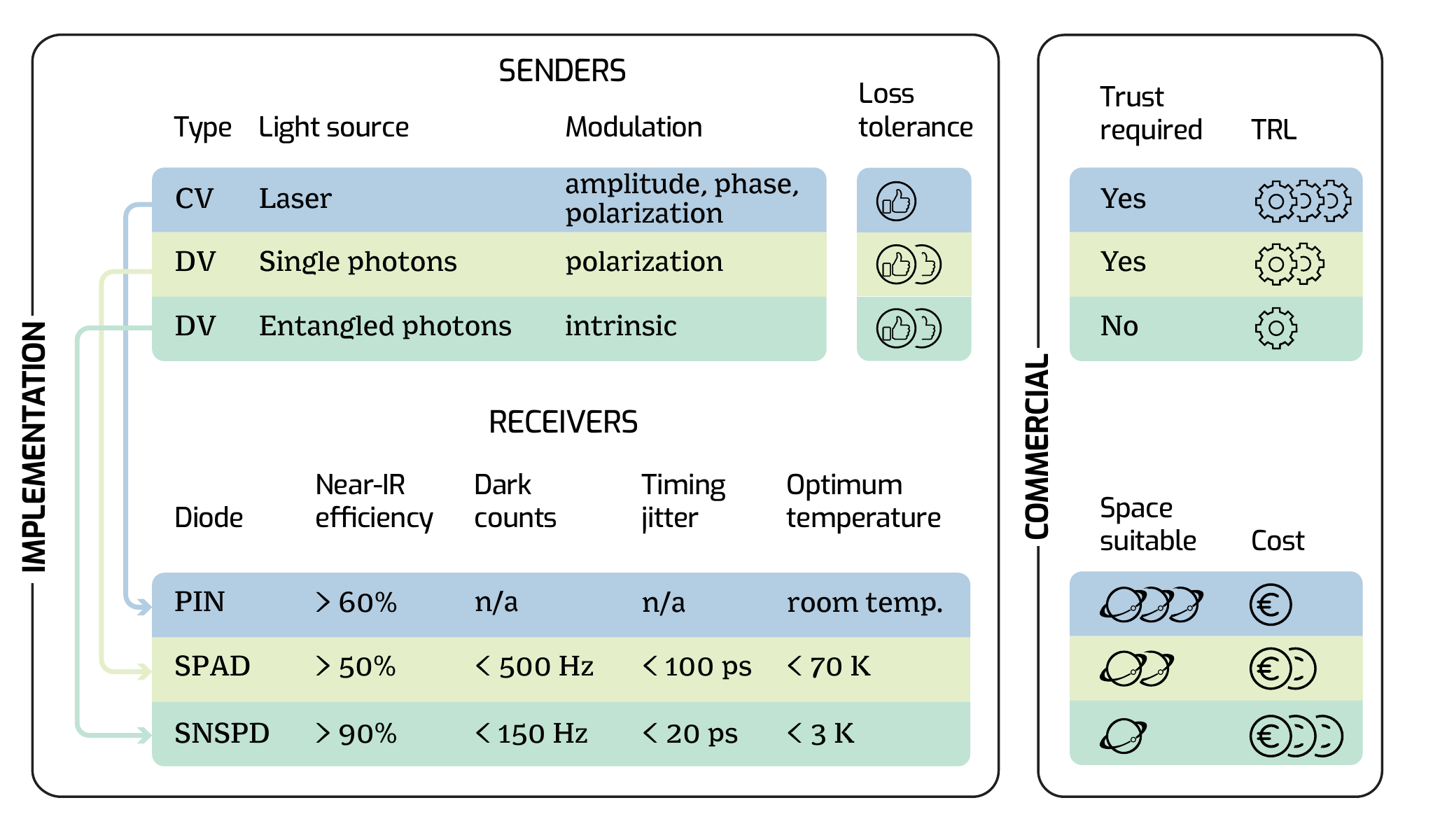}
    \caption{\label{fig:qkdprotocols}  A brief summary of implementation and commercial aspects of different QKD protocols.
    Icons illustrate relative value for a parameter, with higher number corresponding to a higher value of the respective parameter.
    TRL stands for technology readiness level which estimates technical maturity of the equipment required for protocol implementation. 
    For further details regarding receiver parameters see \cite{bronzi2016, esmaeilzadeh2017, huang2016long, zhang2020long}. 
    Figure adapted from \cite{khan2018satellite}.}
    \label{fig:table}
    \end{figure}

Both families can be further divided with respect to the realization scheme, be it prepare-and-measure or entanglement-based. The former is designed for the sender to prepare an optical signal and actively alter its state according to the prescribed key mapping. In the entanglement-based scheme instead, the trusted parties share an entangled state and conduct independent measurements on a received subsystem obtaining knowledge about the state of a remote subsystem. Regardless of the preparation, the secure key stems from the correlated data on both sides that is processed -- typically one-way -- using authenticated classical communication. A short summary of implementation and commercial aspects of different protocols is shown in Fig.~\ref{fig:table}.
	
The most straightforward satellite QKD link configuration is for satellite to act as a trusted node. The satellite assumes the role of one of the trusted parties 
in a prepare-and-measure QKD protocol and establishes independent secure keys with individual ground stations. The satellite stores all the keys and, upon request for connection between two ground stations, it broadcasts a bit-wise parity of keys established with respective stations. Given that both keys are independent secret strings, and their bit-parity is a uniformly random binary sequence, the announcement of the latter does not reveal any actual information to unauthorized third parties. On the other hand, the key stored at the ground station and the broadcast sequence are sufficient to infer the key stored at another station~\cite{polnik2020scheduling}.
 
\begin{figure}[t]
        \centering
        \includegraphics[width=.9\linewidth]{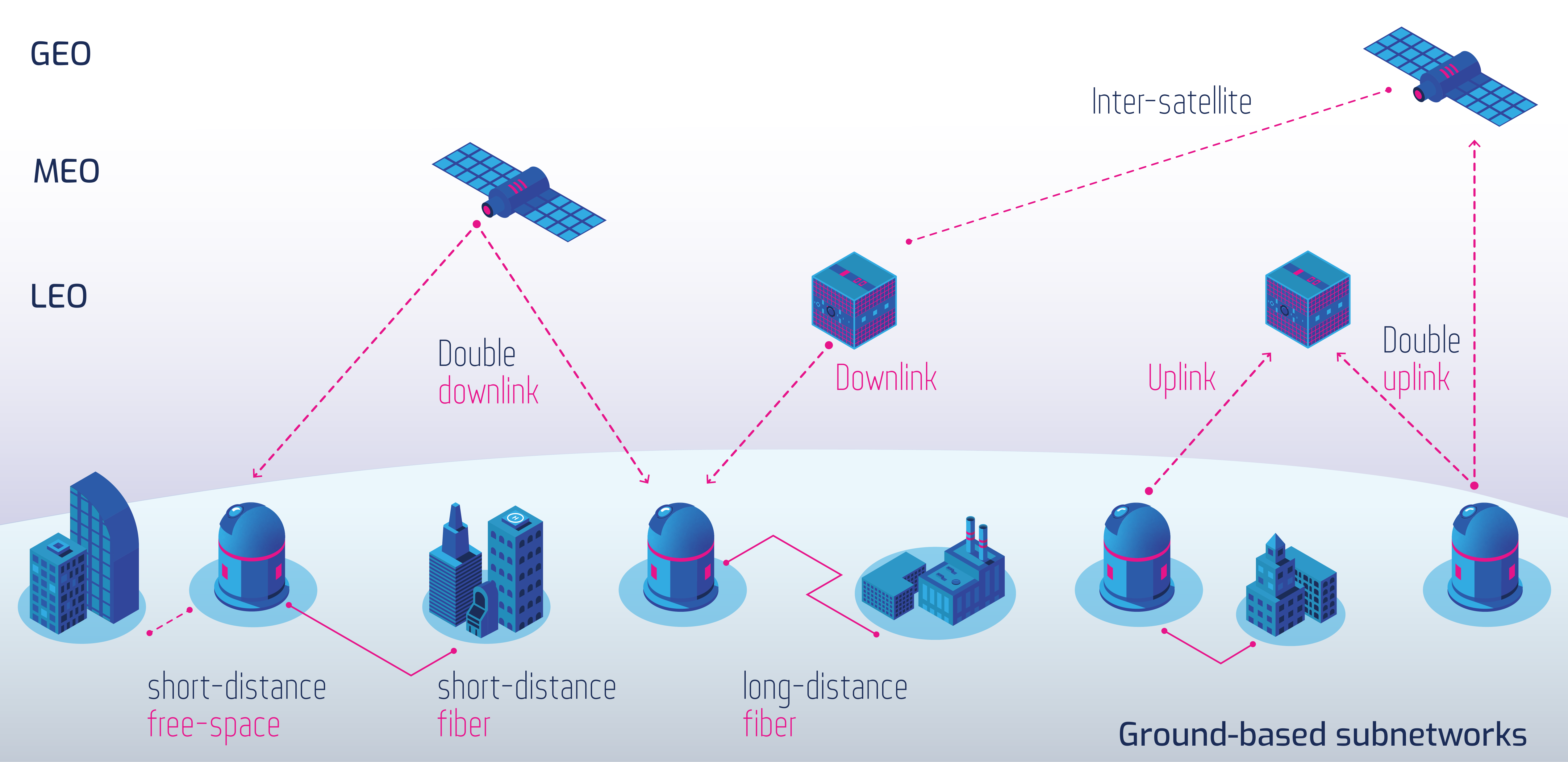}
    \caption{An example of QKD network with various ground (free-space, fiber) and satellite (uplink, downlink) based links allowing to establish a secure key between any pair of trusted nodes. }
        \label{fig:network}
    \end{figure}

Depending on the role assumed by the satellite, the connection can be established as an uplink or downlink~\cite{bonato2009feasibility}, see Fig.~\ref{fig:network}. In the former, the ground station prepares and sends quantum signals to the overflying receiver (also referred to as ground-to-space link), while in the latter the quantum signals are prepared and sent from the satellite to the ground station (space-to-ground link). The main difference between uplink and downlink is the amount of atmospheric attenuation. In the uplink scenario, the atmospheric effects (occurring in the troposphere and lower parts of the stratosphere) such as absorption, scattering, and scintillation contribute to beam spot spreading, deformation and wandering which at higher altitudes are enforced due to further diffraction-induced spreading~\cite{vasylyev2019satellite}. In the downlink scenario, the beam first travels through the vacuum enduring only diffraction-induced spreading and the atmospheric influence occurs only at the very end of the path resulting in lower overall signal loss. Thus, one can expect $10-20$\,dB of additional loss at near-infrared (near-IR) wavelengths in the uplink compared to the downlink under the same conditions~\cite{aspelmeyer2003long, liorni2019satellite}.
The downlink can also be simulated using powerful light sources at the ground station and retroreflectors on the satellite. The mean photon number of the light pulses emitted from the ground station is scaled~\cite{vallone2015experimental} according to the uplink channel transmittivity so that the mean photon number of the pulse reflected (and modulated) from the satellite is one or less, i.e.~in accordance with requirements for the source of the decoy-state QKD protocol~\cite{lo2005decoy}. 

Naturally, the trusted party at ground stations will be flexible and can upgrade or replace components, while the hardware configuration at  satellites must be space-qualified and will not be adapted or changed after the launch. 
This flexibility is especially important for continuous security maintenance of the prepare-and-measure protocol setup and prevention of potential side channel attacks. More specifically, theoretical models used for security analysis may not necessarily faithfully describe actual equipment used in a practical QKD setup, thus opening side channels and loopholes in security that can be exploited by a malicious third party. Moreover, active attacks aim to intervene and temper with the operation of QKD setups creating new vulnerabilities and side channels. Closing loopholes and the protection of the operation require suitable countermeasures. In the best case scenario, these imply improving the models and adapting the security proofs and/or post-processing, all of which would require the modification of the software component of the implementation. Some of the issues can only (or more efficiently) be fixed by modifying or introducing additional physical components to the setup. Examples of two distinct solutions to the same security threat, posed by the photon-number-splitting attack on multiphoton signal pulses in QKD protocols as BB84~\cite{brassard1984quantum}, are the introduction of decoy states that can directly detect such an attack \cite{lo2005decoy}, and the SARG04 modification of data processing at the cost of a reduced performance of the protocol~\cite{scarani2004quantum}. Furthermore, active attacks such as Trojan-horse~\cite{jain2014trojan}, faked-state~\cite{makarov2005faked}, detector efficiency mismatch~\cite{makarov2006effects}, wavelength-dependency~\cite{li2011attacking}, and many more can be prevented hardware-wise, by adjusting the design of the protocol's setup, including additional isolators, filters and detectors, and monitoring the nominal performance. 
However, as already noted, the hardware on the QKD satellite node cannot be updated upon deployment, which leaves the protocol vulnerable to potential novel attacks. Even the viability for future software update can potentially leave the backdoor to the system and compromise the protocol. On the other hand, conducting an active attack requires precise targeting of distant and moving satellite-based trusted station and presents a significant challenge for an adversary with today's technology~\cite{sun2015effect, makarov2016creation}. Besides, trusted stations can verify absence of complex devices within the direct line of sight between them~\cite{vergoossen2019satellite}.
 
If the trusted parties employ entanglement-based QKD protocols~\cite{ekert1991quantum, bennett1992quantum}, the requirement to trust the satellite node can be alleviated. In such configuration, the trusted parties reside at ground stations, while the satellite carries a source of entangled states and establishes a double downlink with the respective stations. Upon successful collection and measurement of both entangled subsystems, the measurement results at the two ends of the communication link will be correlated. In the BBM92 DV QKD protocol~\cite{bennett1992quantum}, the sequence of such symmetrical measurements forms a sifted key that, similarly to the one in the prepare-and-measure BB84 protocol~\cite{brassard1984quantum}, can be distilled and corrected from errors, thus resulting in a secure key. The security of the key is based on the inevitable appearance of errors in the raw key if an unauthorized party tries to become entangled to the signal entangled system. 
In the case of the E91 protocol~\cite{ekert1991quantum}, the security stems from the violation of Bell inequalities. This forces trusted parties to measure in more polarization bases. However, upon successful maximal violation of the inequality Alice and Bob are confident there were no eavesdropping. This is the security foundation for  device-independent (DI)  QKD protocols, as well. Such protocols make no assumptions regarding used equipment, thus eliminating all possible side channels, and only test classical inputs (measurement basis) and outputs (measurement results) of the protocol to verify randomness and integrity of shared binary strings. The implementation of DI QKD protocols is challenging and imposes strict requirements to detection efficiencies to close the detection loophole~\cite{pirandola2019advances}. 
	
Alternatively, measurement-device-independent QKD protocols lift trust assumptions regarding the detection devices only~\cite{lo2012measurement, pirandola2015high}. This is achieved by delegating the Bell measurement to a third party, with the trusted parties preparing and sending the states, but not receiving any. This corresponds to a double uplink with satellite carrying Bell-measurement station and acting as an untrusted node. The announcement of Bell measurement results and consequent post-selection allows to create strong correlations between data sets of remote trusted parties. Combining the protocol with decoy-state method circumvents the multi-photon emission issue of weak coherent sources~\cite{curty2014finite, yin2016measurement}. The main challenges of the implementation are the synchronization of the signals in both uplinks (see Sec.~\ref{subsec:synchronization}), the correction of the Doppler effect due to different apparent velocities of the satellite seen by the two ground stations, and the propagation losses in the links.

Another practical concern for satellite-based QKD is link availability. Aside from weather dependence~\cite{liorni2019satellite}, a foremost factor is orbital altitude of the satellite, either Low Earth Orbit (LEO), Medium Earth Orbit (MEO), or Geostationary Orbit (GEO). LEO is the most common orbit up-to-date due to its proximity to the surface which implies lower diffraction induced losses, lower exposure to ionizing radiation from the Sun, as well as lower launch cost than for orbits of higher altitudes. The downsides of LEO are high satellite speed relative to the ground station, limited line-of-sight window during a flyover and smaller geographical coverage. The first presents a challenge for accurate and fast pointing systems, the second prevents continuous communication and allows to distribute keys only once every one or two hours for a few minutes when the satellite is above 10 degrees of elevation above the horizon~\cite{bedington2017progress, lee2019updated}, and the third reduces the reach of distribution of entangled states. Reaching MEO would support a significantly longer communication window~\cite{larson1999space}, while the satellite at GEO would be permanently accessible. However, increasing the altitude also increases the channel loss and exposure to ionizing radiation. Crucially, it also decreases eclipse fraction of the total orbital period~\cite{larson1999space}, which curtails link access time with minimal background radiation that can be five orders of magnitude lower than the one during day-time operation~\cite{er2005background, gruneisen2015modeling}. Presence of scattered sunlight has confined current QKD tests to night-time operation only, and day-and-night link availability will require advanced acquisition, tracking and pointing systems, filtering, as well as precise temporal synchronization~\cite{liao2017long}. 
CV QKD protocols can be potentially operated during the day due to reliance on coherent detection during which the signal is matched with narrow-band local oscillator that serves as a phase reference for the signal and can efficiently filter out background radiation~\cite{gunthner2017quantum}. However, CV QKD protocols (as well as some DV encoding, such as orbital angular momentum \cite{wang2020satellite}) are sensitive to turbulent fluctuations of the refractive index within the air mass~\cite{dong2010continuous}. Such fluctuations induce untrusted excess noise, proportional to the size of the encoding alphabet and the variance of channel transmittance fluctuations~\cite{usenko2012entanglement}, and bound the range of atmospheric conditions and zenith angles that support secure key distribution. Sub-channel post-selection and data clusterization have been suggested to suppress fading noise influence~\cite{usenko2012entanglement, ruppert2019fading}. Furthermore, squeezed states can provide substantial improvement to the performance of the protocol, although they require optimization in accordance with the shape of transmittance distribution profile~\cite{derkach2020squeezing}. The latter requires accurate channel estimation~\cite{ruppert2019fading}, which also allows for noise suppression via adaptive optics~\cite{chai2020suppressing} or beam-spot size optimization~\cite{usenko2018stabilization}. Overall, feasibility studies show considerable challenges in satellite-based CV QKD~\cite{kish2020feasibility, derkach2021applicability, hosseinidehaj2021composable, dequal2021feasibility, pirandola2020satellite}, yet do not present fundamental limits for its realization.
Modelling of satellite link transmissivity provides an opportunity for preemptive optimization of protocol parameters, but must account for altitude-dependent atmospheric conditions, geographical position of the observer, variations of the slant range and refraction within communication window, etc.~\cite{vasylyev2019satellite}.  Lastly, other relevant effects for all long-range space-based QKD protocols include space-time curvature~\cite{bruschi2014spacetime} and gravity~\cite{pierini2018effects}.

Another point to stress is that the communication time availability determines the raw data block size that can be accumulated and, consequently, the length of the key. Furthermore, the size of the block influences the confidence intervals on estimated security parameters with given composite probability of protocol failure that encapsulates probabilities of successful error correction, parameter estimation, privacy amplification, etc. In other words, the longer time the communication link was established for, the more confident trusted parties can be that actual values of security parameters do not significantly deviate from their most probable values, and therefore that the lower bound on the key rate is correctly assessed~\cite{zhang2017improved}. Various approaches have been developed to ensure correct evaluation of the smooth min-entropy bounds obtained from the finite raw key, such as the exponential de Finetti theorem~\cite{renner2008security, leverrier2017security}, post-selection technique~\cite{christandl2009postselection}, virtual entanglement distillation~\cite{hayashi2012concise}, entropic uncertainty relations~\cite{tomamichel2012tight, furrer2012continuous}, or entropy accumulation~\cite{dupuis2020entropy}. Finite-size effects can be reduced by merging measurement results from different satellite passes to enhance block size, thus creating keys more reliable at the expense of additional time needed for data accumulation~\cite{bourgoin2013comprehensive,Sidhu2020_arxiv_2}.

Finally, the natural progression for space QKD, that eliminates the issue of link unavailability, is the development of global quantum networks. Such networks would consist of satellite constellations that could share a secure key between any two ground terminals (see example in Fig. \ref{fig:network}). Space QKD networks can be deployed for global, targeted or local coverage, and differ in the amount of employed satellites, their orbit types and altitudes, constellation geometry, etc.~\cite{zhang2012internetworking}. Within an {embassy} LEO constellation model, aimed at delivering a message from one ground station to a number of other stations, enabling intra-planar space-to-space links have been shown to drastically increase the key size for all ground stations~\cite{vergoossen2020modelling}. Connecting MEO or GEO relay satellites to LEO network can improve connection stability, handover management and network control, as well as decrease latency~\cite{zhang2012internetworking, biswasa2018iac}. Investigations into networking design, optimization of orbits, inter-satellite links, and QKD protocols are pending and will pave the way towards global quantum-secured communication~\cite{Sidhu2020_arxiv_2}. 

\subsubsection{Quantum enhanced communication}

The indistinguishability of non-orthogonal quantum states underpins the secure transmission of quantum information at the heart of unconditional security in quantum cryptography and QKD~\cite{bennett1992quantum}. Conversely, improved discrimination of multiple non-orthogonal quantum states is important for the efficient readout of encoded information. 
While different signal encodings, such as multilevel encoding~\cite{Betti1995_book} and phase-shift keying schemes~\cite{croke2006maximum,Sidhu2020_PRXQ}, permit increases to channel capacities, their extent depends on how well the signals are resolved. Seminal works by Helstrom and Holevo determined the fundamental limits to non-orthogonal quantum state discrimination~\cite{Helstrom1976, Holevo82}. Although quantum mechanics places fundamental limits to the distinguishability of two quantum states, it  also provides the tools to approach these limits through efficient readout measurements~\cite{bae2015quantum,ivanovic1987how, dieks1988overlap, peres1988how}. 

The choice of detectors and photon encodings used to implement a QKD link depends on the communication protocol used and the environment of the optical link. Development of quantum detectors can improve received data rates that leads to improved key generation rates~\cite{Takemoto2015quantum}. For satellite-based QKD, the receivers generally use single photon avalanche diodes (SPADs) or superconducting nanowire single photon detectors (SNSPDs). For discrete variable protocols, quantum signals are encoded in different photon degrees of freedom. This includes polarisation, frequency, orbital angular momentum (OAM), or spatial modes of qubits or qudits. For space-based applications, optical links usually have highly variable losses owing to large divergences at long-distance propagation and atmospheric turbulences. This makes OAM and spatial mode encodings unsuitable~\cite{krenn2016twisted}. Instead, polarisation or  frequency encodings are a natural choice given their robustness to atmospheric losses~\cite{vallone2015experimental,liao2017satellite,liao2018satellite,jeongwan2018pra,jin2019oe}. For other applications, these encodings have differing performances that may impact the attainable key rates. Similarly, the use of frequency encoding for satellite links has the drawback of requiring a compensation for the Doppler effect due to the Satellite motion~\cite{vallone2016}.

\subsubsection{Timing and synchronization}
\label{subsec:synchronization}

Remote correlation of quantum signals is important for many applications in networked quantum communication. It allows for precise positioning and navigation~\cite{Sidhu2017_PRA, Sidhu2018_arxiv, Sidhu2020_AVS}, distributed quantum computing~\cite{cirac1999distributed,fitzsimons2017private}, distributed quantum sensing~\cite{komar2014network, Sidhu2017_PRA, eldredge2018optimal, polino2019experimental, Sidhu2021_PRX, guo2020distributed}, and applications in fundamental science~\cite{kolkowitz2016gravitational}. Quantum clock synchronisation provides robust means to achieve accurate and secure time transfer by using entanglement based protocols~\cite{jozsa2000quantum, giovannetti2001quantum,yurtsever2002lorentz}. Analogous to classical time synchronisation, arrival times of entangled photon pulses are measured~\cite{wang2016influence}.  With entanglement shared across a network of optical atomic clocks, the intervening medium has no effect on the synchronisation~\cite{jozsa2000quantum}. Entanglement purification operations along each node of the network can remove systematic errors arising from remaining unsynchronised clocks. This reduces the overall complexity required by eliminating the requirement for a common phase reference between each clock~\cite{okeke2018_remote}. 

Synchronising multiple clocks on satellites presents significant challenges owing to high losses, long distances, and their relative motion. Practically, time synchronisation of local clocks at each node is established by encoding time signals into beacon lasers links. This establishes a frequency reference, while a portion of the detected signals can be sacrificed to compensate for the varying path length between different satellites or a link between a satellite and an optical ground station. This is sufficient to counter any relativistic effects. The feasibility of using satellite-based quantum clock synchronisation was demonstrated in Ref.~\cite{wang2016influence}, which accounted for a near-Earth orbiting satellite with atmospheric dispersion cancellation. A satellite-to-ground clock synchronisation that attained a time data rate of 9\,kHz and a time-transfer precision of 30\,ps has been demonstrated in Ref.~\cite{dai2020towards}. Alternatively, a technique based on the detection of quantum state themselves has been proposed and tested for fiber based communication \cite{calderaro2020fast, Agnesi2020simple}. This technique could in principle be extended to free-space and satellite channels. However, long distances, relative motion, and uncertainties in the time of arrival remain significant challenges.

\subsubsection{Deep space communication} 

Optical communication is an integral part of deep-space communication systems that allows for orders-of-magnitude higher data transfer rate compared to techniques utilizing radio frequencies~\cite{hemmati2006deep}. Unlike a link between a ground station and a terminal at near-Earth orbit (LEO, MEO, and GEO), deep-space optical communication links exhibit larger Doppler shifts~\cite{Boroson2004}, larger maximum point-ahead angles~\cite{Andrews2005}, and have extended operation time during small angular separations from the Sun~\cite{Boroson2004, Biswas2006}. 
To satisfy the demand for high data rates, one can either advance existing methods, improving the equipment and signaling techniques \cite{hemmati2011deep}, or employ novel methods for detection~\cite{Mueller2012, Chen2012, becerra2013experimental} and non-classical resources~\cite{Banaszek2020}.  
	
Among the planned projects for the future Lunar Orbital Platform-Gateway (LOP-G) station on lunar orbit, is the Deep Space Quantum Link~\cite{mohageg2018deep}. The goals of the Deep Space Quantum Link are to test the effects of gravity and different inertial reference frames on quantum teleportation, and to establish a space-to-space QKD link between stations on lunar and low Earth orbits, i.e.~the LOP-G and the International Space Station respectively.  
	
\subsubsection{Quantum Internet} 
\label{subsec:quantum_internet}

Space based links provide a natural means to extend the range of quantum communication protocols over global scales. Satellites links, together with ground based fibre networks will form the basis of a global-scale quantum internet. This will support the application of networked quantum information protocols that extend beyond QKD. Specifically, space quantum repeaters with quantum processors will enable a wider variety of protocols including distributed and secure multi-party quantum computing, sensing, and anonymous communications protocols~\cite{kimble2008quantum,Hybrid_qInternet,Sidhu2020_AVS,Gundogan2020_arxiv}.

A road-map to the realisation of a quantum internet requires the development of quantum networks with increased functionality~\cite{wehner2018quantum,sidhu2021advances}. Trusted repeater networks have already been demonstrated in metropolitan areas and between cities~\cite{castelvecchi2018quantum, dynes2019cambridge, zhang2019quantum,joshi2020trusted} and with satellite links~\cite{chen2021integrated}. An extension of this requires entanglement distribution through the network with and without quantum memories. This will enable implementation of clock synchronisation tasks and blind quantum computing. The final development will require error correction capabilities throughout the network. This will permit the operation of high fidelity quantum entanglement distribution and error correction for globally distributed tasks. The step increase in functionality of the network comes at the expense of increased technological difficulty.

\subsection{Quantum Random Number Generation}

A random string of bits that cannot be predicted, and unknown to adversaries, provides a fundamental resource for applications in cryptography. Specifically, the privacy, unpredictability, and randomness of random numbers certify the security of the resulting cryptographic encryption key. Quantum Random-Number Generators (QRNGs) exploit the unpredictability of quantum mechanics to provide enhanced security of this fundamental source of randomness that is independent of the underlying technological implementation~\cite{Gehring2021_NC}. The use of QRNGs can improve the operational trust of devices.

For applications in QKD, secure information-theoretic random numbers are required to drive the transmitter and receiver. This is particularly true in prepare-and-measure schemes where active basis choices are driven by a secure random source~\cite{rusca2019self}. For satellite based QKD, the transmitter rate has to generally be greater than $10^8$\,Hz due to high losses in the optical link. Standard weak coherent pulse decoy state (WCP-DS) schemes require varying number of random bits depending on the operation. Each pulse requires a random bit for the key bit, a bit for basis choice, and the choice of intensity values. There are generally three different intensities requiring two random bits, which overall leads to four random bits per pulse. Additionally, if biased probabilities are implemented (as for example in Efficient BB84~\cite{lo2005efficient}), then the required amount of raw unbiased bits is greater. An arbitrary biased bit can be non-deterministically generated from an average of two unbiased bits~\cite{gryszka2020biased}. For example, the BB84 WCP-DS scheme~\cite{hwang2003quantum, lo2005decoy} with biased basis and optimised decoy state probabilities requires a seven raw unbiased bits to generate one unbiased and three biased bits per pulse. At a source rate of 100 MHz, real-time generation of random numbers requires 700~Mbps of cryptographically secure bits.
In some circumstances, the random numbers can be pre-generated prior to the transmission pass at the cost of large amounts of storage -- 24.4\,GB for a five minute pass with the above source parameters. If the random numbers were generated in real-time, then the storage space could be minimised using a ring buffer and communication from the receiver to identify detections to be copied into storage~\cite{oi2017cubesat}. The size of the ring buffer will then depend on the latency between a receiver detection and on this being communicated to the transmitter. If we consider only a single transmission per orbit, then the off-line random number generation rate is still of the order of 40\,Mb/s.

There are several methods to attain true unpredictability and privacy of random numbers. The first is through a projective measurements on pure quantum states~\cite{pirandola2019advances}. Different QRNGs can be realised by exploiting the quantum uncertainty in photon counting measurements, phase measurements, or quadrature measurements~\cite{Ma2016_NPJQI, Collantes2017_RMP}. An alternative approach is the optical quadrature measurements of the vacuum state by means of a simple homodyne detection~\cite{Gabriel2010_NP, Haw2015_PRA}, which provides a promising route to chip integrability and cost-effectiveness. This approach has demonstrated high generation speeds of 2.9\,Gb/s~\cite{Gehring2021_NC}.

Despite these improvements to QRNGs, there are important challenges that need to be overcome for their use in space networks. First, the cost of QRNG technologies remain high and this makes alternative, less secure approaches attractive. Second, their miniaturisation and space readiness is essential for small quantum satellite missions. Finally, the quality assurance, certification, and standardisation of QRNGs is required. Current security evaluations of cryptographic and security products use the Federal Information Processing Standards Publication 140 standard, which defines the minimum-security level~\cite{fips140-3}. Such a  
standard may inhibit immediate adoption of QRNGs given their current technological readiness.

\subsection{Quantum Sensing of Gravity and Inertial Forces} 
We can define quantum sensing in two ways: quantum limited transduction, which defines the sensitivity of readout; versus sensing, which exploits either superposition or entanglement of the mechanical sensing object itself. Extremely sensitive readout can be met through use of shot-noise limited light, where fluctuations in the optical frequency and intensity are limited by the quantum statistical nature of photons. One can go further, and use squeezed light, where either the phase-noise or amplitude noise is reduced at the expense of increasing the uncertainty in the other quadrature, to further boost the sensitivity. Cooling the mechanical object to its quantum ground state enables the use of quantum superposition and entanglement, which cannot be recreated through classical means ~\cite{rademacher2020quantum}, with strong potential to beat current limits to sensitivity.

The geoid, defined by Earth's gravity field, is a surface of equal gravitational potential. In the absence of tides and currents, it follows a hypothetical ocean surface at rest. A precise model is crucial for understanding the ocean circulation, sea-level change and terrestrial ice dynamics, all of which are affected by climate change. Through gravimetry, one can directly infer information about sub-surface mass distribution, including volcanic activity monitoring~\cite{carbone2017theadded}, ice mass changes~\cite{makinen2007absolute}, subsidence monitoring~\cite{vancamp2011repeated}, and the detection of underground cavities~\cite{romaides2001comparison}. The latter is of interest to the oil and gas industry as well as the construction industry.
 
Free-fall acceleration sensors are known as absolute gravimeters because they give a direct measure of gravity, traceable to metrological standards. Relative gravimeters are masses supported by a spring, e.g. the stiffness of a cantilever, magnetic levitation, or the optical trapping of a nanosphere. One must calibrate relative gravimeters by measuring the stiffness of the spring and placing the instrument in a location with a known gravitational acceleration. Absolute gravimeters are therefore required to calibrate relative ones. Free-fall accelerometers are particularly suited for gravimetry applications aimed at resolving the temporal and spatial fluctuations of gravitational acceleration at the Earth's surface, which can vary roughly between 9.78\,m\,s$^{-2}$ and 9.83\,m\,s$^{-2}$~\cite{menoret2018gravity}.
Gradiometers on the other hand, are devices which can resolve gravity gradients by evaluating the difference between two measurements. For a clamped device such as a cantilever, a gravity gradient can be measured through the pull of gravity acting on two spatially separated masses. For cold atom systems, gradiometers often employ two ensembles of atoms injected into two interferometer paths vertically spaced apart~\cite{Leveque2021, carraz2014spaceborne}. 

In the subsections that follow, we discuss the current state-of-the-art in quantum gravity sensors and inertial measurement units. We focus on cold-atom laboratory devices and optomechanical quantum sensor proposals fit for a space mission. Where possible, we include space-specific feasibility study results. To date, no quantum measurements of gravity or inertial forces have been acquired in space.

\subsubsection{Earth-sensing with cold-atoms} 
Atom interferometry is a precise measurement tool, which 
does not require additional test masses, 
as the atoms are susceptible to accelerations. By exploiting the reduced friction in vacuum  and subsequent drift, atom interferometry can be employed to precisely measure accelerations and rotations~\cite{Berg2015, dutta2016continuous}. Understanding the measurement principle enables its use in various areas~\cite{bongs2019taking}, such as fundamental science~\cite{schlippert2014quantum, rosi2017quantum}, inertial sensors~\cite{geiger2020high}, gravity gradiometers~\cite{Fixler2007, rosi2014, Asenbaum2017}, and gravimeters~\cite{Peters1999}. Especially the latter two are used in Earth observation and sensing.

\begin{figure}[!t]
    \centering
    \includegraphics[width=\linewidth]{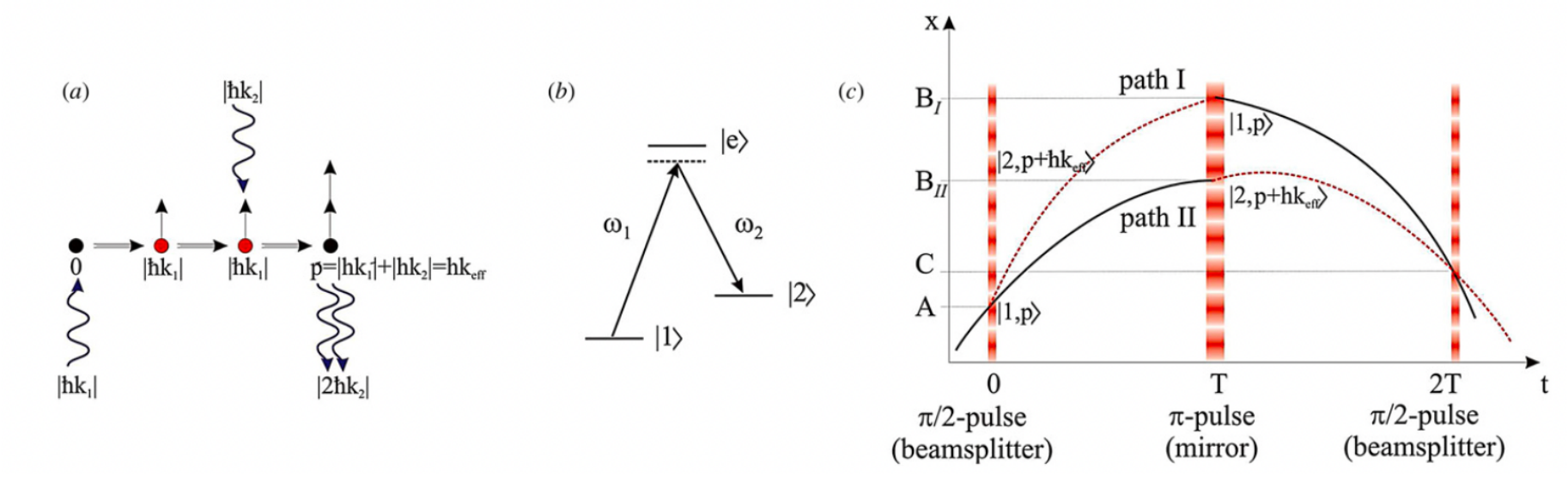}
    \caption{ Atom interferometer scheme for ground based systems following a Mach-Zehnder geometry. Panels (a) and (b) depict the stimulated Raman transition employed for the optical beam-splitters and mirror in panel (c). Panel (c) depicts the interferometer scheme in a ground based fountain, where the atoms trajectories are influenced by gravity. Figure from~\cite{deAngelis2008pic}.
    }
    \label{fig:atomgravi}
\end{figure}

Ground based atom interferometers, such as the gravimetric atom interferometer (GAIN)~\cite{Schmidt2011}, the absolute quantum gravimeter (AQG-A01)~\cite{menoret2017transportable, menoret2018gravity} or the transportable Quantum Gravimeter (QG-1)~\cite{heine2020trans}, rely on the acceleration of atoms in a gravitational field, as depicted for atom fountains in Fig.~\ref{fig:atomgravi}. The underlying principle is that of the Mach-Zehnder interferometer depicted in Fig.~\ref{fig:inter}. 

In order to measure not only gravity gradients but also rotations, the scheme has to be extended to combining four atom interferometers. As described in Ref.~\cite{carraz2014spaceborne}, the combination of the gravity induced phases allows measurements of rotation and gravity gradients.

The space-based gravity gradiometer based on the cold atom interferometry (CAI) mission proposal~\cite{carraz2014spaceborne} and mission study~\cite{Douch2018,Trimeche2019} discuss a 3D gravity gradiometer based on cold atom interferometry accommodated on a dedicated, nadir oriented satellite at low altitude for Earth observation.
The gravity gradiometers are implemented by acquisition of the differential signal from two atom interferometers separated by a distance of 0.5\,m in each of the three axes.
At a cycle rate of about 1\,Hz well collimated ensembles with $10^6$\,atoms enter each Mach-Zehnder-like interferometer based on double Raman diffraction with a total free fall time of 10\,s. With a sensitivity target of  $5\,\mathrm{mE\,Hz}^{-1/2}$ and a white noise at low frequencies, an improvement on the Gravity field and
steady-state Ocean Circulation Explorer (GOCE)~\cite{Rummel2011GOCE} results is expected from simulations.

\begin{figure}[!t]
    \centering
    \includegraphics[width=0.65\linewidth]{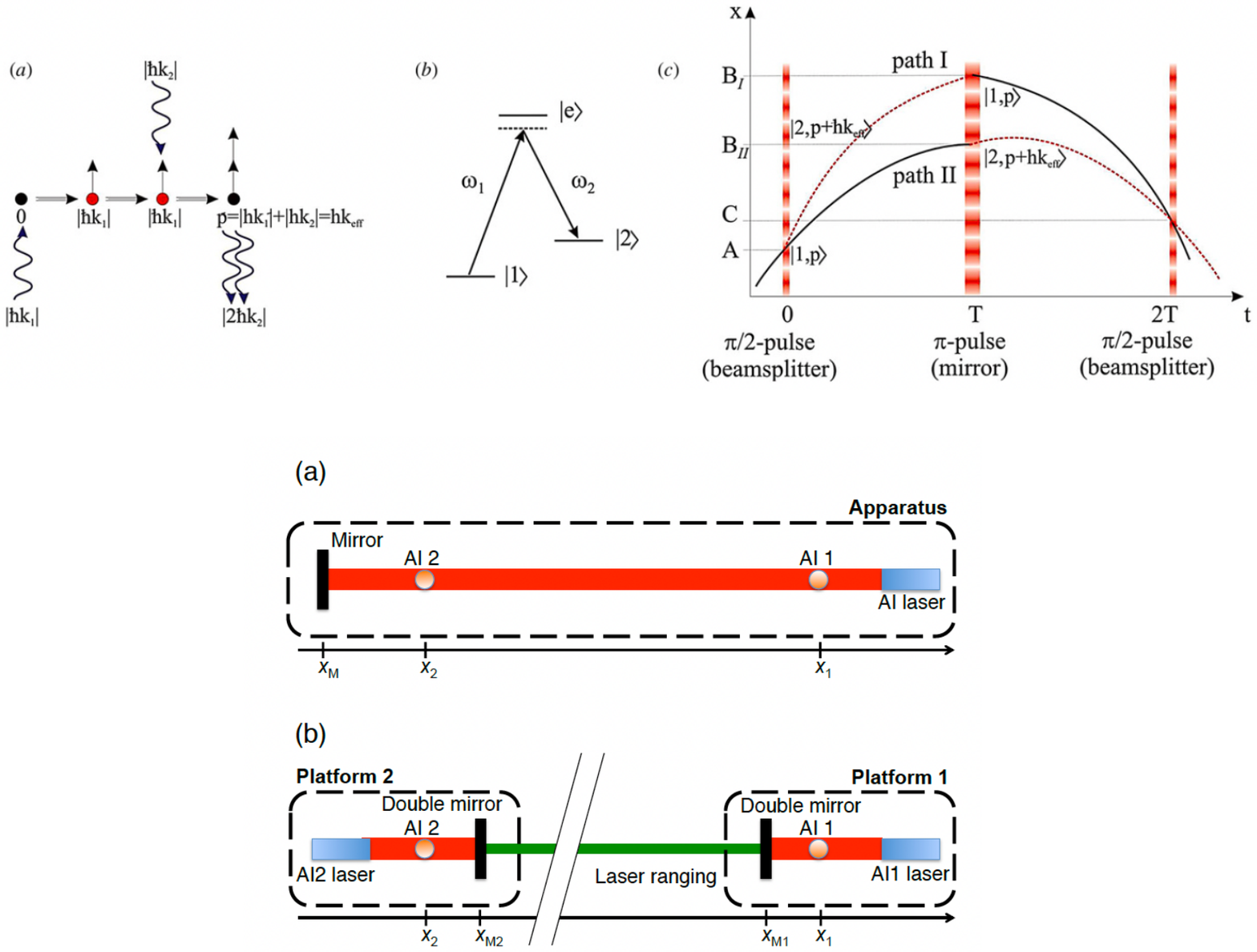}
    \caption{Scheme to map the gravitational field of the Earth using two satellites linked by an optical link. Gravitational gradients induce differences in the distance between the two satellites with the atom interferometers on board of each satellite acting as accelerometers.  Figure from~\cite{chiow2015laser}.
 }
    \label{fig:laserranging}
\end{figure}

Atom interferometrers can be employed in a scheme similar to the Gravity Recovery and Climate Experiment (GRACE~\cite{Tapley2004GRACE} and GRACE Follow-On~\cite{abich2019orbit, Velicogna2020GRACE}). In this scheme, two satellites follow each other with their distance constantly monitored. If the information on the distance between the two satellites is correlated to individual accelerators on both satellites, the gravitational field of the underlying planetary body can be mapped. 
Such a scheme, employing accelerometers based on atom interferometry, has been proposed by Refs.~\cite{Leveque2021, chiow2015laser}. A scheme of the two satellites housing accelerometers and the laser link is depicted in Fig.~\ref{fig:laserranging}.

While the design, payload, and budget of the above described space-based sensing devices are based on existing and available technology, further demonstration activities, dedicated development, and prototyping are recommended for an actual implementation in a mission.

\subsubsection{Earth-sensing with optomechanics}

\begin{figure}[!t]
    \centering
    \includegraphics[width=0.6\linewidth]{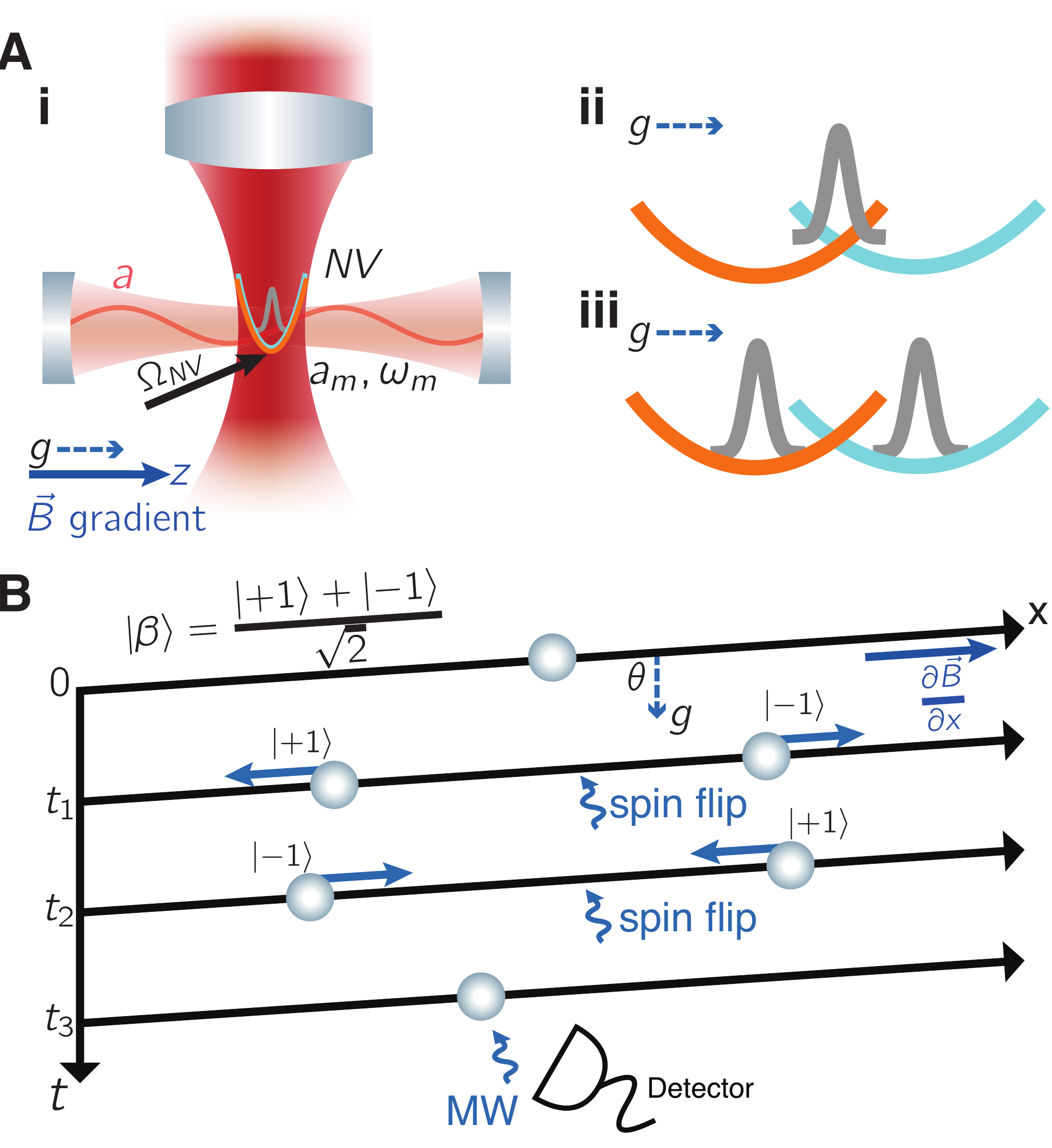}
    \caption{Optomechanical quantum sensing proposals. (\textbf{Panel A}) Spin-oscillator coupling has been proposed as a gravimetry technique, whereby a spatial superposition is created through the interaction of an embedded two-level-system such as a nitrogen-vacancy (NV) centre with an external magnetic field gradient. Variations of (A) have been proposed in~\cite{yin2013large,scala2013matterwave,chen2018highprecision}. A microwave pulse can be applied to split the spin states of the internal NV center in a trapped quantum nanosphere (i), which in turn creates a spatial superposition that can be viewed as the splitting of the optical trap (ii) into a superposition (iii). Note that the cavity is used to prepare the nanosphere in the ground state and can be switched off during (ii) and (iii) (\textbf{Panel B}) Allowing for free evolution increases the spatial size of the superposition, created and probed in a Ramsey-type interferometer, as proposed by~\cite{wan2016free}. Here, the coupled NV-nanosphere superposition is prepared with a microwave (MW) pulse at time $t_{1}$, undergoing free-fall until the spins are flipped at $t_{2}$ to enable matter-wave interferometry at $t_{3}$. Figure from~\cite{rademacher2020quantum}.
    }
    \label{fig:gravimetry}
\end{figure}

Dense macroscopic systems offer an enhanced sensitivity to acceleration when used in a test-mass on a spring set-up such as a quantum cantilever or quantum levitated nanosphere. Predictions for a single levitated quantum nanosphere are reaching acceleration sensitivities $10^{5}$ times higher than a cloud of cold-atoms~\cite{qvarfort2018gravimetry}. Levitated nanospheres can be isolated from environmental decoherence as much as cold-atoms. This isolation is crucial for the long coherence times to perform any kind of matter-wave interferometry and free-fall experiments. The benefit of performing gravimetry in space is the ability to work in a micro-gravity environment with greatly reduced sources of non-gravitational  noises. This is more important for clamped optomechanics experiments over levitated ones. It offers the ability to observe nearly 100\% of Earth’s surface for gravity~\cite{zhamkov2021next}. When conducting free-fall tests, increasing the mass of the test-object does not yield gains in sensitivity due to the equivalence principle, although there is ongoing research to validate this. Nevertheless, the main advantage in levitated optomechanics is the ability to embed spin degrees of freedom within the bulk of the nanosphere which allows generating quantum effects, such as entanglement and superposition, without requiring that the center-of-mass position is cooled to its quantum ground state. This is because the long coherence time of the spin state is used for readout rather than the position, which decoheres faster. This stands in contrast with cold-atom experiments, where the conditions for quantum behaviour are tighter. Hence, it is easier for optomechnical setups to reach increased coherence times, leading to prolonged free-fall times and measurement integration duration. As of writing, 1.4\,$\mu$s is the longest published coherence time for an optically levitated nanometer sized sphere cooled to the quantum ground state \cite{delic2020cooling}.
In Fig.~\ref{fig:gravimetry} we show the state-of-the-art proposals for spin-enabled quantum optomechanical sensing of acceleration which include the measurement of gravity and gravity gradients, and that will be discussed throughout the subsection.

Another benefit of using discrete variables such as spin for readout, rather than continuously measuring the nanoparticle position with light, is that discrete variables allow the use of heralded probabilistic protocols. They also benefit from high fidelity and resilience to background noise or detection losses. Proposals combining levitated particles with two-level systems include levitated nanodiamonds with an embedded nitrogen-vacancy (NV) centre with an electron spin~\cite{yin2013large, scala2013matterwave,chen2018highprecision, bose2017spin, marletto2017gravitationally}, and a superconducting ring resonator coupled to a qubit~\cite{johnsson2016macroscopic}. Here, we consider stationary spatial superpositions of a levitated nanoparticle oscillator with embedded spin, which can be manipulated by microwave pulses and  remains trapped by an optical tweezers throughout the sensing protocol, as illustrated in Fig.~\ref{fig:gravimetry}A. The first pulse introduces Rabi oscillations between the spin eigenvalue states $S_{z}=+1$ and $S_{z}=-1$, such that when a magnetic field gradient is applied the oscillator wavepacket is delocalized. This spin-dependent spatial shift is given by $\pm\Delta z=\frac{g_{\rm{nv}}\mu_{\rm{B}} B_{\rm{z}}}{2m\Omega^{2}}$, where $B_{\rm{z}}$ is the magnetic field gradient along the z-direction, which is the same direction that gravity acts in, $g_{\rm{nv}}\approx 2$ is the Land{\'e} $g$ factor, $\mu_{B}$ is the Bohr magneton and $\Omega$ is the frequency of the harmonic oscillator. This effectively splits the harmonic trapping potential, creating a spatial superposition with equilibrium positions governed by a spin-dependent acceleration. The spin-oscillator system now has states $\ket{+1}$ and $\ket{-1}$ in different gravitational potentials, accumulating a relative gravitational phase difference.
    
 For levitated systems the maximum spatial superposition that can be achieved, either through cooling to the quantum ground state or through manipulation of an embedded spin-state, tends to be much smaller than the physical size of typical nanodiamonds. This makes levitated sensing schemes unfriendly for resolving gravity gradients without the use of surveying or arrays. Releasing the nanoparticle from the trap, such that it undergoes free-fall, allows its wavefunction to evolve, increasing the position spread linearly in time. For detecting transverse accelerations, i.e.~to detect a nearby object placed perpendicular to the force of gravity, one can use a Talbot interferometer scheme proposed in~\cite{geraci2015sensing}. The levitated spin-oscillator Ramsey interferometer scheme shown in Fig.~\ref{fig:gravimetry}A can be modified for free-fall evolution, as shown in Fig.~\ref{fig:gravimetry}B. Due to the long coherence time of spin states, the superposition persists even if the oscillator does not remain in a pure coherent state. The scale of the superposition is controllable through flight time and magnetic field gradient. The measurement time $t_{3}$ is therefore unconstrained, and can be on the order of milliseconds, enabling spatial superpositions spanning~100\,nm, over three orders of magnitude larger than if the nanoparticle was levitated~\cite{wan2016free}, and comparable in scale to the size of the particle.

 The use of ground state cooled optomechanics for gravimetry in space has been proposed for levitated nanospheres to test the possible quantum character of gravity
\cite{carlesso2019testing,krisnanda2020observable} and clamped optomechanical systems \cite{ratzel2018frequency} to test space-time curvuture in black holes.

\subsubsection{Navigation} 

Quantum sensors fit for gravimetry have a dual use for navigation. Indeed, gravimeters are essentially ultrasensitive low-frequency accelerometers, whose signal can be double integrated in time to calculate position travelled with very little error or drift. Quantum measurements of rotation are also feasible, where atoms are used as gyroscopes~\cite{garrido2019compact, dutta2016continuous}. To navigate within a 3D space, 6 degrees of freedom need to be measured using an inertial measurement unit (IMU), which contains 3 accelerometers and 3 gyroscopes, one in each axis.

On Earth, quantum navigation sensors fulfil only a small number of application requirements due to the ease of access to satellite derived positioning, e.g.~using the Global Navigation Satellite Sytem (GNSS). Using classical IMUs such as high-volume manufactured micro-electro-mechanical systems (MEMS) devices made from silicon, where the changing capacitance between moving electrodes is a measure of inertial forces, together with GNSS, is a sufficient solution for the majority of positioning applications. Here, the GNSS signal provides a position fix at 10-50\,Hz rates. However, when GNSS is unavailable, such as in space, there are no longer positional fixes, resulting in an increasing positional error over time even when using military-grade inertial sensors. This is because the thermomechanical noise of the silicon structures in MEMS sensors, combined with electronic noise and drift on the readout, create growing errors of more than 1\,m already after a minute. By using quantum superposition and matter interferometry to derive position, such sources of noise and drift can be eliminated and reset \cite{cheiney2019demonstration}.

\begin{figure}[!t]
    \centering
    \includegraphics[width=1.1\linewidth]{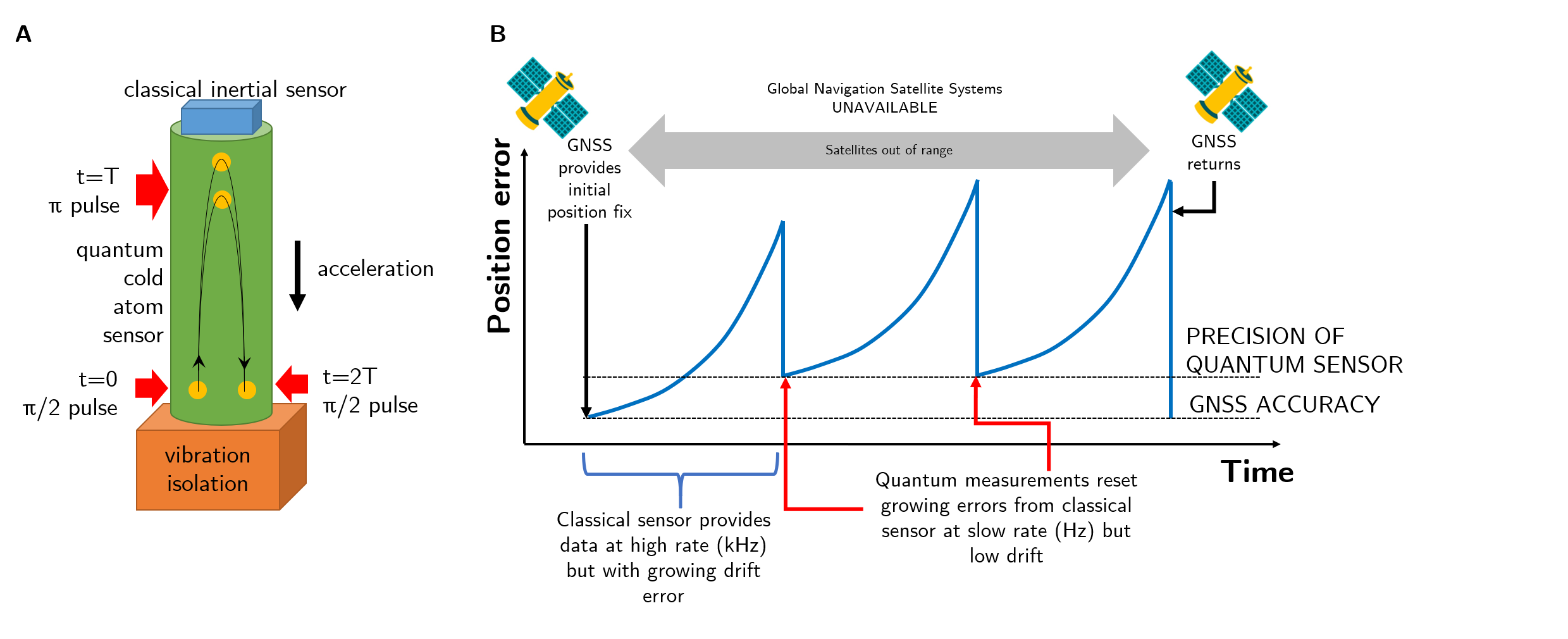}
    \caption{\label{fig:navigationfigure}  \textbf{A.} A typical example of a free-fall quantum cold atom inertial sensor; three precisely timed pulses of light apply momentum to the atoms, placing the atoms into a quantum superposition of two momentum states and recombining them. The first pulse gives half of the atoms an extra momentum kick. This causes one half to travel more quickly through space, splitting the cloud in two. After a time T has passed, a second pulse is used to invert the momentum difference of the two clouds, causing them to move towards each other. Finally, after further time T (t=2T) a third pulse is used to recombine the atom clouds to perform an interferometric measurement. The spatial superposition enables measurement of inertial forces, which can be used for positioning and navigation. To operate the sensor, classical inertial sensors are critical for enabling active vibration isolation of the mounts and base, with a separate inertial sensor used to supplement the cold atom low sampling rate. \textbf{B.} The output of a hybrid quantum sensor, where 'position error' is plotted against time. Here, the initial position would be provided by satellite measurement (Global Navigation Satellite Systems, GNSS) whilst the space vehicle is within range. Then, when GNSS becomes unavailable, it will be the classical inertial sensor that provides the position tracking. Classical accelerometers require performing double integration of the output to derive position, which is why the position error is shown to increase quadratically. The quantum cold atom sensor is able to make a measurement at a lower sampling rate to the classical sensor, and this measurement is used to periodically reset the error of the classical sensor, to the absolute limit of the quantum sensor.
}
\end{figure}

For platforms that need to be positioned rapidly or autonomously, the inertial sensing data should be measured in rates above 100\,Hz which is incompatible with the time of flight used in current free-fall experiments or the pulse sequence needed for Ramsey interferometry. A hybrid flywheel-type operation uses a classical IMU to provide inertial measurements in between each quantum measurement, as shown in Fig.~\ref{fig:navigationfigure}, whilst the quantum sensor resets growing errors. The return of the satellite signal allows for an absolute position fix rather than a relative one. This type of scheme is employed by cold-atom inertial sensor prototypes around the world~\cite{cheiney2018navigation, battelier2020three}. In turn, the quantum measurement, which is less susceptible to drift, is used to reset the growing errors accrued by the IMU. The classical IMU could be also replaced by an optomechanical sensor \cite{richardson2020quantum}, to enable optical and quantum based navigation across a wider bandwidth from DC to 1\,kHz, no longer limited by the sub-Hz to Hz bandwidth of the cold atom sensor.
 
Cold-atom systems enable both high accuracy and precision, where we define precision as the noise floor and accuracy as the degree to which the result of a measurement conforms to the correct value or standard. Quantum sensing of inertial forces, especially when using atomic transitions traceable to international standards, contain less drift and noise which reduces the accumulated error in position caused by double integrating a less noisy acceleration signal. A review of quantum navigation devices can be found here~\cite{geiger2020high}. 

The required sensitivity and drift parameters for navigation attitude determination systems (ADS) depend on the the platform. For a typical cubesat, the inertial sensor system is required to determine the resolution of thrust smaller than 0.01\,mN, resolution of rotation rate smaller than 1\,degrees/s, resolution of attitude determination during manoeuvre to be $\pm$5 degrees with standard deviation of 1$\sigma$, and update frequency higher than 5\,Hz \cite{larsson2016design}. The state of the art for inertial sensors used in space are gyroscopes with 0.15\,h$^{-1}$ bias stability with h being short for hour, and 0.02\,$^\circ\,$h$^{-1/2}$ angular random walk combined with accelerometers with 3\,$\mu$g bias stability and 0.02\,m\,s$^{-1}$/h$^{-1/2}$ velocity random walk \cite{nasa}. It should be noted that the quantum sensors also require navigation algorithms~\cite{cheiney2018navigation}. 
The current achieved parameters for quantum inertial prototypes are shown in Tab.~\ref{tab:comparison} alongside an example of a competitive classical accelerometer or gyroscope that is available commercially. 
\begin{table}[t]
\centering
\begin{tabular}{>{\arraybackslash}m{6cm}|>{\arraybackslash}m{8cm}}
\textbf{Existing System} & \textbf{Achieved sensitivity on ground} \\
\hline
\centering{\textbf{Accelerometers}} & [m\,s$^{-2}\,$Hz$^{-1/2}$] \\
Free-fall cube mirror$^{\dagger}$ &  $1.5\times 10^{-7}$ \textmd{($10^{-9}$\,m\,s$^{-2}$ in 6.25h)} \cite{lacoste} \\
Atom interferometer 2013 & $4.2\times 10^{-8}$ \textmd{$(3\times 10^{-9}$\,m\,s$^{-2}$ in 300\,s)} \cite{hu2013demonstration}\\
Atom interferometer 2016 & $9.6\times 10^{-8}$ \textmd{$(5\times 10^{-10}$\,m\,s$^{-2}$ in 2.8\,h)  {\cite{freier2016mobile}}}\\
On-chip BEC & $5.2\times 10^{-8}$ \textmd{($7.7\times 10^{-9}$\,m\,s$^{-2}$ in 100\,s)} \cite{Abend2016PRL}\\
Commercial atom gravimeter &$5\times 10^{-11} $ \textmd{($1\times 10^{-12}$\,m\,s$^{-2}$ in 40\,min)} \cite{muquans}\\
\centering{\textbf{Gyroscopes}} & [rad\,s$^{-1}$ Hz$^{-1/2}$]\\
Atom interferometer 2018 & $3.3\times 10^{-8}$ \textmd{($3\times 10^{-10}$\,rad\,s$^{-1}$ in 801\,ms)} \cite{savoie2018interleaved}\\
Atom interferometer 2020 & $6\times 10^{-10}$ \textmd{($6\times 10^{-10}$\,rad\,s$^{-1}$ in 1\,s)} \cite{gustavson2000rotation}\\
Commercial fiber optic gyroscope$^{\dagger}$ &  $5.8\times 10^{-7}$ \cite{honeywellhg9900}  \\
\hline
\textbf{Existing System} & \textbf{Achieved accuracy on ground}\\
\hline
\centering{\textbf{Accelerometers}} & [m\,s$^{-2}$]\\
Free-fall cube mirror$^{\dagger}$  & $2\times 10^{-8}$~\cite{lacoste}\\
Atom interferometer (2016) & $3.9\times 10^{-8}${~\cite{freier2016mobile}}\\
Commercial atom gravimeter & $\approx 10^{-12} $ \cite{muquans}\\
\hline
\end{tabular}
\caption{\label{tab:comparison} Table of a select few demonstrated cold atom accelerometers and gyroscopes. To the best of our knowledge, these represent the current best in sensitivity. Also included are classical state of the art sensors available commercially, indicated by ${\dagger}$. Not all sensors report accuracy, which is defined as the trueness of the measurement, as this requires calibrating the measurement against international standards. Note that the quantum systems have not been tested in space yet.}
\end{table}

Aside from exploring uncharted regions of space where there is interference or a total absence of GNSS, quantum assisted navigation is primarily aimed for defence applications where GNSS could be unavailable, jammed, or spoofed~\cite{kramer2014darpa,gamberini2021quantum, coenen2017responsible}.

\section{Proof-of-Principle Experiments and Implementation}\label{ProofOfPrincipleImlementation}

In the foregoing sections, we highlighted that quantum technologies in space offer a vast panorama of uses ranging from fundamental physics tests to technological applications. Quantum technologies in space can advance our knowledge in fundamental physics, by bridging the gap between relativistic physics and quantum physics. They can be enabling technologies, like deep space communication, allowing mankind to further explore the solar system and beyond. In addition, they are attractive for commercial purposes, with satellite-based quantum key distribution and sensing with atom interferometry being a prominent example.

Within this section, we discuss proof-of-principle experiments and implementations of the ideas reviewed in the previous sections. We classify these missions corresponding to the physics on which they are based on, focusing on of cold atoms, photons, and optomechanical systems.

\subsection{Cold Atoms}\label{ProofOfPrincipleImplementation:ColdAtoms}
\begin{figure}[h!]
    \centering
    \includegraphics[width=0.9\linewidth]{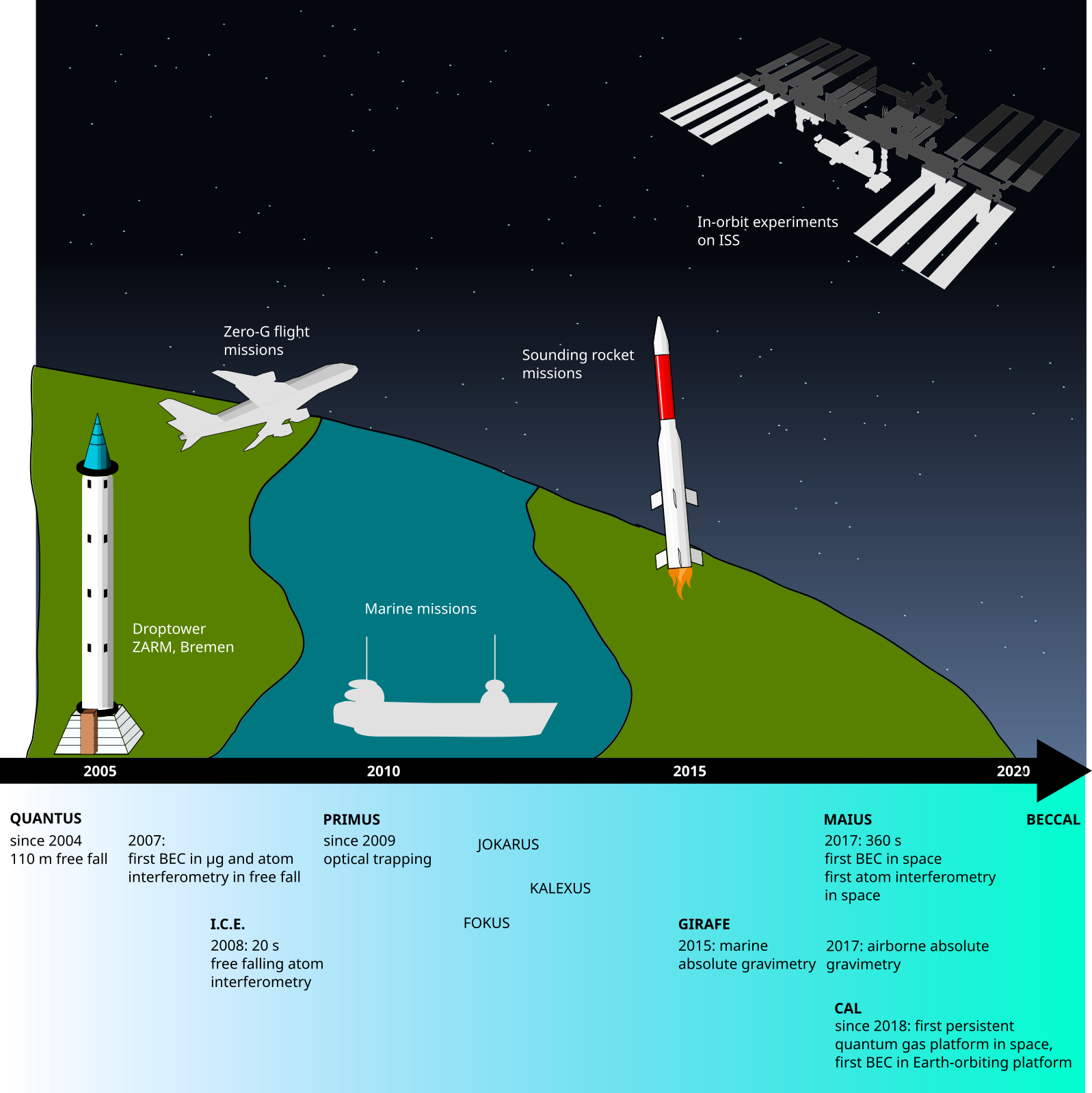}
    \caption{From drop tower experiments at ZARM to the ISS. This timeline depicts important platforms and summarizes important milestones towards cold atom experiments in space. Details about the missions are given in Section~\ref{ProofOfPrincipleImplementation:ColdAtoms}. In 2004 the first cold atom experiments in microgravity environment were performed in a drop tower. Then experiments on aeroplanes and maritime platforms followed. Lastly, sounding rockets and, in 2018,  in-orbit demonstration of cold atom experiments followed. This shows the rapid advancement in the field and the technological maturity of cold atom systems.
    }
    \label{fig:Sec4_ColdAtoms}
\end{figure}

Cold atoms are one of the better studied physical systems utilized to implement quantum technologies in space. They offer a well controlled environment where employing interferometric effects leads to unprecedented precision in sensing applications. Low-gravity environments elevate the precision of these systems even further, mainly due to the increasing free fall times. In addition, ultra-cold atomic condensates can be considered as macroscopic systems showing quantum effects. Thus, in combination with environments where relativistic effects become relevant, they present an attractive playground for testing fundamental physics. 
Cold atom experiments have a long history and have enormously advanced in recent years, going from the laboratory environment to industrial applications and, more importantly, to space implementations. In the following we list projects and missions, which are important milestones in enabling cold atom experiments in space. Figure~\ref{fig:Sec4_ColdAtoms} gives a glimpse into the historical development and summarizes those milestones. 
The ability to operate in harsh environments, like drop towers, ships, aeroplanes or sounding rockets, requires cold atom experiments to be robust and autonomous. Hence, it is an important indicator for their technological maturity and paves the way for commercial use in, for example, Earth observation and in reaching unprecedented parameter regimes for fundamental physics tests.

\subsubsection{Ground-based microgravity projects}\label{Ground-basedmicrogravityprojects}
Weightlessness on ground can be achieved by placing a payload in free fall, thus compensating for the gravitational pull. For this purpose, drop towers have been erected. Among these, the {Drop-Tower Bremen} at the Center of Applied Space Technology and Microgravity (ZARM) of the University of Bremen was chosen as microgravity platform for BEC experiments in weightlessness, as it provides good accessibility and superior quality of microgravity compared to other ground based platforms. At ZARM, an experiment's capsule can either be dropped for 110\,m inside the evacuated vacuum-tube of the drop tower to generate 4.72\,s of microgravity time, or can be launched with a piston-catapult to almost double the microgravity time, $\sim 9.3\,$s. In the future, the {GraviTower Bremen} (GTB Pro) will become a third generation drop tower at ZARM, complementing the Bremen drop tower with its unique catapult system.
The GTB Pro is designed to fit the same proven experiment designs and dimensions as used in the Bremen Drop Tower making both facilities fully compatible.
The initial acceleration and the transition into microgravity are made very smooth by following a sine function limited to $5\,g$. With a repetition time of just three minutes, each flight offers 2.5\,s of microgravity - fully automated all day.
Experiment preparation, automation, tests and flights are carried out in teamwork with the engineers of ZARM and in collaboration with the Bremen Drop Tower.
The interested reader is referred to~\cite{gierse2017fast,konemann2015concept} for further details.

On a smaller scale, a zero-$g$ simulator was designed and built by the French company Symétrie and is operated at LP2N, Bordeaux.
This simulator works by moving a platform, on which the experimental apparatus rests, in a way that mimicks the trajectory of an object launched vertically and in free fall, i.e. a parabola. The platform moves vertically between two granite columns thanks to two carriages with air bearings for a frictionless motion. Linear motors mounted on the sides of the columns are responsible for the accelerations of the moving parts necessary to perform parabolic trajectories. The 0-$g$ simulator can provide up to half a second of weightlessness on every trajectory and, thanks to its very high repetition rate (1 parabola every 12\,s), gives access to a very long accumulated duration of 0-$g$. 

Another very important large-scale facility is the {Einstein Elevator} at HITec (Hannover Institute of Technology).
This large-scale research device is a next-generation drop tower facility with a total height of 40\,m and therefore allows for four seconds of microgravity with residual acceleration of 10$^{-6}g$.
Payloads with up to 1000\,kg, diameter of 1.7\,m and height of 2\,m can be operated with a repetition rate of 300 flights per day thanks to the innovative electromagnetic linear motor drive unit. This is a major improvement in comparison to 3-4 drops possible with the ZARM drop tower.
This motor drive additionally allows for hyper- or hypogravity to generate conditions as they prevail on other celestial bodies, like the Moon or Mars.
The first test operation started in 2019 \cite{firstee} and the interested reader is referred to~\cite{LOTZ2018EE,lotz2013mechanische, lotz2017einstein}.

In the following we report the details on projects employing cold atoms that were developed in ground-based microgravity facilities, instrumental for the realization of cold atom experiments in space.

\begin{itemize}
\item \textbf{QUANTUS} (\textit{QUANTengase Unter Schwerelosigkeit}, English translation: Quantum Gases under Microgravity):\\
The DLR (Deutsches Zentrum für Luft- und Raumfahrt) funded QUANTUS-Project started in 2004 and aims at developing the necessary methods for space-borne microgravity platforms like sounding rockets, experiments on the ISS and on dedicated satellites.
Within QUANTUS, the technology and the physical understanding of these complex experimental apparatuses are developed and preliminary studies for space-borne missions are performed. Using the first generation payload {QUANTUS-1}, this capability has been used for the demonstration of the first BEC in microgravity in 2007~\cite{Zoest2010Science} and the first interferometry experiments with freely falling BECs~\cite{Muentinga2013PRL}.
Sine 2014, the second generation apparatus {QUANTUS-2} is operational at ZARM, featuring a novel compact high-flux BEC source~\cite{rudolph2015high}. This apparatus is more compact with respect to the previous one, allowing using the catapult mode of the Bremen Drop-Tower at ZARM~\cite{Rudolph2015Cover} to double the time in microgravity and to increase the overall data rate. Additionally, it is designed to be able to use Potassium as a second atomic species.
In recent years, using both apparatuses, novel interferometric schemes~\cite{Ahlers2016PRL, Abend2016PRL}, large-momentum beam splitters~\cite{gebbe2019twinlattice} and 3D magnetic delta-kick collimation techniques in microgravity were developed to reduce the kinetic energy of the atomic ensemble down to 38\,pK. 
These developments have enabled the MAIUS and BECCAL missions, which are described in section \ref{Space-basedcoldatomsprojects}.\\
The interested reader is referred to~\cite{gaaloul2010quantum, rudolph2011degenerate, cornelius2018quantus, menoret2017transportable, schubert2016atom, gaaloul2018degenerate, abend2017symmetric, gebbe2017atom}.

\item \textbf{PRIMUS} (\textit{PRäzisionsInterferometrie mit Materiewellen Unter Schwerelosigkeit}, English translation: Precision Interferometry with Matter Waves in Zero Gravity):\\
In addition to the magnetic trap based efforts of the QUANTUS project to utilize ultracold atom technologies in microgravity, since 2009 the potential of optical traps is investigated in the PRIMUS project.
This project initially focused on a test of the weak equivalence principle, leading to outstanding ground-based results~\cite{schlippert2014quantum, albers2020quantum}. Lately, PRIMUS widened its spectrum of interests to more general questions concerning cooling and phase transitions.
Within PRIMUS, a compact experimental setup was realized~\cite{Kulas2016} to apply optical trapping in the drop tower of Bremen. 
Further milestones where the implementation of a single beam optical dipole trap in microgravity, successful evaporative cooling in weightlessness, and the advancement to a crossed beam configuration by actively stabilizing the trapping beam’s pointing.
In principle, the absence of gravity should increase the dimension of evaporation since the trap is not tilted anymore~\cite{Ketterle1996Eva}. Previous studies did not observe this effect for magnetically tilted traps~\cite{Hung2008Acc}. The nonexistence was confirmed in PRIMUS for evaporative cooling in microgravity and could be explained by the anharmonicity of the traps~\cite{vogt2020evaporative}.\\
The interested reader is referred to~\cite{schlippert2014quantum, albers2020quantum, Kulas2016, vogt2020evaporative}.

\item \textbf{I.C.E.} (\textit{Interférometrie atomique à sources Cohérentes pour l'Espace}, English translation: Coherent Source Atomic Interferometry for Space) on ground:\\
Since 2018, the I.C.E. experiment is able to perform experiments in the laboratory thanks to the unique, purpose made Einstein elevator on which the experimental apparatus is installed. This device allows to produce Bose-Einstein condensates with forty thousand Rubidium-87 atoms at a temperature of 35\,nK in weightlessness~\cite{Condon2019}.\\
The interested reader is referred to~\cite{Condon2019} and section \ref{section.air-marine}.

\item \textbf{DESIRE} (Dark Energy search by Interferometry in the Einstein Elevator)
Based on pre-studies~\cite{hamilton2015atom, chiow2018multiloop}, in 2021 a new NASA/JPL-DLR project started aiming for a significant increase in sensitivity for the search of dark matter, which can be reached with atom interferometers using free-falling, compact, low-energy wave packets in interaction with a test mass on macroscopic time scales of several seconds.
To reach these time scales the apparatus will be operated in the Einstein-Elevator in Hannover.
The DESIRE project utilizes the BEC interferometer of the MAIUS-1 mission, which will be modified for its use in the Einstein-Elevator.\\
The interested reader is referred to~\cite{hamilton2015atom, chiow2018multiloop, becker2018space, lachmann2021ultracold}.

\subsubsection{Air/Marine-borne cold atoms projects}\label{section.air-marine}
Another way to achieve weightlessness are parabolic flights in an aircraft by alternating upward and downward arcs interspersed with level flight. 
In April 2015, Novespace began operating its third aircraft, the Airbus A310 Zero G to provide a microgravity environment for scientists to conduct research without going into space.

Not directly benefiting from weightlessness, but still relevant for research on cold atoms in space, is the GIRAFE project~\cite{bidel2018absolute, bidel2020absolute}, where cold atom interferometers are operated during flight or on a ship. We thus include it below. 

\item \textbf{I.C.E.} (\textit{Interférometrie atomique à sources Cohérentes pour l'Espace}, English translation: Coherent Source Atomic Interferometry for Space):\\
The CNES-funded I.C.E. operated an atom interferometer for inertial sensing in reduced gravity on board the NOVESPACE Zero-G plane.
During a 20 seconds-lasting ballistic parabolic flight residual acceleration on the order of 10$^{-2}g$ are achieved.
Within I.C.E., Ramsey fringes have been obtained in 2008 operating an atom interferometer using a series of two Raman transitions within cold Rubidium-87 atoms~\cite{Stern2009ice}.
In 2011, the first airborne operation of a horizontally measuring high-resolution cold-atom inertial sensor, both at 1\,$g$ and in reduced gravity has been reported~\cite{geiger2011detecting}.
This measurement technique has been then advanced to a vertical mode and measurements of the acceleration along the vertical and horizontal axis with one-shot sensitivities of 2.3$\times$10$^{-4}g$ have been achieved~\cite{Battelier2016ice}.
The measured loss of contrast was attributed to the high level of vibrations on-board the aircraft and the large rotation rates during a parabolic flight. A first on-board operation of simultaneous Rubidium-87 and Potassium-39 interferometers in the weightless environment was demonstrated in 2016.
In this parabola campaign, I.C.E. demonstrated its capability of operating a dual-quantum sensor and with this measured the E\"otv\"os parameter -- see also discussion in Sec.~\ref{Fundamental} -- with systematic-limited uncertainty of $3.0\times10^{-4}$ in reduced gravity~\cite{barrett2016dual}.
This constituted the first test of the equivalence principle in a free-falling vehicle with quantum sensors.

During the last years, I.C.E. was upgraded and is now operated at a 3-m high 0\,g-simulator built by the French company Symétrie, which gives access to microgravity in the laboratory (see section \ref{Ground-basedmicrogravityprojects}).\\
The interested reader is referred to~\cite{Stern2009ice, geiger2011detecting, Battelier2016ice, AntoniMicollier2017ice}.

\item \textbf{GIRAFE} (\textit{Gravim\`etre Interf\'erom\'etrique de Recherche \`a Atomes Froids, Embarquable}; English translation: Shipborne cold atom research interferometric gravimeter):\\
In the scope of a funding program of the French DGA and CNES, since 2006 the company ONERA designed and built an absolute marine gravimeter based on atom interferometry called GIRAFE.
GIRAFE was tested multiple times (in October 2015 and January 2016) at sea on an oceanographic survey vessel and demonstrated a superior performance compared to classical technology~\cite{bidel2018absolute}.
Subsequently, the GIRAFE instrument was adapted for airborne measurements for surveying areas where gravity is poorly resolved by ground or satellite measurements, as for example in coastal and mountainous areas. 
In April 2017, in an airborne campaign above Iceland, GIRAFE was compared with other conventional airborne gravimeter and inertial sensors and showed differences with a standard deviation ranging from 3.3 to 6.2 mGal and a mean value ranging from -0.7 mGal to -1.9 mGal~\cite{bidel2020absolute}.\\
The interested reader is referred to~\cite{bidel2018absolute, bidel2020absolute}.

\subsubsection{Space-based cold atoms projects}\label{Space-basedcoldatomsprojects}
Since 2017, cold atom experiments in space have been performed.
Here, the microgravity platforms in use are parabolic sounding rocket missions and the International Space Station (ISS).
A next interesting platform under investigation for cold atoms research is represented by CubeSats.

\item \textbf{MAIUS} (MAteriewellen-Interferometer Unter Schwerelosigkeit, English translation: Matterwave Interferometry under Microgravity):\\
The DLR funded MAIUS missions are the continuation of the aforementioned QUANTUS project.
Their aim is to bridge the gap between laboratory or drop tower systems and future orbital missions by implementing cold atoms, BECs, and atom interferometry on sounding rockets.
In total, three sounding rocket missions are planned with the first one -- MAIUS-1 -- was successfully launched in 2017 \cite{becker2018space}.
This constitutes a major advancement over the aforementioned projects in drop towers or airborne, as it is not only the first setup to undergo environmental qualification but also operated autonomously in the harsh environment of an unmanned sub-orbital spacecraft. 

During the maiden flight in 2017 the first BEC in space has been demonstrated and its collective dynamics were analyzed~\cite{becker2018space}.
Additionally, important manipulation techniques, like internal state preparation, have been performed and autonomously optimized during the parabolic flight. Finally, first atom interferometry experiments in space have been conducted~\cite{lachmann2021ultracold}.

After MAIUS-1, the second generation rocket payload MAIUS-B is in its setup phase. It reached its critical design review at the end of 2018~\cite{elsen2019pathway}.
This apparatus features, additionally to Rubidium-87, also Potassium-41 atoms and aims at performing sequential atom interferometry experiments and mixture studies with both species during the second rocket mission MAIUS-2 currently scheduled for 2021.
The launch of a third and final mission -- MAIUS-3 -- is currently planned for 2022 and aims at demonstrating simultaneous atom interferometry in space, paving the path for tests of the universality of free fall using atoms (cf. Sec.~\ref{Fundamental}).\\
The interested reader is referred to~\cite{seidel2015atom, grosse2015mechanical, lachmann2018creating, becker2018space, christ2017towards, rudolph2015high, spindeldreier2016fpga}.

\item \textbf{CAL} (Cold Atom Laboratory):\\
The NASA-funded Cold Atom Laboratory (CAL) was developed by NASA’s Jet Propulsion Laboratory  and utilizes a compact atom chip-based system to create ultracold mixtures and degenerate samples of Rubidium-${87}$, Potassium-${39}$, and Potassium-${41}$~\cite{elliott2018nasa}.
It was launched to the ISS in 2018 and operates as a multi-user facility to provide the first persistent quantum gas platform in space for an international group of investigators with broad applications in fundamental physics and inertial sensing.
Up to date, experiments transferring Rubidium-87 BEC in ultra-shallow traps are realized based on adiabatic decompression ramps. Fast and perturbation‑free transport with a micrometer-level control of the atomic positions is realized based on shortcut-to-adiabaticity protocols~\cite{Corgier2018fast}.
Moreover, a space atom laser~\cite{meister2018space} and a radio frequency bubble experiment are conducted.
Finally, delta-kick collimation drastically reducing the free expansion rate of the atomic clouds has been performed on the ISS.
At the beginning of 2020, the new science module SM3 has been installed on the ISS, adding the possibility to perform atom interferometry experiments with CAL.\\
The interested reader is referred to~\cite{aveline2018nasa, williams2017opportunities}.

\item \textbf{BECCAL} (Bose-Einstein Condensate and Cold Atom Laboratory):\\
Built upon the heritage not only of the projects QUANTUS, MAIUS and CAL, but also of JOKARUS and KALEXUS (see section \ref{Atom-basedClocks}), the NASA- and DLR-funded BECCAL will serve as the next-generation multi-user and -purpose facility aboard the ISS~\cite{frye2021bose}.
The apparatus is designed to operate with cold and condensed ensembles of different isotopes of Rubidium and Potassium.
Hence, BECCAL will enable the study of scalar and spinor degenerate gasses as well as mixtures thereof.
BECCAL supports atom interferometry for fundamental physics and studies for future quantum sensors.
Additionally, arbitrary shaped red- and blue-detuned potentials will be possible to implement, allowing for versatile trapping and anti-trapping configurations. 
Throughout its lifetime, BECCAL will perform a variety of experiments and serve as a pathfinder for future missions.\\
The interested reader is referred to~\cite{frye2021bose}.

\item \textbf{CASPA} (Cold Atom Space PAyload):\\
In the Innovate UK and Engineering and Physical Sciences Research Council (EPSRC) funded project CASPA, a British consortium led by the company Teledyne e2v is designing a system for autonomous cold atom experiments in space. The project's aim is the elevation of the Technology Readiness Level (TRL) of required subsystems and the tackling of challenges in building compact cold atom interferometers for space use. After a careful analysis of the environment's impact on the cold atom system, a prototype 6U CubeSat that is capable of trapping Rubidium atoms was built, qualified, and tested. CASPA has a weight of 4\,kg, fits into four units of a CubeSat and consumes 12\,W of electrical power. It can cool down atoms to the order of 100\,$\mu$K.
An actual cold atom sensor based on the heritage of CASPA is planned to be built~\cite{devani2020gravity}.\\ 
The interested reader is referred to~\cite{devani2020gravity}.

\end{itemize}

\subsection{Atom-based Clocks}\label{Atom-basedClocks}
Another implementation of quantum technologies in space are atom-based clocks. These are atomic and optical clocks which provide increased accuracy and precision as frequency and time references and are nowadays the reference in modern time keeping systems. In conventional systems, primarily radio frequency based references have been employed. For example, the current Galileo system is based on different such references. Optical systems, based on atomic transitions or fixed lengths, promise higher accuracy and precision~\cite{Hollberg2005, batori2020gnss}. 

Atomic clocks with hot atoms are widespread and offer compact and robust setups. As the systems depend on the width of atomic lines, their accuracy and precision can be improved by reducing internal and external temperatures. Another improvement to atomic clocks is based on atom fountains, which increase the interrogation time for the atoms. Optical clocks, which are atom-based clocks with transition energies at optical frequencies, promise even higher precision. Hence, different concepts of generating a stable frequency reference exist~\cite{Hollberg2005, batori2020gnss}. Some optical systems exploit atomic, see for example~Ref.~\cite{Falke2014Strontium}, or molecular transitions~\cite{Schuldt2016, doeringshoff2019}, while others rely on fixed distances~\cite{sanjuan2019}.

Ground-based systems can make use of controlled environments, such as operating at cryogenic temperatures, or large volumes to generate the desired frequency stability. In space-based systems budgets are limited and the system needs to operate reliably without interference~\cite{Schuldt2016, doeringshoff2019, schkolnik2017jokarus}. Consequently, several missions are deployed to test concepts. While atomic-clocks might have more stringent demands in space, the microgravity environment can also improve performances. In addition to dedicated scientific missions, commercial systems have been used to measure fundamental principles. The next sections are dedicated to outline a couple of those missions.

\subsubsection{Scientific Experiments}

Scientific missions revolve around measuring the gravitational redshift caused by  Earth and  Sun, and probing  special relativity. In addition, these systems aim at enabling next generation gravity missions and future global navigation satellite systems (see also Sec.~\ref{Fundamental} and Sec.~\ref{Applications}). Furthermore, clocks are necessary to perform high precision experiments in space, such as the gravitational waves antenna LISA. These aims and goals are similar for all the scientific missions surrounding optical frequency references, either ongoing or planned, and that are reported in the following~\cite{lammerzahl2004experiments}.

\paragraph{Sounding Rocket Missions}
A first step towards optical clock operation in space are sounding rocket missions. Here, three of the major developments in recent years are listed:

\begin{itemize}

    \item \textbf{KALEXUS} (Kalium Laser-Experimente unter Schwerelosigkeit; English translation: Potassium Laser Experiment under Microgravity):\\
    Within the KALEXUS mission, two extended cavity diode lasers alongside an optical preparation stage were launched on a sounding rocket mission. With this mission, the technological readiness of the Potassium laser system, the automatic frequency stabilization of the lasers, and the switching between redundant systems during flight was demonstrated in 2016~\cite{dinkelaker2017autonomous}.
    
    \item \textbf{FOKUS} (Faserlaserbasierter Optischer Kammgenerator unter Schwerelosigkeit; English translation: Optical Frequency Comb metrology Under Microgravity):\\
    Within FOKUS, the technological readiness of optical frequency combs for deployment in space was demonstrated. For this purpose, a frequency comb alongside a diode laser and an optical preparation stage were mounted to a sounding rocked and launched into space in 2015~\cite{lezius2016space}. 
    
    \item \textbf{JOKARUS} (Jod-Kammresonator unter Schwerelosigkeit; English translation: Iodine, Comb, and Resonator Under Microgravity)
    Within the JOKARUS mission, a Iodine frequency reference, an optical frequency comb, and the optical and electrical preparation stages were launched to space on board of a sounding rocket at the end of 2017. This campaign served to demonstrate the miniaturization and deployment of key technologies in space~\cite{doeringshoff2019, schkolnik2017jokarus}
    
    \end{itemize}

\paragraph{Space-based Platforms}
A prominent example of using frequency references in orbit for fundamental research is the experiment on the Galileo satellites executed in 2014: 

\begin{itemize}
    \item \textbf{Galileo Satellites}: 
    In August 2014, two Galileo~\cite{herrmann2018galileo} satellites were launched, and -- due to an error during the launch -- orbited Earth on eccentric orbits. Such an eccentricity allowed for measurements of the gravitational redshift similar to the measurements of Gravity Probe A~\cite{vessot1979test}, see also the related discussion in Sec.~\ref{Fundamental}.  
\end{itemize}
    
Following the success of measuring the weak equivalence principle with the MICROSCOPE mission~\cite{touboul2020microscope} and the need for more precise optical frequency references in space, different experiments and missions have been devised: 

\paragraph{Flown Missions}
\begin{itemize}
        \item \textbf{CACES} (Cold Atomic Clock Experiment in Space):\\
    This is a mission started in 2011 under the Chinese manned space program. It set out to test laser cooling and manipulation of atoms in orbit. It is based on laser-cooled Rubidium-87 atoms in an atomic fountain for stabilization of frequencies to a level of $2 \times 10^{-16}$ on ground. The system was launched into space aboard the Chinese space laboratory Tiangong-2 in September 2016. It has proven long-term in-orbit operation of cold-atom clocks under various environmental effects such as varying gravity levels, magnetic fields and radiation~\cite{liu2018orbit,ren2020development}.
    
    \item \textbf{DSAC} (Deep Space Atomic Clock):\\
    NASA founded DSAC is developed as a step towards independent spacecraft navigation in deep space as opposed to relying on communication to the ground. It houses a Mercury ion atomic clock and it has been launched in June 2019~\cite{nasa2019dsac, nasa2020dsac}. 
    
\end{itemize}

\paragraph{Currently Prepared Missions}
In order to execute optical clocks in space, the following missions are currently prepared. They are targeting both fundamental questions and technology development: 

\begin{itemize}
    \item \textbf{ACES} (Atomic Clock Ensemble in Space):\\
    This is an ESA-coordinated mission that will prepare a system of two atomic clocks in the Columbus module of the ISS. The two clocks are the laser-cooled Caesium atomic clock PHARAO (Projet d'Horloge Atomique par Refroidissement d'Atomes en Orbite), developed by CNES, an active Hydrogen maser, the SHM (Space Hydrogen Maser), developed by Spectratime and a precise time and frequency transfer system~\cite{cacciapuoti2009space}. The clock ensemble is expected to have a stability of $2 \times 10^{-16}$ in the short term. Currently, the launch for the ensemble is expected for 2021. The technical goal of this mission is testing the performance of a new generation of atomic clocks in space and the scientific goals are to perform fundamental physics tests, as for example an improved measurements of the gravitational redshift, the search for anisotropies of the speed of light, and for space-time variations of physical constants. In addition, the two atomic clocks will allow mapping the Earth's gravitational potential by measuring the differential gravitational redshift~\cite{augelli2016aces}. The ACES mission will also allow comparing the time reference with ground-based ones and the ultra-precise time-reference distribution~\cite{meynadier2018atomic}. 
    
    \item \textbf{COMPASSO} :\\
    COMPASSO~\cite{Compasso} is a DLR mission that aims at demonstrating necessary optical frequency and link technologies for future missions, such as BOOST, LISA, Next Generation Gravity Mission, and GNSS. In particular, the mission will be the first in-orbit verification of optical clocks. The payload consists of optical and radio frequency references, and a bidirectional optical link to compare the in-orbit measurements to frequency references on ground. This shall demonstrate the possibility to operate optical links to other space-based platforms and, in addition, will allow investigating atmospheric effects in space-to-ground clock distribution. The latter especially includes the comparison to a GNSS signal to demonstrate the readiness of optical technologies for GNSS applications. Currently, it is planned that COMPASSO will be operated from the end of 2024 on the ISS aboard the Bartolomeo platform.  
    \end{itemize}
 
\paragraph{Proposed Missions} 
 Based on the current technology additional missions have been proposed, some of which are listed below:
    
    \begin{itemize}
    
        \item \textbf{BOOST} (BOOst Symmetry Test):\\
    This is a DLR mission to conduct a Kennedy Thorndike experiment in space~\cite{gurlebeck2018boost}, which aims at improving the current boundaries set by ground based experiments~\cite{tobar2010kennedy} (see also Sec.\ref{Fundamental}). Two optical references based on two different operational principles are compared. Here, an optical cavity is compared to an Iodine frequency reference. The targeted stability is better than $10^{-15}$ at orbit time, which takes around $90$ minutes.
    
    \item \textbf{IVORY} (In-orbit Verification of stabilized Optical and microwave Reference sYstems):\\
    The DLR and Airbus Defence and Space joint project IVORY aims at verification of optical technologies, as optical clocks, frequency combs and space-to-ground laser communication, on the ISS aboard the Bartolomeo platform \cite{IVORY}. It includes a Rubidium clock and a microwave reference system. 
    
    \item \textbf{SOC} (Space Optical Clocks):\\
    The DLR and ESA funded project SOC aims at operating optical lattice clock systems on the ISS. The systems include Ytterbium and Strontium optical clocks. With these atoms cooled and trapped in a magneto-optical trap, a frequency stability of $10^{-16}$ is targeted~\cite{schiller2017space}.

\end{itemize}

\subsubsection{Commercial Missions
}

Currently, the main commercial application of atomic clocks is in the Global Navigation Satellite Systems (GNSS).
Starting with the initially military American Global Positioning System (GPS) in 1978, four GNSS and several Regional Navigation Satellite Systems have been established.
The latest GNSS is the European Galileo system \cite{galileognss}.
Each current generation Galileo Satellite contains two passive Hydrogen maser atomic clocks and two secondary Rubidium atomic clocks. The use of other atomic clocks on future Galileo generations is currently under evaluation.
Nowadays, the precise position and timing data provided by GNSS is indispensable for a modern economy and plays a key role for future developments such as autonomous vehicles.
Several future commercial applications of quantum technologies in space are being studied. These include the use of atomic clocks as reference for deep space navigation\cite{ely2018using} and as sensors for the measurement of Earth's gravitational field\cite{muller2018high}.

\subsection{Photons
}

Transmission of quantum signals over long distances has been demonstrated through a series of satellite-based experiments and feasibility studies. The first proposals to implement satellites for this application emerged in the 1990s~\cite{Hughes1999_SPIE}. Since this initial proposal, numerous feasibility studies and demonstrations of satellites QKD have been made. They include free space QKD over high altitude ranges~\cite{kurtsiefer2002step}, feasibility of quantum communications in space~\cite{Rarity2002_NJP, aspelmeyer2003long}, and a record breaking inter-island key exchange over 144\,km~\cite{schmitt2007experimental, ursin2007entanglement}. The feasibility of space links was realised through experiments that exchanged single photons from a low Earth orbit (LEO) satellite to ground by exploiting retroreflectors aboard the spacecraft~\cite{villoresi2008experimental,vallone2015experimental}. These experiments recorded small quantum bit error rate, which provided a concrete proof for satellite-based quantum communications. The transmission of quantum photonic signals has also been increased through use of Medium Earth orbit (MEO) satellites or higher orbits, up to the current single-photon exchange limit of 20,000\,km~\cite{Dequal2016,calderaro2018towards}. 

Recently, the QUESS experiment involving the Chinese LEO satellite Micius became the first space-based quantum communication mission to be launched and has made further developments~\cite{liao2017satellite, yin2017satellite}. It has demonstrated entanglement distribution to two ground stations separated by $\sim1200$\,km~\cite{liao2017satellite}, ground-to-satellite quantum teleportation over distances of up to 1400\,km~\cite{ren2017ground}, and the realisation of a hybrid quantum communication network with a total quantum communication distance of 4600\,km~\cite{chen2021integrated}.

Despite these demonstrations, establishing long term reliable ground and satellite links remains the principle challenge in satellite QKD. A notable development in space systems is the rise of in-orbit demonstrations with small satellites and CubeSats for rapid and less-costly space systems developments. This increase is partly driven by miniaturisation and increasing robustness of quantum components. In addition, constellations of small satellites offer the possibility of a cost effective approach to improving coverage and ground-satellite link reliability compared to traditional satellites. CubeSat missions in 2015~\cite{tang2016generation} and 2019~\cite{villar2020entanglement} have performed in-orbit demonstrations of miniaturised quantum photon pair sources. 
We report a timeline of missions that have demonstrated key milestones or feasibility studies towards global satellite-based QKD in Fig.~\ref{fig:satellite_mission_timeline}~\cite{sidhu2021advances}. This includes recently proposed mission that aim to integrate space and terrestrial segments to step closer to a globally quantum networking. 

{Given the recent and comprehensive review works on space quantum communication missions, here we do not further elaborate on the subject and refer the interested reader to the dedicated reviews Refs.~\cite{bedington2017progress,lee2019updated,sidhu2021advances}.}

\makeatletter
\pgfarrowsdeclare{center*}{center*}
{
  \pgfarrowsleftextend{+-.5\pgflinewidth}
  \pgfutil@tempdima=0.4pt%
  \advance\pgfutil@tempdima by.2\pgflinewidth%
  \pgfarrowsrightextend{4.5\pgfutil@tempdima}
}
{
  \pgfutil@tempdima=0.4pt%
  \advance\pgfutil@tempdima by.2\pgflinewidth%
  \pgfsetdash{}{+0pt}
  \pgfpathcircle{\pgfqpoint{4.5\pgfutil@tempdima}{0bp}}{4.5\pgfutil@tempdima}
  \pgfusepathqfillstroke
}

\pgfarrowsdeclare{centero}{centero}
{
  \pgfarrowsleftextend{+-.5\pgflinewidth}
  \pgfutil@tempdima=0.4pt%
  \advance\pgfutil@tempdima by.2\pgflinewidth%
  \pgfarrowsrightextend{4.5\pgfutil@tempdima}
}
{
  \pgfutil@tempdima=0.4pt%
  \advance\pgfutil@tempdima by.2\pgflinewidth%
  \pgfsetdash{}{+0pt}
  \pgfpathcircle{\pgfqpoint{4.5\pgfutil@tempdima}{0bp}}{4.5\pgfutil@tempdima}
  \pgfusepathqstroke
}
\makeatother

\tikzset{pics/nodetwo/.style n args={2}{
	code = {%
		\node[fill=white,opacity=40,draw=brown,rounded corners] at (#1,1) {\LARGE \textcolor{brown!80!black}{#2}};
}
}}

 
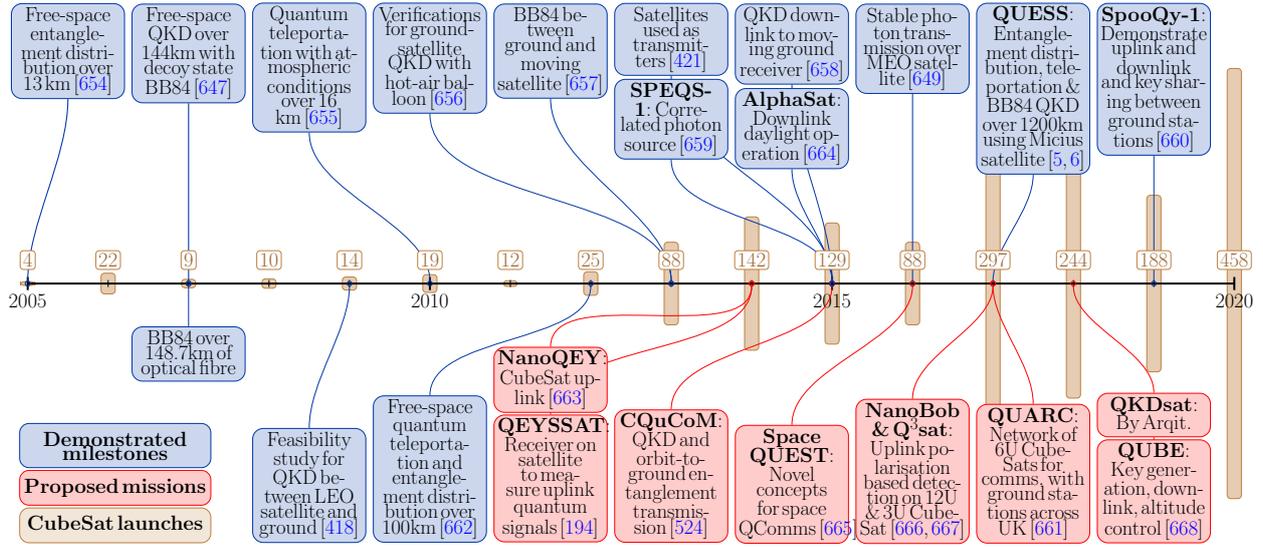
\begin{figure*}[t!]
\pgfkeys{/pgf/number format/set thousands separator={}}
\centering
        \resizebox{1\columnwidth}{!}{%
\begin{tikzpicture}

\def\w{0.3}
\def\scale{25}
\def\barcolor{brown}
\def\d{17} 
\pgfmathsetmacro \unit {\d/5}
\pgfmathsetmacro \twod {2*\d}
\pgfmathsetmacro \threed {3*\d}

\filldraw[color=\barcolor,fill opacity=0.4,draw,rounded corners=.15cm] (0+\w,2/\scale) rectangle (0-\w,-2/\scale);  
\filldraw[color=\barcolor,fill opacity=0.4,rounded corners=.15cm] (\unit+\w,11/\scale) rectangle (\unit-\w,-11/\scale);  
\filldraw[color=\barcolor,fill opacity=0.4,rounded corners=.15cm] (2*\unit+\w,4.5/\scale) rectangle (2*\unit-\w,-4.5/\scale);  
\filldraw[color=\barcolor,fill opacity=0.4,rounded corners=.15cm] (3*\unit+\w,5/\scale) rectangle (3*\unit-\w,-5/\scale);  
\filldraw[color=\barcolor,fill opacity=0.4,rounded corners=.15cm] (4*\unit+\w,7/\scale) rectangle (4*\unit-\w,-7/\scale);  
\filldraw[color=\barcolor,fill opacity=0.4,rounded corners=.15cm] (5*\unit+\w,9.5/\scale) rectangle (5*\unit-\w,-9.5/\scale);  
\filldraw[color=\barcolor,fill opacity=0.4,rounded corners=.15cm] (6*\unit+\w,3/\scale) rectangle (6*\unit-\w,-3/\scale);  
\filldraw[color=\barcolor,fill opacity=0.4,rounded corners=.15cm] (7*\unit+\w,12.5/\scale) rectangle (7*\unit-\w,-12.5/\scale);  
\filldraw[color=\barcolor,fill opacity=0.4,rounded corners=.15cm] (8*\unit+\w,44/\scale) rectangle (8*\unit-\w,-44/\scale);  
\filldraw[color=\barcolor,fill opacity=0.4,rounded corners=.15cm] (9*\unit+\w,71/\scale) rectangle (9*\unit-\w,-71/\scale);  
\filldraw[color=\barcolor,fill opacity=0.4,rounded corners=.15cm] (10*\unit+\w,64.5/\scale) rectangle (10*\unit-\w,-64.5/\scale);  
\filldraw[color=\barcolor,fill opacity=0.4,rounded corners=.15cm] (11*\unit+\w,44/\scale) rectangle (11*\unit-\w,-44/\scale);  
\filldraw[color=\barcolor,fill opacity=0.4,rounded corners=.15cm] (12*\unit+\w,148.5/\scale) rectangle (12*\unit-\w,-148.5/\scale);  
\filldraw[color=\barcolor,fill opacity=0.4,rounded corners=.15cm] (13*\unit+\w,122/\scale) rectangle (13*\unit-\w,-122/\scale);  
\filldraw[color=\barcolor,fill opacity=0.4,rounded corners=.15cm] (14*\unit+\w,94/\scale) rectangle (14*\unit-\w,-94/\scale);  
\filldraw[color=\barcolor,fill opacity=0.4,rounded corners=.15cm] (15*\unit+\w,229/\scale) rectangle (15*\unit-\w,-229/\scale);  


\draw[line width=2.4pt] (0,0)--(\threed,0);
\foreach \j in {0,...,15}{\draw[black,line width=1pt] (\j*\unit,.15)--(\j*\unit,-.15);}
\foreach \i in {0,...,3}{
\draw[black,line width=2.4pt] (\i*\d,.28)--(\i*\d,-.28) node[below]{\Huge \textbf{\pgfmathparse{2005 + \i*5}\pgfmathprintnumber{\pgfmathresult}}};}


\node[fill=RoyalAzure!20,draw=RoyalAzure,line width = 1.2pt,rounded corners=.45cm,inner sep=4pt,text width=4.5cm,align=center] at (0.5*\unit,9.9) (Peng) {\Huge Free-space entanglement distribution over 13 km~\cite{Peng2005_PRL}};
\draw[-centero,line width = 1.3pt,RoyalAzure] (Peng) to[out=270,in=90,looseness=0.8]  (0,0) ;

\node[fill=RoyalAzure!20,draw=RoyalAzure,line width = 1.2pt,rounded corners=.45cm,inner sep=4pt,text width=4.5cm,align=center] at (2*\unit,9.8) (2006) {\Huge Free-space QKD over 144km with decoy state BB84~\cite{schmitt2007experimental}};
\draw[-centero,line width = 1.3pt,RoyalAzure] (2006) to[out=270,in=90,looseness=0.8]  (2*\unit,0) ;

\node[fill=RoyalAzure!20,draw=RoyalAzure,line width = 1.2pt,rounded corners=.45cm,inner sep=4pt,text width=4.5cm,align=center] at (2*\unit,-3) (2007) {\Huge BB84 over 148.7km of optical fibre};
\draw[-centero,line width = 1.2pt,RoyalAzure] (2007) to[out=90,in=270,looseness=0.8]  (2*\unit,0) ;

\node[fill=RoyalAzure!20,draw=RoyalAzure,line width = 1.2pt,rounded corners=.45cm,inner sep=4pt,text width=4.5cm,align=center] at (3.5*\unit,-8.6) (Bonato) {\Huge Feasibility study for QKD between LEO satellite and ground~\cite{bonato2009feasibility}};
\draw[-centero,line width = 1.2pt,RoyalAzure] (Bonato) to[out=90,in=270,looseness=0.8]  (4*\unit,0);

\node[fill=RoyalAzure!20,draw=RoyalAzure,line width = 1.2pt,rounded corners=.45cm,inner sep=4pt,text width=4.5cm,align=center] at (3.5*\unit,9.2) (Jin) {\Huge Quantum teleportation with atmospheric conditions over 16 km~\cite{Jin2010_NP}};
\draw[-centero,line width = 1.2pt,RoyalAzure] (Jin) to[out=270,in=90,looseness=0.8]  (5*\unit,0);

\node[fill=RoyalAzure!20,draw=RoyalAzure,line width = 1.2pt,rounded corners=.45cm,inner sep=4pt,text width=4.5cm,align=center] at (5*\unit,9.6) (Wang) {\Huge Verifications for ground-satellite QKD with hot-air balloon~\cite{wang2013direct}};
\draw[-centero,line width = 1.2pt,RoyalAzure] (Wang) to[out=270,in=90,looseness=0.8]  (8*\unit,0);

\node[fill=RoyalAzure!20,draw=RoyalAzure,line width = 1.2pt,rounded corners=.45cm,inner sep=4pt,text width=4.5cm,align=center] at (6.5*\unit,9.9) (Nauerth) {\Huge BB84 between ground and moving satellite~\cite{Nauerth2013_NP}};
\draw[-centero,line width = 1.2pt,RoyalAzure] (Nauerth) to[out=270,in=90,looseness=0.8]  (8*\unit,0);

\node[fill=RoyalAzure!20,draw=RoyalAzure,line width = 1.2pt,rounded corners=.45cm,inner sep=4pt,text width=4.5cm,align=center] at (8*\unit,10.4) (Vallone) {\Huge Satellites used as transmitters~\cite{vallone2015experimental}};
\draw[-centero,line width = 1.2pt,RoyalAzure] (Vallone) to[out=270,in=90,looseness=0.8]  (10*\unit,0) ;

\node[fill=RoyalAzure!20,draw=RoyalAzure,line width = 1.2pt,rounded corners=.45cm,inner sep=4pt,text width=4.5cm,align=center] at (9.5*\unit,10.2) (Bourgoin) {\Huge QKD downlink to moving ground receiver~\cite{Bourgoin2015_OE}};
\draw[-centero,line width = 1.2pt,RoyalAzure] (Bourgoin) to[out=270,in=90,looseness=0.8]  (10*\unit,0);

\node[fill=RoyalAzure!20,draw=RoyalAzure,line width = 1.2pt,rounded corners=.45cm,inner sep=4pt,text width=4.5cm,align=center] at (8*\unit,7) (SPEQS) {\Huge \textbf{SPEQS-1}: Correlated photon source~\cite{Ling_webpage2020}};
\draw[-centero,line width = 1.2pt,RoyalAzure] (SPEQS) to[out=270,in=90,looseness=0.8]  (10*\unit,0);

\node[fill=RoyalAzure!20,draw=RoyalAzure,line width = 1.2pt,rounded corners=.45cm,inner sep=4pt,text width=4.5cm,align=center] at (11*\unit,10) (Dequal) {\Huge Stable photon transmission over MEO satellite~\cite{Dequal2016}};
\draw[-centero,line width = 1.2pt,RoyalAzure] (Dequal) to[out=270,in=90,looseness=0.8]  (11*\unit,0);

\node[fill=RoyalAzure!20,draw=RoyalAzure,line width = 1.2pt,rounded corners=.45cm,inner sep=4pt,text width=4.5cm,align=center] at (12.5*\unit,8.3) (QUESS) {\Huge \textbf{QUESS}: Entanglement distribution, teleportation \& BB84 QKD over 1200km using Micius satellite~\cite{yin2017satellite,liao2017satellite}};
\draw[-centero,line width = 1.2pt,RoyalAzure] (QUESS) to[out=270,in=90,looseness=0.8]  (12*\unit,0);

\node[fill=RoyalAzure!20,draw=RoyalAzure,line width = 1.2pt,rounded corners=.45cm,inner sep=4pt,text width=4.5cm,align=center] at (14*\unit,8.7) (SpooQySats) {\Huge \textbf{SpooQy-1}: Demonstrate uplink and downlink and key sharing between ground stations~\cite{Bedington2016_EPJ}};
\draw[-centero,line width = 1.2pt,RoyalAzure] (SpooQySats) to[out=270,in=90,looseness=0.8]  (14*\unit,0);


\node[fill=red!20,draw=red,line width = 1.2pt,rounded corners=.45cm,inner sep=4pt,text width=4.5cm,align=center] at (12.5*\unit,-8.1) (QUARC) {\Huge \textbf{QUARC}: Network of 6U CubeSats for comms. with ground stations across UK~\cite{Mazzarella2020_C}};
\draw[-centero,line width = 1.2pt,red] (QUARC) to[out=90,in=270,looseness=0.8]  (12*\unit,0);

\node[fill=red!20,draw=red,line width = 1.2pt,rounded corners=.45cm,inner sep=4pt,text width=4.5cm,align=center] at (6.5*\unit,-8.3) (QEYSSAT) {\Huge \textbf{QEYSSAT}: Receiver on satellite to measure uplink quantum signals~\cite{jennewein2014qeyssat}};
\draw[-centero,line width = 1.2pt,red] (QEYSSAT) to[out=90,in=270,looseness=0.8]  (9*\unit,0);

\node[fill=RoyalAzure!20,draw=RoyalAzure,line width = 1.2pt,rounded corners=.45cm,inner sep=4pt,text width=4.5cm,align=center] at (5*\unit,-7.9) (Yin) {\Huge Free-space quantum teleportation and entanglement distribution over 100km~\cite{Yin2012_N}};
\draw[-centero,line width = 1.2pt,RoyalAzure] (Yin) to[out=90,in=270,looseness=0.8]  (7*\unit,0);

\node[fill=red!20,draw=red,line width = 1.2pt,rounded corners=.45cm,inner sep=4pt,text width=4.5cm,align=center] at (6.5*\unit,-4.1) (NanoQEY) {\Huge \textbf{NanoQEY}: CubeSat uplink~\cite{Jennewein2014_SPIE_2}};
\draw[-centero,line width = 1.2pt,red] (NanoQEY) to[out=90,in=270,looseness=0.8]  (9*\unit,0);

\node[fill=red!20,draw=red,line width = 1.2pt,rounded corners=.45cm,inner sep=4pt,text width=4.5cm,align=center] at (8*\unit,-8.2) (CQuCoM) {\Huge \textbf{CQuCoM}: QKD and orbit-to-ground entanglement transmission~\cite{oi2017cubesat}};
\draw[-centero,line width = 1.2pt,red] (CQuCoM) to[out=90,in=270,looseness=0.8]  (10*\unit,0);

\node[fill=RoyalAzure!20,draw=RoyalAzure,line width = 1.2pt,rounded corners=.45cm,inner sep=4pt,text width=4.5cm,align=center] at (9.5*\unit,6.6) (ALPHASAT) {\Huge \textbf{AlphaSat}: Downlink daylight operation~\cite{elser2017quantum}};
\draw[-centero,line width = 1.2pt,RoyalAzure] (ALPHASAT) to[out=270,in=90,looseness=0.8]  (10*\unit,0);

\node[fill=red!20,draw=red,line width = 1.2pt,rounded corners=.45cm,inner sep=4pt,text width=4.5cm,align=center] at (9.5*\unit,-8.55) (SpaceQUEST) {\Huge \textbf{Space QUEST}: Novel concepts for space QComms~\cite{armengol2008quantum}};
\draw[-centero,line width = 1.2pt,red] (SpaceQUEST) to[out=90,in=270,looseness=0.8]  (11*\unit,0);

\node[fill=red!20,draw=red,line width = 1.2pt,rounded corners=.45cm,inner sep=4pt,text width=4.5cm,align=center] at (11*\unit,-8) (NanoBob) {\Huge \textbf{NanoBob \& Q$^3$sat}: Uplink polarisation based detection on 12U \& 3U CubeSat~\cite{kerstel2018nanobob,neumann2018q}};
\draw[-centero,line width = 1.2pt,red] (NanoBob) to[out=90,in=270,looseness=0.8]  (12*\unit,0);

\node[fill=red!20,draw=red,line width = 1.2pt,rounded corners=.45cm,inner sep=4pt,text width=4.5cm,align=center] at (14*\unit,-5.6) (QKDsat) {\Huge \textbf{QKDsat}: By Arqit.};
\draw[-centero,line width = 1.2pt,red] (QKDsat) to[out=90,in=270,looseness=0.8]  (13*\unit,0);

\draw[line width = 1.2pt,red] (QKDsat) to (14*\unit,-8);
\node[fill=red!20,draw=red,line width = 1.2pt,rounded corners=.45cm,inner sep=4pt,text width=4.5cm,align=center] at (14*\unit,-8.85) (QUBE) {\Huge \textbf{QUBE}: Key generation, downlink, altitude control~\cite{Haber2018_Qubeproceedings}};

\pic {nodetwo={0}{\Huge{4}}}; \pic {nodetwo={1*\unit}{\Huge{22}}}; \pic {nodetwo={2*\unit}{\Huge{9}}}; \pic {nodetwo={3*\unit}{\Huge{10}}};
\pic {nodetwo={4*\unit}{\Huge{14}}}; \pic {nodetwo={5*\unit}{\Huge{19}}}; \pic {nodetwo={6*\unit}{\Huge{12}}}; \pic {nodetwo={7*\unit}{\Huge{25}}};
\pic {nodetwo={8*\unit}{\Huge{88}}}; \pic {nodetwo={9*\unit}{\Huge{142}}}; \pic {nodetwo={10*\unit}{\Huge{129}}}; \pic {nodetwo={11*\unit}{\Huge{88}}};
\pic {nodetwo={12*\unit}{\Huge{297}}}; \pic {nodetwo={13*\unit}{\Huge{244}}}; \pic {nodetwo={14*\unit}{\Huge{188}}}; \pic {nodetwo={15*\unit}{\Huge{458}}};

\begin{scope}[xshift=2.5cm, yshift=0.0cm]
\node[fill=RoyalAzure!20,draw=RoyalAzure,line width = 1.2pt,rounded corners=.4cm,inner sep=4pt,text width=7.8cm,align=center, text height = 0.8cm,text depth = 0.8 cm] at (1.2,-6.9) {\textbf{\Huge Demonstrated milestones}};

\node[fill=red!20,draw=red,line width = 1.2pt,rounded corners=.4cm,inner sep=4pt,text width=7.8cm,align=center, text height = 0.8cm,text depth = 0.4 cm] at (1.2,-8.7){\textbf{\Huge Proposed missions}};

\node[fill=\barcolor!20,draw=\barcolor,line width = 1.2pt,rounded corners=.4cm,inner sep=4pt,text width=7.8cm,align=center, text height = 0.8cm,text depth = 0.4 cm] at (1.2,-10.3){\textbf{\Huge CubeSat launches}};
\end{scope}

\end{tikzpicture}%
}
\caption{ Timeline of key milestones in field demonstration and feasibility studies towards the developments of satellite-based QKD. Notice that the increase in the number of missions involving CubeSats reflects their growing importance in satellite-based global quantum communications. {Figure adapted from~\cite{sidhu2021advances}.}
}
\label{fig:satellite_mission_timeline}
\end{figure*}

\subsection{Optomechanics
}

\begin{figure}[h!]
    \centering
    \includegraphics[width=0.9\linewidth]{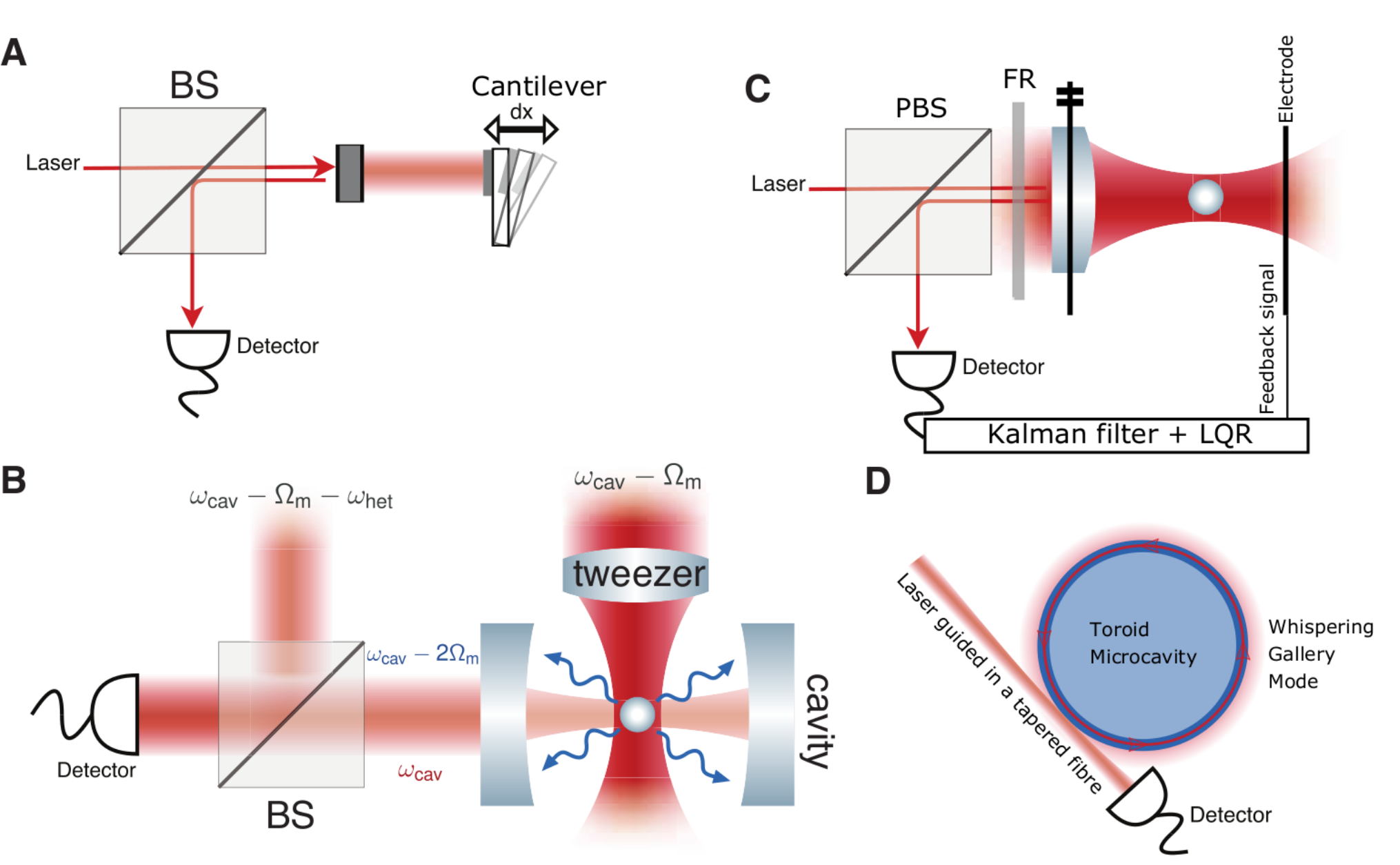}
    \caption{Schematic representation of different optomechanical systems for quantum ground state preparation. (A) Optomechanical cooling of the motion of a cantilever using radiation pressure. The mirror placed on the cantilever is akin to the end-mirror of a Fabry-Perot resonator. The motion of the cantilever detunes the frequency of the resonator and thereby reduces the extent of the oscillator's position deflection. (B) Cavity cooling via coherent scattering of a levitated nanoparticle in a optical tweezer setup, as demonstrated in~\cite{delic2020cooling}. The optical cavity is solely filled with the scattered light of the nanoparticle. 
    (C) Schematic setup showing an active feedback cooling scheme of a levitated nanosphere to its c.o.m. motional groundstate via optimal control techniques. An electric field is modulated via the electrode using the signal returned by the Kalman filter and scaled by the linear-quadratic regulator, as shown in~\cite{magrini2020optimal}. This field acts on the charged nanoparticle.
    (D) Depiction of an optomechanical interaction between a frequecncy detuned laser guided in a tapered fibre close-by a whispering gallery mode resonator to control and cool the mechanical motion of the structure, as demonstrated in~\cite{anetsberger2008ultralow,li2016simultaneous}.}
    \label{fig:schematic}
\end{figure}

\begin{figure*}[t!]
    \centering
    \includegraphics[width=\linewidth]{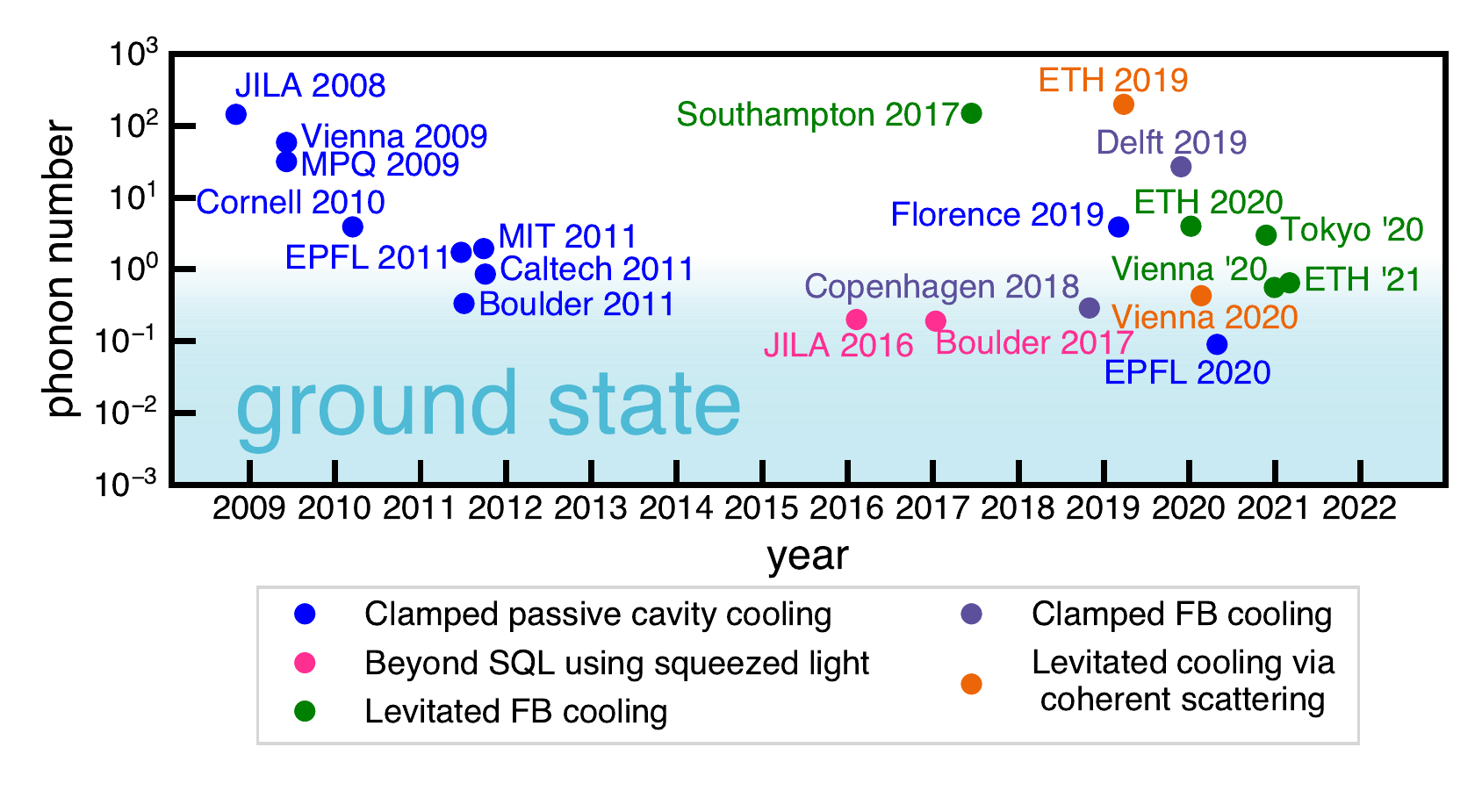}
    \caption{ Experimental results for cooling of macroscopic systems over the years. Minimum phonon occupation number is plotted against the date of publication. Blue data points represents experiments relying only on passive cavity cooling: JILA 2008~\cite{teufel2008dynamical}, Vienna 2009~\cite{groblacher2009demonstration}, MPQ 2009~\cite{schliesser2008resolved}, Cornell 2010~\cite{oconnell2010quantum}, MIT 2011~\cite{schleiersmith2011optomechanical}, EPFL 2011~\cite{riviere2011optomechanical}, Caltech 2011~\cite{chan2011laser}, Boulder 2011~\cite{teufel2011sideband}, Florence 2019~\cite{chowdhury2019calibrated} and EPFL 2020~\cite{qiu2020lasercooling}. Pink data points are results using squeezed light to surpass the standard quantum limit imposed on cavity cooling: JILA 2016~\cite{peterson2016laser} and Boulder 2017~\cite{clark2017sideband}. Purple data points present results using a feedback cooling scheme: Copenhagen 2018~\cite{rossi2018measurement} and Delft 2019~\cite{guo2019feedback}. Orange data points show recent results of cooling levitated nanoparticles using coherent scattering in a cavity: ETH 2019~\cite{windey2019cavity} and Vienna 2020~\cite{delic2020cooling}. Green data points show recent data of a nanoparticle feedback cooled in an optical tweezer using no cavity for cooling or read-out purposes: Southampton 2017~\cite{vovrosh2017parametric}, ETH 2020~\cite{tebbenjohanns2020motional}, Tokyo '20~\cite{kamba2020recoil}, Vienna '20~\cite{magrini2020optimal} and ETH '21~\cite{tebbenjohanns2021quantum}.}
    \label{fig:quantumoptomechanics}
\end{figure*}

Following the first motional ground state cooling of a clamped optomechanical resonator in 2010~\cite{oconnell2010quantum}, the technological advances in laser cooling and trapping have allowed larger objects to enter the quantum regime. Many clamped systems have reached close to, or have entered, the quantum regime~\cite{aspelmeyer2014cavity} as shown in Fig.~\ref{fig:quantumoptomechanics}. Examples include a nano-drum~\cite{teufel2011sideband}, a silicon nanobeam~\cite{chan2011laser} as shown in Fig.~\ref{fig:schematic}(A), a membrane~\cite{peterson2016laser}, a whispering gallery mode microresonators~\cite{schliesser2014cavity, schliesser2008resolved} as shown in Fig.~\ref{fig:schematic}(D), a membrane where both coherent light and squeezed microwave fields are used for cooling~\cite{clark2017sideband}, and a millimeter sized membrane that also acts as a photonic crystal~\cite{rossi2018measurement}.

Operating in the quantum regime with {free} or {levitated} particles allows the generation of macroscopic quantum states that are less coupled to their environment than clamped systems, which greatly enhances the coherence time of the quantum states~\cite{millen2020optomechanics}. A macroscopic quantum state is created by cooling the center of mass (c.o.m.) motion of the nanosphere at trapping frequency $\Omega$ provided by an optical tweezers, optomechanical cavity, ion trap or magnetic field~\cite{matsko2020mechanical}, where the ground state condition requires the average phonon occupancy to be less than one at this trapping frequency. Typical c.o.m.~oscillation frequencies of levitated nanoparticles range from 10-200\,kHz while the mass varies from $10^{-19}$\,kg to $10^{-16}$\,kg~\cite{delic2020cooling,millen2020optomechanics}. 

 \FloatBarrier
 
To understand the feasibility of performing measurements with a quantum nanoparticle, we consider its positional spread. This grows approximately linearly in time when the particle is released from the levitating potential. Considering typical parameters for a levitated nanoparticle of mass $m = 10^{-18}\,$kg and $\Omega_m = 2\pi\times 10^5\,$rad/s, this yields zero point displacement fluctuations on the order of $\sigma_{\rm{zpf}} \approx 10^{-11}\,$m, requiring hundreds of seconds of expansion until the quantum position spread is as large as the particle with a radius of $r=35\,$nm, a reasonable definition of a macroscopic quantum state.

In order to ascertain when the ground state condition has been reached, a narrow cavity resonance linewidth $\kappa$ is preferred as it enables  to reach the resolved-sideband regime, assuming that the mechanical frequency $\Omega_{m}$ is larger than $\kappa$. This allows read-out of energy transfer between the optical and mechanical modes in an anti-Stokes/Stokes process\footnote{Conversely, selectively pumping the anti-Stokes process enables cooling of the mechanical oscillator \cite{aspelmeyer2014cavity}. 
}~\cite{aspelmeyer2014cavity}. The sideband asymmetry allows for measurement of the number of phonons and is considered a more accurate method of thermometry than read-out of the mechanical mode's power spectral density. Although a cavity is not necessarily required to reach the ground state, it provides resonant enhancement in read-out and interaction strength, thereby reducing the number of photons needed to interact with the mechanical oscillator and improving the signal to noise ratio~\cite{tebbenjohanns2020motional}.

\subsubsection{Ground based quantum state preparation}

A range of passive and active cooling methods to achieve quantum ground state preparation of macroscopic objects using the optomechanical coupling are described in multiple review papers~\cite{millen2020optomechanics,aspelmeyer2014cavity}, with many techniques such as side-band resolved cooling derived from the cold atom community~\cite{ritsch2013cold}.

In 2020, the c.o.m.~motion of a 143\,nm diameter silica nanosphere levitated by an optical tweezers within an optical cavity was cooled to its zero point energy, which corresponds to an average phonon occupancy smaller than one, using the cavity optomechanical interaction together with a coherent scattering scheme~\cite{delic2020cooling}. This experimental set-up uses a technique called coherent scattering~\cite{vuletic2000laser,windey2019cavity} and is shown in Fig.~\ref{fig:schematic}(B). The tweezers reduces the chances of losing the particle when reaching ultra high vacuum, as compared to direct trapping by the optical cavity field.  
The coupling strength is at its highest when the particle is held at the cavity node. The optical cavity is not pumped separately, rather the trapping optical tweezers' frequency is stabilized relative to the cavity resonance using a weak beam which minimally interacts with the nanoparticle. Light scattered out of the tweezers field by the nanosphere then bounces off the cavity mirrors and interacts coherently with the oscillator again. Pumping of the cavity using only light scattered by the nanoparticle is a key feature. Consequently, each photon populating the cavity mode interacts with the particle, increasing the optomechanical coupling rate. As a result the quantum cooperativity of the experiment, namely the ratio of the optomechanical coupling strength and the product of the optical and mechanical decay rates, is well above 1000. To put this into perspective, a system with a quantum cooperativity bigger than one has been a long pursued goal in levitated optomechanics and is the benchmark for entering the quantum backaction regime~\cite{millen2020optomechanics}. A high cooperativity is also known in cold-atom physics to produce a constant cooling rate for cavity assisted molecule cooling in dynamical potentials~\cite{ritsch2013cold}. Compared to the general cavity cooling scheme~\cite{delic2020levitated}, the estimated improvement in cooperativity is $10^5$-fold~\cite{delic2019cavity} due to the coherent scattering procedure, with the added benefit of a reduced cavity drive power.

At the end of 2020, the ground state cooling of the c.o.m.~motion of a 143\,nm diameter levitated nanosphere using optimal control, schematically shown in Fig.~\ref{fig:schematic}(C), was announced~\cite{magrini2020optimal}. This type of cooling differs from the passive scheme employed in coherent scattering through the use of active feedback loops to generate the damping forces. 
A combination of an optical tweezers trap and a state-estimation feedback algorithm is used, enabling cooling of the 105\,kHz c.o.m.~in one direction with a final average phonon occupancy of $n=0.56 \pm 0.02$ quanta. Crucial for the success of this ground-state cooling scheme were the Heisenberg limited con-focal position detection and a combined implementation of a Kalman filter together with a linear-quadratic regulator~\cite{kwakernaak1972linear,bouten2008separation} determining the optimal feedback output control. The optimised detection of the particle's motion in the back-scattering plane of the optical tweezers allows to follow the particle's position with an uncertainty that is 1.3 times the size of the zero-point motion fluctuation. Additionally, the identification of important external noise sources and photon/information loss mechanisms of the experimental setup enabled to provide a high confidence in the accuracy of the model parameters of the employed Kalman-Bucy filter. 
Interestingly, the authors point out that the ground-state of a levitated particle can be reached even with a simple derivative filter using the correct gain settings~\cite{magrini2020optimal}.

As explained in Sec.~\ref{Applications}, many quantum sensing proposals that utilise spin-coupling do not require the mechanical oscillator to be in the ground state. Instead, low phonon occupancy ($n<10$) is sufficient as, even at this regime, the zero point motion emerges as a sizable contribution to the dynamics. In 2020, this was shown using a 136\,nm diameter nanoparticle trapped only by an optical tweezers, and cooled using active velocity damping~\cite{tebbenjohanns2020motional}. Also notable is the absence of a cavity which reduces the configuration overheads and allows for less obstructed measurements. For measurement protocols, it removes timing constraints posed by the response time of the cavity, possibly enabling faster pulse sequences and sampling rates. The c.o.m.~mode was cooled to an average occupation of $n=4$ phonons at frequency 50\,kHz. A major benefit of this scheme is the use of backscattering to detect the oscillation along the cooling axis, which allows for cooling to the ground state provided that the laser noise on the detector is sufficiently low.

\subsubsection{Space Feasibility studies}

\begin{figure}[t!]
     \centering
     \includegraphics[width=\linewidth]{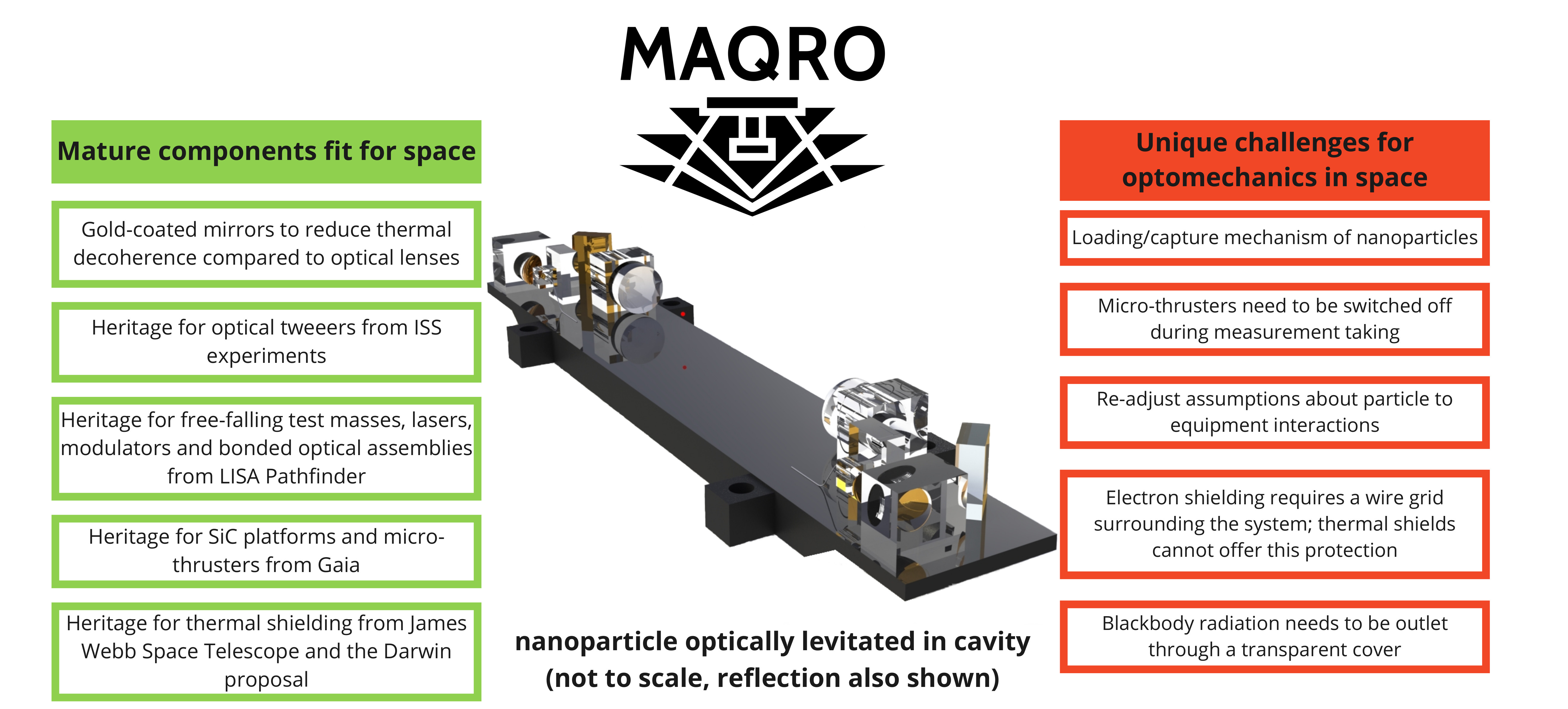}
     \caption{ The MAQRO project aims to bring quantum optomechanics into space, focussing on levitated nanospheres within an optical cavity (shown in the center). After completing an ESA CDF study, the components that are fit for space and the unique challenges to overcome for macro-scale particles in space were identified. Universal challenges in sending any experiment, quantum or classical, to space are summarised in the next Sec.~\ref{WishList}. Shown is the prototype of a fiber-coupled high-finesse cavity on a Silicon-Carbide (SiC) base plate for MAQRO, being developed as part of ULE-Cavity-Access (FFG project number 854036). The levitated nanoparticle (not to scale) is shown in red (particle reflection also shown).}
     \label{fig:maqroschematic}
 \end{figure}

The advantages of performing quantum optomechanics experiments in space is a reduction in ground-based noise sources such as seismic noise and changes to Earth's gravitational field. Vibrations, gravitational field-gradients, and decoherence through interaction with the environment fundamentally limit ground-based macroscopic quantum superpositions. This is particularly important for sensing, for example to eliminate the bulky and complex stabilisation platforms required for gravitational wave detection. Furthermore, many fundamental tests of physics require a micro-gravity environment ($\leq 10^{-9}g$), long free-fall times (100\,s), and a large number of repetitions ($10^4$) per measurement, which are more easily fulfilled with a space-based setup~\cite{CDFMAQRO2019}. Over the years, the mission scenario MAQRO has been developed with these aims~\cite{kaltenbaek2012macroscopic, kaltenbaek2013maqro, kaltenbaek2013testing, kaltenbaek2013final, kaltenbaek2014maqro, kaltenbaek2016macroscopic, hechenblaikner2014cold, kaltenbaek2015optomechanical, kaltenbaek2015testing, kaltenbaek2016towards, zanoni2016thermal, kaltenbaek2017concepts}. Some aspects, especially the thermal shielding and how cold one can get in space, were studied numerically in detail. A publication related to the debrief of the ESA CDF study on the MAQRO related Quantum Physics Payload Platform (QPPF) has been published in January 2019~\cite{CDFMAQRO2019}. In Fig.~\ref{fig:maqroschematic}, it is shown the core levitated optomechanics experiment platform, along with all the mission design considerations, indicating the technology maturity and projects that have arisen to solve certain technological challenges.

It is important for any new platform technology to consider legacy components and methods that are already present in space. For example, the MAQRO mission proposal benefits from technological heritage from the James Webb Space Telescope (JWST) \cite{Lightsey2012a} and the DARWIN mission \cite{Leger2007} proposal with respect to methods for passive radiative cooling, and from LISA \cite{baker2019laser} with respect to outgassing to space and the implementation of stably bonded optical assemblies \cite{Armano2009}. In addition, optomechanical experiments share very similar components to those on-board the ISS and LISA Pathfinder. Optical tweezers already present on the ISS as part of the Light Microscopy Module~\cite{Resnick2001}, which was the first optical tweezers deployed in a microgravity environment with military specifications, are crucial building blocks for future space-based optomechanics missions. The LISA Pathfinder mission established to be capable of keeping large test-masses (cube of 46\,mm size and 1.928\,kg mass) in free-fall with unprecedented precision~\cite{PhysRevLett.116.231101}. The mission found excess noise at lower frequencies from forces acting on the surface of the spacecraft such as spontaneous out-gassing, virtual leak pressure effects, electrostatic noise from fluctuating small-scale surface charges~\cite{armano2017charge}, or other short range forces~\cite{armano2021sensor}. Additional effects characterised include mass depletion~\cite{schlappi2010influence} and the generation of false acceleration due to electrostatic noise~\cite{armano2017charge}. For optimum stability, LISA Pathfinder highlighted that noise from control voltages, electrostatic potentials, and laser intensity needs to be reduced such that it causes displacement changes no greater than 1\,fm\,s$^{-2}$/Hz$^{1/2}$~\cite{armano2018beyond}. In 2021 the LISA Pathfinder achieved an in-flight measurement, with background stabilisation of 32$\times 10^{-15}$ms$^{-2}$/Hz$^{1/2}$ \cite{armano2021sensor}. In previous tests, the reduction of the different noise sources enabled a stability measured at close to 1\,mHz of the LISA Pathfinder mission along the three axis as X: $5\times 10^{-15}$ms$^{-2}$/Hz$^{1/2}$, Y and Z: $4\times 10^{-14}$ms$^{-2}$/Hz$^{1/2}$, while the angular acceleration noises are respectively $3\times 10^{-12}$rad s$^{-2}$/Hz$^{1/2}$ and  $3\times 10^{-13}$rad s$^{-2}$/Hz$^{1/2}$ \cite{armano2019lisa}. 

Not all components or methods are interchangeable across the photonics, cold-atom and optomechanics platforms. Unique for optomechanics is the need to trap, control, and cool macro-sized objects, which are substantially larger than single atoms and more prone to environmental decoherence than photons. The technical challenges unique to a levitated optomecahnics platform to operate in space relate to the levitated test particles, their confinement, and enhancing the coherence time~\cite{CDFMAQRO2019}. 
The test particles must be dielectric and highly transparent and also uncharged with uniform shape. The levitated optomechanics community uses the St{\"o}ber process to prepare silica nanoparticles which is an example of a liquid phase sol-gel process~\cite{stober1968controlled}. A molecular precursor (typically tetraethylorthosilicate) is first reacted with water in an alcoholic solution causing resulting molecules to join together to build larger structures. The reaction produces silica particles with diameters ranging from 50 to 2000\,nm, depending on the preparation conditions~\cite{levy2015sol-gel}. Recently, efforts have been made to fabricate nanostructures using clean room processes on silicon wafers~\cite{kuhn2015cavity}. Since levitated nanospheres are bulk objects, they are prone to light absorption and internal heating that is harder to control than for atoms and photons. Decoherence associated to blackbody radiation should be smaller than $10^{-10}$ms$^{-2}$/Hz$^{1/2}$, requiring thermal stability at the milli-Kelvin level~\cite{CDFMAQRO2019}.
Ideally, the same nanoparticle would be used repeatably over several thousand runs. However, because repeated optical interaction with the particle would significantly increase the internal temperature of the particle, and because the position uncertainty of the particle after an extended period of free fall will unavoidably be larger than the waist of the cavity mode, a new particle needs to be used for each data point. Since each run may take up to 100\,s, it is critical to ensure that unpredictable movements of the test particle with respect to the cavity mode are kept to a minimum. For that reason, the microthrusters will need to be turned off during measurements, and the spacecraft will drift freely~\cite{CDFMAQRO2019}. Gravity gradients, uncertainty in the mass model and variations in the solar radiation pressure will then limit the maximum measurement time to a few 100\,s~\cite{CDFMAQRO2019}.
During free fall, the test particles must remain limited to a confined region to allow an optical measurement of their position after extended free-fall times. This places limitations on the photon occupation numbers of $\leq 10$ along the cavity and $\leq 10^4$ perpendicular to it, to ensure suitably coherent and quantum measurements~\cite{CDFMAQRO2019}. In the base-line design for MAQRO and QPPF, the length of the optical cavity is 97\,mm, requiring a minimum finesse of $3 \times 10^4$ to achieve cooling close to the quantum ground state. To ensure at least partially coherent evolution, and the formation of a discernible interference pattern, the QPPF study concluded that that should be an 80\% probability for less than two collisions occurring during the measurement period. The required vacuum can be sustained by the use of a reactive material, namely a getter material, placed inside the enclosure of the optical bench. Indeed, when gas molecules hit the material, they will combine with it chemically or by absorption~\cite{CDFMAQRO2019}. 
A specific prerequisite for levitated nanoparticle experiments on ground and in space is a loading mechanism for the nanoparticles. In this context, a number of nanoparticle loading mechanism have been demonstrated. They range from nebulising the nanoparticles suspended in a solution near the trapping site and electrospraying to directly launching them from a surface via a piezoelectric transducers or laser-induced acoustic desorption~\cite{ashkin1971optical,Millen2015cavity,kuhn2015cavity,bullier2020characterisation,grass2016optical}. More recently, recapture methods have been demonstrated~\cite{hebestreit2018sensing} which will be of particular interest for matter-wave interferomertry using levitated nanoparticles in space due to the possibility of recycling the same nanoparticle over and over again and thereby circumventing the resource problem of a finite number of nanoparticles to load. In the initial MAQRO proposal, it was suggested to store the test particles close to the optical trap on the surface of piezoelectric devices, and to release the particles by exciting surface acoustic waves \cite{kaltenbaek2012macroscopic}. Currently, MAQRO proposes to use a hollow core photonic crystal fiber (HCPCF) to transfer particles to the experiment sight. Proof-of-principle demonstrations on ground were successful\cite{grass2016optical}. For MAQRO, it was proposed that the HCPCF, in combination with a segmented linear Paul trap, is used to transfer nanoparticles to the experimental region outside the spacecraft from a buffer gas chamber inside the spacecraft~\cite{kaltenbaek2016macroscopic}. The storage of particles requires further thought due to heating increasing the internal test particle's temperature (which must not be significantly higher than environment temperature). For this, several strategies are considered, from the use of low absorption materials, the use of each particle only once, the use of non-optical methods for trapping, or filling the HCPCF with a buffer gas for sympathetic cooling - none of these have been demonstrated yet.

\section{The ecosystem of space technology development
}\label{WishList}
Developing quantum systems for space requires structured, long-term planning to ensure that systems operate autonomously in challenging environments for several years. The persistent quality of each component is thus critical for the integrity of experiments.  Furthermore,  the cost-to-performance ratio, where space qualifying a quantum device can bring a clear gain~\cite{chase2003identifying}, needs to be taken into account for projects that require conspicuous investments. Indeed, similarly to the challenge of commercialising quantum technology, sending quantum devices into space requires extensive scientific and technical   investments and a mature infrastructure that can financially support these operations~\cite{rasanen2021path}. 

This section summarizes some of the main challenges and advancements that quantum technologies need to face to be successfully translated into space from the perspective of both the academic and the industrial sector.  The most relevant are

\begin{center}
\begin{itemize}
   \item \textbf{Environmental Control}
   \begin{itemize}
     \item Maintaining vacuum and environmental temperature
     \item Mechanical Stability/Vibration Isolation/Alignment
       \end{itemize}
     \item \textbf{Component Development}
   \begin{itemize}
     \item Miniaturization
     \item Optics
       \end{itemize}
       \item \textbf{Automation}
   \begin{itemize}
     \item Power consumption
     \item Electronics \& Control
       \end{itemize}
       \item \textbf{Quality Assurance/Environmental Testing}
   \item \textbf{Collaborative Ecosystem}
   \end{itemize}
 \end{center}
The following list, though not complete, applies to every space mission involving quantum technologies on board.

\subsection{Environmental Control}
The space environment is not homogeneous, with variations in temperature, vacuum pressure, and radiation.  For a ground based experiment to become space ready, it is necessary to translate all the subcomponents to space-hardened ones, developing new equipment where necessary.  

\subsubsection{Maintaining vacuum and environmental temperature}
Setting up stable conditions for an experiment in space is very difficult due to the extreme thermal conditions the spacecraft will be facing when traversing through outer space~\cite{kaltenbaek2012macroscopic,hechenblaikner2014cold}. The solution is to add a passive shielding which reduces the impact of solar radiation on the thermal stability. Using a passive solution is particularly better for a spacecraft than active cooling which would require cryostats~\cite{zanoni2016thermal}. Indeed, cryostats and dilutions refrigerators, which are typically used on ground in quantum experiments, cause vibrations, especially when one needs to reach low temperatures~\cite{triqueneaux2006design,branco2014athena}. {Additionally, keeping liquid helium in a container in a gravity free environment is a challenge on its own because dewars (Thermos bottles) function due to gravity, enabling the liquid and gas to naturally separate. In the zero gravity of space, a different method of separation is needed, which has been solved by the Gravity Probe B using a porous plug that releases the evaporating helium while retaining the superfluid liquid helium~\cite{buchman2000gravity,wang2015porous}.} The mission used a sponge mechanism inside a porous plug to prevent liquid helium sloshing and to control the pressure in the dewar~\cite{everitt2015gravity,einsteinuniverse}. Finally, due to resource constraints in space, a limited supply of liquid Helium in an open cooling loop can drastically curtail the lifetime of a space mission and unaccounted heating sources can cause depletion even faster, leading to a potential abrupt end of the mission. Similar resource constraints apply to closed loops cooling systems on a space craft due to the combination of the high power consumption and the limited energy budget for a space mission's lifetime~\cite{dipirro2016superfluid,duband2015space}.

An additional environmental challenge in space is the achievable vacuum pressure. The inevitable out-gassing of spacecraft components due to the interaction with thermal radiation~\cite{hechenblaikner2014cold} may ruin the required mission pressure. However, the thermal radiation out-gassing can be mediated to some extent by using radiation shields~\cite{zanoni2016thermal}, appropriate vacuum packaging~\cite{schwindt2016highly,rushton2014contributed,birkl2007micro} and enough getter material to outlast the mission's lifetime in fully sealed devices~ \cite{boudot2020enhanced, rushton2014contributed}. Current state of the art techniques are considered to allow an achievable and stable vacuum pressure of $10^{-10}-10^{-11}$\,mbar in space for low volume. {It may be possible to achieve large volumes of high vacuum by exploiting the wake behind a rapidly moving spacecraft in low Earth orbit, the so-called wake-shield concept~\cite{wuenscher1970unique,melfi1976molecular,strozier2001wake}.} Note that vacuum pressures below $10^{-12}$\,mbar, i.e. extreme high vacuum (XHV), are a two-fold challenge as the sheer measurement of these pressures is a difficult task in itself~\cite{redhead1999extreme,yoshimura2007vacuum}. As a reference, on Earth $1.2\times 10^{-18}$\,mbar is the best vacuum pressure that has been achieved~\cite{sellner2017improved}.
 
\subsubsection{Mechanical Stability/Vibration Isolation/Alignment} 
Mechanical instability arises from two sources; {forces applied to the spacecraft from the external environment and movements of objects within}. For environmental sources of vibration, atmospheric drag and solar wind can cause unwanted instabilities. The micro-propulsion system that controls the fine motion of a spacecraft also creates vibrations. Depletion of nitrogen from the cold gas micro-propulsion system produces a deterministic gravitational drift typically of several hundreds of fm\,s$^{-2}$/day~\cite{armano2018beyond}

Vibrations created by components of the experiment itself, e.g. cryostats, require attention~\cite{schmoranzer2019cryogenic}. However, there are existing solutions, used for Gravity Probe B and the Planck mission, where closed-loop and low-vibration cryostats were demonstrated. It should be noted that closed-loop cryostats also offer a longer lifetime compared with open loop ones. Although the helium leakage is unavoidable and amplified when there are sources of heat within the chamber, closed-loop recycles the Helium in a more conservative manner~\cite{vorreiter1980cryogenic}. 

Force noise caused by vibrations can be very detrimental for successfully observing coherent evolution of large-mass particles. The MAQRO mission, and its evolution in the QPPF, constitutes at present the prototypical proposal for such experiments in space. Partially to avoid potential limitations introduced by vibration noises, MAQRO has been planned for an L2 orbit where reduced contributions from environmental noise are possible~\cite{lu2019review,mcknight2019examination}. The constraints on acceptable levels of vibration noise are similar to those achieved by LISA Pathfinder \cite{CDFMAQRO2019}. One of the main drivers of these strong constraints if the fact that one cannot monitor the particle position while it is evolving freely. 

Finally, the lack of access to experiments once launched requires detailed understanding of how components and mounts change dimension from the initial alignment, usually conducted at atmospheric pressure and at room temperature, to the final environmental conditions, e.g. cryostat temperatures or high vacuum. Out-gassing, flexure stresses, and warping can all cause misalignment that is not easy to correct once the experiment is sealed. 

\subsection{Component Development} 
The unique and harsh environment of space requires components to be matured to greater lengths with respect to ground-based applications. Due to the cost of launch, sub-components should be lightweight and compact. In the context of performing quantum experiments in space, there are further demands on the lasers and optics as these enable the preparation and/or measurement of quantum properties. Engagement of component manufacturers is crucial as advancements and improvements are often needed in both performance as well as in addressing the components robustness to launch and space conditions.  

\subsubsection{Miniaturisation}
The size and mass of every component within a spacecraft contributes to a total `cost'. To reach the furthest orbits, such as geostationary transfer orbit, a high initial velocity is required which favours lower payload mass. The Tsiolkovsky rocket equation~\cite{turner2008rocket}, which defines the conservation of momentum for rockets, sets a maximum allowable mass. This equation is cast as different energy balances, for example, the energy expenditure against gravity (often called \textit{delta v} or the change in rocket velocity), the energy available in the rocket's propellant (often called exhaust velocity or specific impulse), and the propellant mass fraction (how much propellant is needed compared to the total rocket mass). The energy expenditure against gravity is specified by where one wants to go. As an example, for LISA to reach geostationary orbit implies a mass limit of 500\,kg.

In preparation for a surge in quantum experiments conducted in space, there are a variety of space saving and mass reducing efforts underway. For example, huge strides have been made on the sub-components required for atomic interferometry through a recent upgrade of CAL on board the ISS; CAL is the first atom interferometer to operate in space and it is approximately the size of a dishwasher~\cite{aveline2020observation}. Other examples include miniaturised cold atom parts and vacuum chambers~\cite{kim2017matter,elliott2018nasa},
;the miniaturisation of vapour cells has seen rapid progress thanks to the use of Rubidium cell frequency standards in satellite systems ~\cite{batori2020gnss}; and additive manufacturing for space missions that are being explored~\cite{rinaldi2021additive}. 

Finally, in recent years miniaturisation is becoming less of a barrier as the cost of launch reduces. Space-X has effectively lowered the cost of launch by an order of magnitude~\cite{jones2018recent}, with cubesats opening opportunities for low-cost access to low orbits with less size restrictions going from 6 units (6U) to 24U.

\subsubsection{Optics}
Lasers, detectors and optical components interact with quantum experiments to probe, prepare and reset quantum states. Components used in ground based experiments are not automatically suitable for space and often require specialist involvement. For example, the Ferdinant-Braun Institut builds the lasers for BECCAL and MAIUS missions~\cite{bawamia2019semiconductor} and CERN has been involved in miniaturising radiation tolerant Indium Phosphide lasers as photonic integration circuit components~\cite{alt2016photonic,zhao2018indium}. Furthermore, in order to guarantee the long-term stability of the laser systems during the space missions' lifetime, frequency locking the laser is a necessity. To this extent, several techniques have been established including mode-locking ultra-short femtosecond pulses to control the spectral width and pulse shape~\cite{li2020intelligent,mondin2017laser}.

Some laser wavelengths require substantial development, such as 200\,nm wavelength UV sources. In general, space ready laser systems tend to be at the telecoms and infra-red wavelengths~\cite{gregory2017tesat}, such as those already used by LISA Pathfinder. Such lasers are high performance and have narrow linewidth, with reduced laser noise from active feedback methods referenced to a stable frequency source and often operating at the shot-noise limit whereby classical fluctuations in the laser output are suppressed. For other quantum experiments, distinctly different light sources are required such as space qualified single photon light sources which generate radiation with a photon-number distribution that has a mean of one and variance of zero~\cite{sun2004space, yang2019spaceborne, cheng2015space}.

Other innovations in maturing optics for space include a roadmap for space-qualified delay lines in the DARWIN proposal~\cite{Leger2007}, used to balance optical paths to nanometer accuracy. NASA has also developed many Earth orbiting Laser Detection and Ranging (LiDAR) instruments for analysing climate change such as ATLAS (Advanced Topographic Laser Altimeter System) and GEDI (Global Ecosystem Dynamics Investigation), both launched in 2018~\cite{ott2020space}. 
In particular, ATLAS' laser has thousand times longer lifetime than previous LiDAR missions. However, during its development approximately 20\% of the Q-switch crystals had bond failure, highlighting the need for extensive subcomponent screening tests and verification when constructed from commercial parts. 

Reference~\cite{ott2020space} displays an extensive list of optoelectronics components that have been tested for space by ``Parts, Packaging and Assembly Technologies (Code 562)'' branch of NASA's Goddard Space Flight Center, including lasers, photodiodes and LEDs. The branch previously helped the James Webb Space Telescope Optics team with qualifying surface mount LEDs for use in cryogenic temperatures. They have also conducted extensive testing on optical fibers to qualify their robustness to radiation, vibration and cryogenic temperatures. 

{Developments in telescope design can also be of benefit to quantum communications, including deployable optics to enable larger apertures to be launched, freeform surfaces allowing more complex, compact optical designs, and new materials with better strength and thermal properties. For transmitters, larger apertures allow for smaller beam widths and lower link losses. Ground-based receivers will benefit from more widespread availability of adaptive optics to compensate for turbulence and allow for single mode coupling, reducing losses and cutting down on stray light, important for improving signal-to-noise for space to ground quantum communication.}

\subsection{Automation}
Regardless of the specific tasks considered, once in space experiments should be capable of run autonomously. Even when based on the ISS, requiring intervention from the ground is costly and extraordinary maintenance tasks come with unexpected overheads and skill requirements.
Thus,  with few exceptions, the need for access, maintenance, operation, and repair of quantum experiments conducted in space should be minimised. This is not only to protect the safety and security of any on-board personnel, but to enable experiments to run autonomously without any human intervention. For example, BECCAL, on board the ISS, is a fully automated system and the high voltage lines and free-space laser beams are guarded by three levels of containment for the safety of ISS astronauts, as detailed in the NASA document SSP51721 on payload safety policy and requirements for the ISS~\cite{andrews_2019}. Below we highlight unique challenges in automating quantum experiments.

\subsubsection{Power consumption}
The complexity of preparing, maintaining and measuring quantum states requires high power consumption. LISA Pathfinder uses a solar array with power output of $\sim680$\,W. 
Power budgets are usually expressed as the Orbit Average Power (OAP) minus the average power used, where a positive power budget is preferred unless load can be removed to recover from a negative power budget. Power storage typically occurs in batteries with custom designed power management and distribution systems, and solar power generation as the predominant method used by 85\% of all nanosatellite form factor spacecraft. Solar cell efficiencies hover around 30\% and primary-type batteries comprise of silver Zinc devices or Lithium based batteries. Secondary type batteries include Nickel and Lithium based portable devices~\cite{nasapower}.

\subsubsection{Electronics \& Control}
Although some electronics can be sourced off-the-shelf, calm current operation may not be guaranteed, with bespoke requirements such as fast switching times not commercially available. Going against the need for miniaturisation, large electronic chips may be preferable for avoiding radiation issues -- like cosmic rays damaging small form factor transistors~\cite{hoeffgen2020investigating} -- balanced by the more flexible opportunity to build in redundancy by moving to smaller device geometries placed in different locations.

Electronics requirements can be quite different to drawing based, plug and play components such as vacuum chambers, with added cost in manufacture, e.g. it can cost up to a million USD to develop an application-specific integrated circuit (ASIC) for low-power and ruggardised operation. Control software is also important. Finally, to perform Fourier transforms, power spectrums, and store data require large amounts of memory and storage. For example, the James Webb Space Telescope can only store 58.8 Gbytes of science data where downlinks for data transfer occur in 4-hour contacts, twice per day, transmitting at least 28 Gbytes~\cite{jwstpower}. Efforts to space harden off-the-shelf memory -- like the development of a 3D memory cube design with 24 NAND flash dies~\cite{sidana2019flashrad} -- are underway.

\subsection{Quality Assurance/Environmental Testing}
The most important step for any component to be deemed space ready is passing quality assurance and environmental testing. The quality assurance stage can stagnate or prevent a project from continuing. Dependability and safety are considered in the product assurance stage, which requires specialist understanding. Vibration testing is required for any component with movable parts, applying a range of random and sinusoidal vibrations with electrodynamic shakers~\cite{paris2015vibration}. Radiation hardening or shielding may be required more for satellite components than for experiments on board the ISS due to existing shields. Access to the test facilities incurs a cost, but many engineering departments within universities will have electrodynamic shakers that can be programmed to apply appropriate forces specified by the launch providers. Effects of charges are more detrimental for the optomechanics platforms, but can also effect dielectrics, for example, due to cosmic radiation~\cite{Kaletenbaek2012techreport}. Discharging procedures have been explored but are still not at a maturity level suitable for space~\cite{frimmer2017controlling,ranjit2015attonewton}, and may not be sufficient for removing any dipole moments~\cite{Moore2014}. Surfaces close to trapping potentials can also skew the potential, creating biases in position~\cite{diehl2018optical,winstone2018direct}. 

The European Cooperation for Space Standardization (ECSS, \href{https://ecss.nl/}{https://ecss.nl/}) is an initiative established to develop a coherent, single set of user-friendly standards for use in all European space activities. ECSS was created under the auspices of ESA in 1989. The European Space Components Information Exchange System (ESCIES, \href{https://escies.org/}{https://escies.org/}) is a repository for EEE parts information hosted by ESA, on behalf of the Space Components Steering Board, as part of the European Space Components Coordination. The European Preferred Parts List (EPPL) is a list of preferred and suitable components to be used by European manufacturers of spacecraft hardware and associated equipment. It is part of the ECSS system which releases an updated parts list every year, for example, EPPL Issue 41 corresponds to the year 2020~\cite{eppl}. The IEEE (\href{https://www.ieee.org/standards/}{https://www.ieee.org/standards/}) also develops industry standards and welcomes participation from the broad academic and industrial community. ISO (the International Organisation for Standardisation, \href{https://www.iso.org/}{https://www.iso.org/}), originally founded in 1947 to set the future of international standardisation, works in a similar way.

\subsection{Collaborative ecosystem}
Expanding the diversity of quantum experiments being performed in the optimal conditions that space and micro-gravity offer has clear benefits for the progression of our understanding of the Universe and for generating economic and innovative outputs. It will take a global effort and deeper cross engagement with existing communities in space research to realise the tangible benefits offered by quantum technologies.

A relevant effort towards coordination is represented by the COST Action ``Quantum Technologies in Space'' (QTSpace, \href{http://www.qtspace.eu/}{http://www.qtspace.eu/}), the first pan-European community comprising academic and industrial partners, addressing the definition, study and development of Quantum Technologies (QT) for Space. QTSpace organized regular meetings to identify the fundamental questions to be addressed, the scientific and technological requirements, and the principles to be demonstrated, to share knowledge and promote cooperation. It coordinated the writing of the ``Intermediate Strategic Report for ESA and the National Space Agencies'' (\href{http://qtspace.eu:8080/sites/testqtspace.eu/files/QTspace_Stretegic_Report_Intermediate.pdf}{link to the document}) and a ``Policy White Paper on Quantum Technologies for Space'' (\href{http://www.qtspace.eu/?q=whitepaper}{link to the document}), which contain the roadmaps for the development of QT for Space, and are now used by Space Agencies and Policy Makers in Europe for the definition of new programs for research and technological development.

Consortia in between academia and industry, like the Horizon 2020 project Optomechanical Technologies and the European Photonics Industry Consortium (EPIC), allow for early exposure to commercial supply chains, standardisation, industrial co-developers and alternative funding streams. The longevity of consortia relies on consistent funding - for example, EPIC is funded through membership fees, with industry members charged proportionally to their size. Relying on research grant funding is not sufficient to maintain thriving communities due to the life-cycle of grants and the academic workforce. 

The type of funding required for quantum experiments in space should be diverse, with incentives for companies to participate, and long overarching project aims replacing small standalone demonstrations. The responsibility for the success of the mission is best managed by a mission lead who is able to manage technical and programme risks, especially quality assurance. 

Different types of community engagement are valuable for ensuring the longevity of quantum activities in space. For example, quantum technology is an export controlled by many countries, and there are calls for responsible and ethical development~\cite{coenen2017responsible}. One can take inspiration from existing initiatives such as the European Centre for Space Law (ECSL), founded by ESA in 1989 to improve space law research, education and practice in Europe. The Consultative Committee for Space Data Systems (CCSDS) is an international voluntary consensus organization of space agencies and industrial associates interested in mutually developing standard data handling techniques to support space research, including space science and applications \cite{lafferranderie2011astronauts}.

Lastly, training a new generation of quantum engineers is important for crossing technical barriers in applying quantum research to technological and novel environments. Combining efforts that already exist for quantum and space outreach and education could be an effective mechanism for generating a skilled workforce. As an example, Qiskit was developed as an open source software development kit for IBM's quantum computers and IBM also runs a Quantum Computing Summer school annually. Along these lines, there are student programs at NASA Langley and the ISSET (International Space School Educational Trust).

\section{Summary and Outlook}\label{Conclusions}
The communion between quantum technologies and space science is bound to have strong impact on our understanding of the physical world, both at the fundamental and applied level. We have shown that space offers novel avenues for applying and developing quantum technologies based, mainly, on three different physical platforms: cold atoms, photonics, and optomechanical systems. Some of these technologies have already shown readiness for operating in space, while others are at earlier stages and in search for validation. 

The widespread use of quantum technologies in space is becoming a reality nowadays. Quantum sensors have been employed in fundamental studies of Einstein's theory of general relativity and in the first demonstration of the potential of satellite quantum communication. In the last couple of decades, proposals aimed at exploring fundamental questions at the boundary between quantum physics and relativity in space have sprout driven by the possibility to exploit the metrological advantages offered by quantum mechanics. At the same time, the applications of quantum technologies for communication -- bearing with it the possibility of implementing quantum cryptographic protocols --, the possible advantages for navigation systems, and the promise of establishing a quantum internet in the future, have propelled the investigation into quantum technologies in space. 

Crucial milestones have already been reached. In 2017~\cite{becker2018space}, the first demonstration of Bose--Einstein condensation and interferometry in a sounding rocket ushered the path that has led to have a cold atom interferometry laboratory on-board the International Space Station~\cite{aveline2020observation}. Atomic clocks are being employed in the Global Navigation Satellite System (GNSS), they have been used and envisaged for fundamental tests of general relativity, and are expected to offer the possibility of independent spacecraft navigation in deep space without relying on communication to the ground. The Chinese-led mission QUESS, with the LEO satellite Micius, was the first space-based quantum communication mission to be launched, allowing to demonstrate entanglement distribution, ground-to-satellite quantum teleportation, and the realisation of a hybrid quantum communication network over thousands of kilometer distances. And the astounding results of the LISA Pathfinder mission, enabled through the use of quantum technologies, have laid the ground for the more ambitious LISA, on the path to observing the gravitational waves sky in an uncharted range of parameters and opening an entirely new observational window on the early Universe.

That the time is ripe for this communion between quantum physics and space science is also testified by the several calls for new ideas and proposals sported by the main space agencies around the world where projects involving quantum technologies and quantum physics research have been more and more considered. This is further fuelled by the large investments of all the major world's economies in the commercialization of quantum technologies and their application to face the challenges of the digital era. 

The path ahead for quantum technologies in space is an exciting one but it does not lack challenges. In this review we have identified some of such challenges: from environmental control to component development and from automation to quality assurance. Tackling them calls for a collaborative environment involving the academic as well as private sector and space agencies.

\section*{Acknowledgments}

The present review is one of the major outcomes of the coordination activities of the COST Action QTSpace, whose goal is to promote research and collaborations for the development of Quantum Technologies for Space applications, and for the study of fundamental physics in Space. All the authors acknowledge partial support from COST Action QTSpace (CA15220).

A. Belenchia acknowledge support from the Deutsche Forschungs-gemeinschaft  (DFG,  German  Research  Foundation)  projectnumber BR 5221/4-1, the MSCA project pERFEcTO (GrantNo.~795782).

G. Gasbarri acknowledge support from the Spanish Agencia Estatal de Investigaci\'{o}n, project PID2019-107609GB-I00, Spanish MINECO FIS2016-80681-P (AEI/FEDER, UE), Generalitat de Catalunya CIRIT 2017-SGR-1127, from QuantERA grant "C'MON-QSENS!", by Spanish MICINN PCI2019-111869-2, and the Leverhulme Trust (RPG-2016- 046)

W. Herr acknowledges financial support from the German Space Agency (DLR) with funds provided by the Federal Ministry for Economic Affairs and Energy (BMWi) due to an enactment of the German Bundestag under Grant No. DLR 50WM1952 “QUANTUS-V Fallturm” and from “Niedersächsisches Vorab” through the “Quantum- and Nano-Metrology (QUANOMET)” initiative within the project QT3 as well as through “Förderung von Wissenschaft und Technik in Forschung und Lehre“ for the initial funding of research in the new DLR-SI. Additionally W. Herr acknowledges support by the Deutsche Forschungsgemeinschaft (DFG, German Reasearch Foundation) – project-ID 434617780 – SFB 1464 TerraQ and under Germany’s Excellence Strategy – project-ID 390837967 – EXC-2123 QuantumFrontiers.

Y.L. Li acknowledges financial support from a Royal Academy of Engineering Intelligence Community Postdoctoral Fellowship Award: ICRF1920-3-10. 

M. Rademacher acknowledges funding from the EPSRC Grant No. EP/S000267/1.

R. Kaltenbaek acknowledges support by the Austrian Research Promotion Agency (projects 854036 and 865996) and by the Slovenian Research Agency (research projects N1-0180, J2-2514, J1-9145 and P1-0125).

L. Woerner acknowledges financial support through the newly funded DLR institutes, DLR-QT and DLR-GK and the DLR 'Wettbewerb der Visionen' through the Federal Ministry for Economic Affairs and Energy (BMWi). 

A. Xuereb supported by the European Union's Horizon 2020 research and innovation programme under grant agreement No 732894 (FET-Proactive HOT) and grant agreement No 101004341 (QUANGO), by the NATO Science for Peace and Security programme under grant agreement No G5485 (SEQUEL), and by the Julian Schwinger Foundation (TOM).

M. Paternostro is supported by the DfE-SFI Investigator Programme (grant 15/IA/2864), the Royal Society Wolfson Research Fellowship (RSWF\textbackslash R3\textbackslash183013) and the Leverhulme Trust Research Project Grant (grant nr.~RGP-2018-266).

A. Bassi acknowledges financial support from the INFN, the University of Trieste and the support by grant number (FQXi-RFP-CPW-2002) from the Foundational Questions Institute and Fetzer Franklin Fund, a donor advised fund of Silicon Valley Community Foundation.

M. Carlesso, H. Ulbricht,  M. Paternostro and  A. Bassi are supported by the H2020 FET Project TEQ (Grant No. 766900).

M. Carlesso and M. Paternostro acknowledge support from UK EPSRC (grant nr.~EP/T028106/1).

I. Derkach and V.C. Usenko acknowledge support from European Union’s Horizon 2020
research and innovation
programme project CiViQ (grant agreement no. 820466) and from the
project 19-23739S of the Czech
Science Foundation.

J.S. Sidhu and D.K.L. Oi acknowledge the support of EPSRC via the Quantum Communications Hub through grant number EP/T001011/1. D.K.L. Oi also acknowledges EPSRC grant EP/T517288/1.

\section*{List of acronyms and abbreviations}\label{sec:listAcronyms}
\addcontentsline{toc}{section}{\nameref{sec:listAcronyms}}

\subsection*{Experiments, Missions and Proposals}

\begin{longtable}{r|l|l}
\endfirsthead
\hline
\textbf{Acronym} & \textbf{Meaning} &\textbf{Section}\\
\hline
\endhead
\hline\hline
\textbf{Acronym} & \textbf{Meaning}&\textbf{Section}\\
\hline
ACES       & Atomic Clock Ensemble in Space & \ref{Fundamental}, \ref{ProofOfPrincipleImlementation}                                        \\
AEDGE      & Atomic Experiment for Dark   matter and Gravity Exploration   & \ref{Fundamental}         \\
ALIA       & Advanced Laser Interferometer   Antenna     & \ref{Fundamental}                            \\
ASTROD     & Astrodynamical Space Test of   Relativity using Optical Devices    & \ref{Fundamental}     \\
BBO        & Big Bang Observer                                                & \ref{Fundamental}       \\
BECCAL     & Bose Einstein Condensate and Cold Atom Laboratory    & \ref{Fundamental}  , \ref{ProofOfPrincipleImlementation}                  \\
BOOST      &  BOOst Symmetry Test                                 & \ref{Fundamental}, \ref{Applications}, \ref{ProofOfPrincipleImlementation}                                       \\
CACES   &Cold Atomic Clock Experiment in Space&\ref{ProofOfPrincipleImlementation}\\
CAI        &   Cold Atom Interferometry                                                  &  \ref{Applications}                   \\
CAL        & Cold Atom Laboratory                               & \ref{Fundamental},  \ref{ProofOfPrincipleImlementation}                   \\
CASPA   & Cold Atom Space PAyload  &\ref{ProofOfPrincipleImlementation}\\
CE         & Cosmic Explorer                                      & \ref{Fundamental}                  \\

COFROS     & Compact Optical Frequency References On a Satellite &\ref{ProofOfPrincipleImlementation}
\\
DECIGO     & DECi-hertz Interferometer   Gravitational wave Observatory    & \ref{Fundamental}            \\

DSAC       & Deep Space Atomic Clock                                  &\ref{ProofOfPrincipleImlementation}              \\
DSQL       & Deep Space Quantum Link                            & \ref{Fundamental}                    \\
EGE        & Einstein Gravity Explorer                            & \ref{Fundamental}                  \\
EPTA       & European Pulsar Timing Array                        & \ref{Fundamental}                   \\
ET         & Einstein Telescope                                  & \ref{Fundamental}                   \\
FOKUS      & \textit{Faserlaserbasierter Optischer Kammgenerator unter Schwerelosigkeit}   &\ref{ProofOfPrincipleImlementation}                                                                    \\
GAIA       &  Global Astrometric Interferometer for Astrophysics &\ref{ProofOfPrincipleImlementation}                                                                     \\
GAUGE      & GrAnd Unification and Gravity   Explorer      & \ref{Fundamental}                         \\
GIRAFE     &  Gravim\`etre Interf\'erom\'etrique de Recherche \`a Atomes Froids &\ref{ProofOfPrincipleImlementation}                                                                     \\
GNSS       & Global Navigation Satellite   Systems                 &\ref{Fundamental}, \ref{Applications}, \ref{ProofOfPrincipleImlementation}                 \\
GOCE       &  Gravity field and
steady-state Ocean Circulation Explorer                              &\ref{Applications}              \\
GPA        & Gravity Probe A     &\ref{Fundamental}                                                   \\
GPS        & Global Positioning System                                  &\ref{ProofOfPrincipleImlementation}            \\
GRACE-FO   & Gravity Recovery And Climate   Explorer Follow-On   &\ref{Fundamental}                   \\
GSAT       &   Galileo SATellite &\ref{Fundamental}                                                                     \\
GTB Pro    & GraviTower Bremen &\ref{ProofOfPrincipleImlementation}                                                     \\
HLVK       & Hanford Livingston VIRGO KAGRA       &\ref{Fundamental}                                  \\
I-SOC       &  ISS Space Optical Clocks   &\ref{Fundamental}                                                                    \\
ICE        &   {Interf\'erometrie atomique \`a sources Coh\'erentes pour l'Espace}    &\ref{Fundamental}, \ref{ProofOfPrincipleImlementation}                                                                 \\
IPTA       & International Pulsar Timing   Array    &\ref{Fundamental}                                \\
ISS        & International Space Station       &\ref{Fundamental}, \ref{ProofOfPrincipleImlementation}                                     \\
IVORY      & In orbit Verification of   stabilized Optical and mw Reference sYstems &\ref{ProofOfPrincipleImlementation} \\
JOKARUS    &  Jod-Kammresonator unter Schwerelosigkeit   &\ref{ProofOfPrincipleImlementation}                                                                   \\
KAGRA      &  Kamioka Gravitational Wave Detector     &\ref{Fundamental}                                                                 \\
KALEXUS    &    Kalium Laser-Experimente unter Schwerelosigkeit    &\ref{ProofOfPrincipleImlementation}                                                                \\
LEO        & Low-Earth orbit                                                       &\ref{Applications} \\
LIGO       & Laser Interferometer Gravitational-Wave Observatory                       &\ref{Fundamental}                     \\
LISA       & Laser Interferometer Space   Antenna                  &\ref{Fundamental}, \ref{ProofOfPrincipleImlementation}, \ref{WishList}                 \\
LLR        & Laser Lunar Ranging                &\ref{Fundamental}                                         \\
LOP-G      & Lunar Orbital Platform Gateway        &\ref{Applications}                                 \\
MAIUS      &    MAteriewellen-Interferometer Unter Schwerelosigkeit &\ref{Fundamental}, \ref{ProofOfPrincipleImlementation}                                                                   \\
MAQRO      & MAcroscopic Quantum ResOnator     &\ref{Fundamental}, \ref{ProofOfPrincipleImlementation}                                                                                          \\
MICROSCOPE &   Micro-Satellite a tra\^in\'ee Compens\'ee pour l'Observation du Principe d'Equivalence    &\ref{Fundamental}                                                                                                                         \\

NANOGRAV   & North American Nanohertz   Observatory for GAvitational waVes      &\ref{Fundamental}    \\
NGGM       & Next Generation Gravity Mission    &\ref{ProofOfPrincipleImlementation}                                                                                                                        \\
PHARAO     &   Projet d’Horloge Atomique par Refroidissement & \ref{Fundamental}, \ref{ProofOfPrincipleImlementation}                                                                     \\
PPTA       & Parkes Pulsar Timing Array      &\ref{Fundamental}                                       \\
PRIMUS     &    PR\"azisionsInterferometrie mit Materiewellen Unter Schwerelosigkeit &\ref{ProofOfPrincipleImlementation}                                                                   \\
Q-WEP      & Atom Interferometry Test of the Weak Equivalence Principle in Space       &\ref{Fundamental}              \\
QEYSSAT    & Quantum EncrYption and Science   SATellite    &\ref{Fundamental}                         \\
QPPF       & Quantum Physics Payload platForm   &\ref{Fundamental}, \ref{ProofOfPrincipleImlementation}                                    \\
QTEST      &  Quantum Test of the Equivalence principle and Space Time   &\ref{Fundamental}                                                                   \\
QUANTUS    &  QUANTen Gase Unter Schwerelosigkeit     &\ref{Fundamental}, \ref{ProofOfPrincipleImlementation}                                                                 \\
QUESS      & QUantum Experiments at Space   Scale    &\ref{Fundamental}                               \\
SAGAS      & Search for Anomalous Gravitation   using Atomic Sensors     &\ref{Fundamental}           \\
SAGE       & Space Atomic Gravity Explorer    &\ref{Fundamental}                                      \\
SAI        & Space Atom Interferometer     &\ref{Fundamental}, \ref{ProofOfPrincipleImlementation}                                         \\
SKA        & Square Kilometer Array          &\ref{Fundamental}                                           \\
SOC        & Space Optical Clock                   &\ref{Fundamental}, \ref{ProofOfPrincipleImlementation}                                 \\
STE-QUEST  & Space-Time Explorer and QUantum   Equivalence principle Space Test  &\ref{Fundamental}, \ref{ProofOfPrincipleImlementation}   \\
STEP       & Satellite Test of the   Equivalence Principle  &\ref{Fundamental}                                                                                                              \\
                                                    
\hline\hline
\end{longtable}

\subsection*{Other used acronyms}

\begin{longtable}{r|l|l}
\endfirsthead
\hline
\textbf{Acronym} & \textbf{Meaning} &\textbf{Section}\\
\hline
\endhead
\hline\hline
\textbf{Acronym} & \textbf{Meaning}&\textbf{Section}\\
\hline
BEC        & Bose Einstein Condensate       &     \ref{Fundamental}, \ref{Applications}, \ref{ProofOfPrincipleImlementation}                                   \\
BIV        & Bell's Inequalities Violation  &        \ref{Fundamental}                                 \\
CDF        & Concurrent Design Facility                         &             \ref{Fundamental}, \ref{ProofOfPrincipleImlementation}                                 \\
CNES       &     Centre national d'\'etudes spatiales                     &           \ref{Fundamental}, \ref{ProofOfPrincipleImlementation}                                   \\
c.o.m.        & Center Of Mass             &      \ref{ProofOfPrincipleImlementation}                                      \\
COW        & Colella Overhauser Werner                      &  \ref{Fundamental}                      \\
CSL        & Continuous Spontaneous   Localization            &      \ref{Fundamental}                \\
CV         & Continuous Variable                            &  \ref{Applications}                      \\
DE         & Dark Energy                           &  \ref{Fundamental}\\
DLR        &  Deutsches Zentrum für Luft- und Raumfahrt                            &   \ref{Fundamental}, \ref{ProofOfPrincipleImlementation}, \ref{WishList}                                       \\
DP         & Diosi-Penrose             &        \ref{Fundamental}                                     \\
DV         & Discrete Variable                               &        \ref{Applications}               \\
EEP        & Einstein Equivalence Principle                  &       \ref{Fundamental}                \\
ESA        & European Space Agency                                 &   \ref{Fundamental}, \ref{ProofOfPrincipleImlementation}, \ref{WishList}              \\
GEO        & GEostationary Orbit                                  &        \ref{Applications}          \\
GR         & General Relativity                             &  \ref{Fundamental}                      \\
GRW        & Ghirardi Rimini Weber                                   &   \ref{Fundamental}            \\
GW         & Gravitational Waves                                   &    \ref{Fundamental}             \\
IMU        & Inertial Measurement Unit                      &        \ref{Applications}                \\
JWST        & James Webb Space Telescope                      &        \ref{ProofOfPrincipleImlementation}                \\
LEO        & Low Earth Orbit          &     \ref{Applications}, \ref{ProofOfPrincipleImlementation}                                         \\
LIV         & Lorentz Invariance Violation                          &        \ref{Fundamental}                   \\
LLI        & Local Lorentz Invariance                           &        \ref{Fundamental}            \\
LPI        & Local Position Invariance                             &        \ref{Fundamental}         \\
MEMS       & Micro- Electro- Mechanical   Systems                   &        \ref{Applications}        \\
MEO        & Medium Earth Orbit                                 &      \ref{Applications}, \ref{ProofOfPrincipleImlementation}              \\
MW         & Micro Wave               &       \ref{Applications}                                       \\
NASA       & National Agency Space   Administration                 &       \ref{Fundamental}, \ref{ProofOfPrincipleImlementation}         \\
NV         & Nitrogen Vacancy                                &         \ref{Applications}              \\
OAM        & Orbital Angular Momentum                               &       \ref{Applications}         \\
QKD        & Quantum Key Distribution                           &       \ref{Fundamental}, \ref{Applications}, \ref{ProofOfPrincipleImlementation}             \\
QRNG       & Quantum Random Number Generator                        &       \ref{Applications}         \\
SHM        & Space Hydrogen Maser   &\ref{Fundamental}                                                                    \\
SM         & Standard Model          &          \ref{Fundamental}                                     \\
SM3        &   Science Module 3                         &               \ref{Fundamental}, \ref{ProofOfPrincipleImlementation}                             \\
SME        & Standard Model Extension                             &       \ref{Fundamental}           \\
SNSPD      & Superconducting Nanowire Single   Photon Detectors   &      \ref{Applications}            \\
SPAD       & Single Photon Avalanche Diode                         &         \ref{Applications}        \\
STEP       & Satellite Test of the   Equivalence Principle  &        \ref{Fundamental}               \\
WCP-DS     & Weak Coherent Pulse Decoy State                        &         \ref{Applications}       \\
WEP        & Weak Equivalence Principle                   &        \ref{Fundamental}                  \\
WIMP       & Weakly Interacting Massive   Particle     &    \ref{Fundamental}                         \\
ZARM       &   Zentrum f\"ur Angewandte Raumfahrttechnologie und Mikrogravitation                        &     \ref{Fundamental}, \ref{ProofOfPrincipleImlementation}                                   \\
\hline\hline
\end{longtable}

\end{document}